\documentclass[aps,prx,reprint,showpacs,superscriptaddress,longbibliography]{revtex4-1}
\usepackage[english]{babel}
\usepackage{amsmath,amssymb,bbm,graphicx,color,comment,txfonts}
 \usepackage{braket}
\usepackage{bbold}
\usepackage[bookmarks=true,colorlinks,citecolor=blue,urlcolor=blue]{hyperref}
\usepackage{dsfont}
\usepackage[normalem]{ulem}

\newcommand{\mic}[1]{\textcolor{black}{ #1}}

\DeclareMathOperator{\tr}{tr}

\newcommand{\eq}[2]{\begin{eqnarray}\label{#1} #2 \end{eqnarray}}

\begin{document}

\date{\today}

\title{Effective Theory for the Measurement-Induced Phase Transition of Dirac Fermions}
\author{M. Buchhold}
\affiliation{Institut f\"ur Theoretische Physik, Universit\"at zu K\"oln, D-50937 Cologne, Germany}
\author{Y. Minoguchi}
\affiliation{Vienna Center for Quantum Science and Technology, Atominstitut, TU Wien, 1040 Vienna, Austria}
\author{A. Altland}
\affiliation{Institut f\"ur Theoretische Physik, Universit\"at zu K\"oln, D-50937 Cologne, Germany}
\author{S. Diehl}
\affiliation{Institut f\"ur Theoretische Physik, Universit\"at zu K\"oln, D-50937 Cologne, Germany}
\begin{abstract}
A wave function \mic{subject to unitary time evolution and} exposed to measurements undergoes pure state dynamics, with deterministic unitary and probabilistic measurement-induced state updates, defining a quantum trajectory. For many-particle systems, the competition of these different elements of dynamics can give rise to a scenario similar to quantum phase transitions. To access \mic{this competition} despite the randomness of single quantum trajectories, we construct an $n$-replica Keldysh field theory for the ensemble average of the $n$-th moment of the trajectory projector. A key finding is that this field theory decouples into one set of degrees of freedom that heats up indefinitely, while $n-1$ others can be cast into the form of pure state evolutions generated by an effective non-Hermitian Hamiltonian. This decoupling is exact for free theories, and useful for interacting ones. In particular, we study locally measured Dirac fermions in $(1+1)$ dimensions, which can be bosonized to a monitored interacting Luttinger liquid at long wavelengths. For this model, the non-Hermitian Hamiltonian corresponds to a quantum Sine-Gordon model with complex coefficients. A renormalization group analysis reveals a gapless critical phase with logarithmic entanglement entropy growth, and a gapped area law phase, separated by a Berezinskii-Kosterlitz-Thouless transition. The physical picture emerging here is a \mic{ measurement-induced pinning of the trajectory wave function into eigenstates of the measurement operators, which succeeds upon increasing the monitoring rate across a critical threshold.}
\end{abstract}

\maketitle
\section{Introduction}
In quantum mechanics, there are two fundamentally distinct dynamical evolutions: First, a pure state can evolve deterministically according to the Schrödinger equation, with dynamics generated by a Hamiltonian operator  $\hat H$. Second however, it can be updated in a stochastic fashion, when the quantum system is subject to observation. In the case of a strong, projective measurement of an observable $\hat M$, the wave function abruptly collapses into one of the eigenstates $\ket{m}$ of $\hat M$, with a probability determined by the overlap of the state before the measurement with $\ket{m}$. If the Hamiltonian commutes with the measurement operator, after a single collapse into a certain state $\ket{m}$ the system will be confined to it indefinitely. In contrast, if $[\hat H, \hat M]\neq 0$, generically the competition of $\hat H$ and $\hat M$ will not allow the evolution to come to rest. 

In a many-body context, such a competition of two operators can give rise to fundamental macroscopic phenomena such as quantum phase transitions. In this case, two non-commuting terms, e.g. kinetic and potential energy, each separately stabilize ground states with macroscopically distinct properties. While finite systems can only undergo a gradual change of properties upon tuning the ratio of competing energy scales, in the thermodynamic limit a phase transition separating  qualitatively distinct phases of matter will occur.  In this light, it is a natural question whether a many-body system undergoing competing Hamiltonian and measurement dynamics likewise may undergo an abrupt change in behavior, and which quantities may host this information. This question was answered in the affirmative in \cite{Skinner2019,Fisher2018,Jian2020,Bao_2020} for projectively measured random unitary circuits. In these setups, the dynamics of a one-dimensional spin chain is generated either by randomly selected quasilocal entangling unitary gates, or by quasilocal measurements. Since the sets of operators of the entangling gates and the measurements do not commute, a competition is realized, with a strength tuneable via the ratio of applied unitaries per time unit vs. applied measurements per time unit.

 For Haar random unitaries, moderate size numerics \cite{Chan2019,Skinner2019,choi2019} and analytical studies \cite{Jian2020,Bao_2020} have been conducted. Choosing time evolution randomly from the Clifford- and the measurement operators from the Pauli-group, efficient numerical analysis of even large systems is possible \cite{Li18,Li2019b,ippoliti2020}. The two extreme cases are clearly distinct in their dynamics: The unitarily evolving circuit is characterized by unbounded growth of the entanglement entropy with system size (volume law)~\cite{Calabrese_2005,Hosur_2016,Nahum2017,Zhou19,Kim2013,jonay2018,Akhtar20,Bertini2020}. On the other hand, for any random initial state local measurements of the spins e.g. in the $z$-basis will collapse the state into a pure product state of some configuration of the $z$ projection of the spins, thus characterized by a saturation of the entanglement entropy to an area law behavior. It was shown in Refs.~\cite{Nahum2017,Fisher2018} that the respective entanglement growth averaged over the ensemble of trajectories is a good witness for the phase transition between volume and area law growth at a finite competition ratio between unitary and measurement dynamics. 

This discovery has sparked significant research on the nature of this transition, its proper description, and its generality in terms of models hosting such behavior~\cite{Zabalo2020,zhang2020,biella2021manybody,Tang2020,fuji2020}. The entanglement growth quantifier suggests a physical picture in terms of information scrambling vs. information localization. This has been made more precise in \cite{choi2019,fan2020,Fidkowski20,li2020stat}, which explain the robustness of the volume law phase within a quantum error correction picture, where the fast information spreading is protected from errors realized as the readout of the measurements. Giving up the focus on pure state evolution Ref.~\cite{Gullans2019}
characterizes combined unitary and measurement dynamics by its potential to purify a maximally mixed initial state. Alternatively to random circuits, models of fermions hopping on a one-dimensional lattice and exposed to local density measurements have been suggested~\cite{Cao2019}. These do not show a volume to area law transition \cite{Cao2019}, but rather a transition from a critical phase with a logarithmic scaling of the entanglement entropy to an area law \cite{Alberton20} (the transition can also be found in non-unitary circuit models \cite{chen2020,li2020conformal,jian2020criticality}). For free fermion models, a general correspondence between non-unitary circuit dynamics and unitary but random Hamiltonian dynamics in $(d+1)$ dimensions has been shown to enable a classification of the measurement dynamics in terms of symmetries \cite{jian2020criticality}. 
These results suggest a finer structure in the phenomenology of measurement-induced phase transitions. 

In this work, we approach the measurement-induced phase transition from the perspective of non-equilibrium quantum statistical mechanics, asking in particular what the proper degrees of freedom are to capture it. \mic{Once this is done, we discuss how the previously established phenomenology of the transition, e.g., the behavior of the entanglement entropy and the purification dynamics, can be understood from a statistical mechanics viewpoint.} To this end, we develop a replica field theory approach for a minimal model of Dirac fermions in one spatial dimension, undergoing continuous measurements. This model enables  an alternative representation via  bosonization at long wavelength, amounting to a measured non-linear Luttinger liquid. In terms of complexity, the model is thus comparable to free fermion problems with disorder; upon integrating out the measurement noise, the problem becomes formally, and for all practical means, an interacting one. In our analysis, we are guided by the physical picture of a pinning or localization transition upon increasing the measurement strength: On the level of a single, pure state quantum trajectory, the delocalization due to kinetic energy competes with pinning of fermions due to the measurement-induced collapse of the wave function.

Care has to be exercised when taking the ensemble average. The binary measurement outcomes on each point in space lead to an extensive \textit{configurational} entropy. The averages that are usually considered, which are linear in the state, correspond at long times to infinite temperature states and mask the transition in the trajectory ensemble. \mic{In contrast, averages of non-linear functions of the state, such as the celebrated entanglement entropy or connected correlation functions $C_{O_i,O_j}=\langle O_i O_j\rangle-\langle O_i\rangle\langle O_j\rangle$ for some observables $O_{i,j}$ (the second contribution is nonlinear in the state), do witness a transition. We demonstrate this on our concrete model: The trajectory averaged $n$-replica state hosts one structureless 'hot' mode, but also $n-1$ 'cold' modes, which undergo a quantum phase transition in (1+1) dimensions. The hot mode absorbs all of the configurational entropy, and exclusively determines any linear average, such as $\langle O_i O_j\rangle$. All linear averages are therefore featureless. Nonlinear averages, however, such as $C_{O_i,O_j}$, also depend on the cold modes, which are free of configurational entropy, and thus reveal the features of the measurement-induced evolution.} We expect the structures revealed here to be useful for even broader classes of measurement-induced phase transitions.

Applied to the problem at hand, it allows us to make progress in terms of the bosonized theory. In particular, we distill an effective non-Hermitian  sine-Gordon model for density fluctuations, and show that it undergoes a Berezinskii-Kosterlitz-Thouless (BKT) pinning transition. Its gapless phase is protected by current conservation. The pinning induced by measurements, into the eigenstates of the measurement operators, drives a gap opening. As witnesses of the transition, we compute the 2-replica correlation functions showing algebraic and exponential behavior in the gapless and in the gapped phases, respectively. Furthermore, we compute the entanglement entropy in the weakly and strongly monitored regimes, showing logarithmic growth and saturation with system size $L$, respectively. A measurement rate dependent effective central charge, appearing as the coefficient of the $\log(L)$ growth term, is found in the gapless phase. This scenario is consistent with previous numerical results for a related model of monitored lattice fermions~\cite{Cao2019,chen2020,Alberton20,bao2021symmetry}. A power law decay of the effective central charge $c(\gamma)\sim \gamma^{-\kappa}$ is confirmed. The value $\kappa\ge0.5$ is bounded from below but non-universal, continuously varying with system parameters. We believe that such a pinning mechanism, and the BKT universality class itself, could underlie broader classes of measurement-induced phase transitions in $1+1$ dimensions. 

Our field theory approach complements previous quantum information theoretic approaches to measurement-induced phase transitions discussed above. We also note that mappings to statistical mechanics models have been proposed previously based on a tensor network approaches to the dynamics~\cite{Bao_2020,Jian2020,Skinner2019,nahum2020long}. What sets our approach apart from these works is that we directly identify the physically relevant degrees of freedom -- the above mentioned 'cold' modes made of monitored fermions -- which are at the root of the transition. The behavior of entanglement entropies then constitutes an important hallmark of it, but is not the only signature -- as exemplified by our analysis of replica correlation functions, likewise serving to detect the transition.

We identify several observables that have been determined in previous works, and which confirm our findings. In addition, we provide potential quantifiers, which establish a relation to other setups, including relaxation dynamics and a purification time scale~\cite{Gullans2019}.

{\it Structure of the paper:} In the following section, we introduce our setup and provide a summary of our main results.
In Section~\ref{sec:thback}, we discuss general properties of the measurement model and draw the temporal and spatial continuum limit, introducing a field theory for continuously monitored Dirac fermions. In Section~\ref{sec:RepFielTheory}, a two-replica field theory for a general continuous monitoring setup is derived and then applied to our concrete setup. It provides the basis for the identification of the relevant degrees of freedom at the measurement-induced phase transition. Then, in Section~\ref{sec:RG}, we perform a renormalization group analysis of the phase transition. In Section~\ref{sec:nkeldy}, the replica theory is extended to an arbitrary number of replicas $n$ in a Keldysh real-time path integral framework. This serves as a basis for the computation of higher order correlation functions, and we include a discussion of the Rényi entropy and the von Neumann entanglement entropy. We conclude in Section~\ref{sec:conclusio}. Several technical details of our analysis are presented in the appendix.

\section{Extended Synopsis and Key Results}

\textit{Setup} -- Our model for the measurement dynamics in one spatial dimension is based on the stochastic Schrödinger equation for the trajectory wave function $\ket{\psi_t}$ \cite{Dalibard92,Dum92,Molmer1992}, or alternatively the conditioned trajectory projector $\hat\rho^{(c)}_t = \ket{\psi_t}\bra{\psi_t}$ \cite{Belavkin_1989,Carmichaelbook,Jacobs_2006}, which amounts to a continuous time description of measurements. \mic{Throughout this work, we focus on continuous-time measurements. The phenomenology of a measurement-induced phase transition, which we establish for such continuous measurements, however,  includes the case of projective measurements. In that case, the continuous-time evolution can be understood as an effective, coarse grained description, which is valid on time scales on which many projective measurements have been performed. We discuss this aspect in more detail below. }

The competition between unitary dynamics  and measurements is set by the dimensionless parameter \mic{
\begin{equation}
    g=\gamma/J, \text{ with }\left\{\begin{array}{cl}J & \text{: kinetic energy} \\ \gamma & \text{: measurement rate}\end{array}\right.. 
\end{equation} }Here $J$ describes the kinetic energy of the Hamiltonian, and $\gamma$ the rate at which measurements are performed. This formulation is able to capture the collapse of the wave function characteristic of physical measurements. Previous work on measurement-induced phase transitions has mainly focused on the case of projective (discrete) measurements \cite{Skinner2019,Fisher2018,nahum2020long, Bao_2020, bao2021symmetry}; however, the continuous measurement scenario is smoothly connected to the latter. 

Phase transitions generally are rooted in a competition of incompatible terms. In the context of the measurement-induced phase transitions, the competition is realized by the entangling unitary dynamics, and a disentangling measurement dynamics. In contrast, \mic{both continuous as well as discrete (projective)} measurements of the local density lead to the same long time limit. These two types of measurement dynamics can be smoothly interpolated to one another~\cite{Jacobs_2006,bookJacobs}, but since there is no competition realized along this interpolation axis, one should expect no additional phase transition. Indeed, recent works
~\cite{Schomerus2019,Romito2020} have established that the transition persists upon interpolating between the limiting cases of strong projective and weak continuous measurements in random circuits. For our purpose of constructing an analytically tractable model for the competing dynamics, the continuous time formulation proves advantageous.

\textit{Quantifying the phase transition} -- 
\mic{As one may expect quite naturally,} continuously measuring a set of local, mutually commuting operators $\hat O_j$, e.g., the local particle density on some lattice site $j$, pins the system onto their eigenstates. Competing with this is the unitary dynamics, which leads to the spreading of the local operators via particle propagation, and therefore to a depinning from eigenstates of the measurement operators. \mic{The crux of the matter is what type of behavior follows from this competition and when are the two limiting cases of eigenstate (de-)pinning observed?}
The competition of two non-commuting operators driving a phase transition must thus be encoded in the evolution of the pure state wave function, similar to a quantum phase transition in a ground state problem. Importantly however, the pure state considered here corresponds to a random variable, due to the stochastic nature of measurements. To acquire information independent of this randomness, suitable ensemble averages need to be considered. Taking the statistical average over the trajectory ensemble (which we denote with $\overline{.\phantom{l}.\phantom{l}.}$) in the way familiar from quantum statistical physics for instance yields 
\begin{eqnarray}
\overline{\braket{\psi_t | \hat O_i \hat O_j|\psi_t}} = \overline{\tr [ \hat O_i \hat O_j \hat\rho_t^{(c)} ]} = \tr [ \hat O_i \hat O_j \overline{\hat\rho_t^{(c)} ]} ,
\end{eqnarray} 
where $\overline{\hat\rho_t^{(c)}}$ is the trajectory averaged density matrix. For a measurement dynamics, however, $\overline{\hat\rho_t^{(c)}}=\frac{1}{\mathcal{N}}\mathds{1}$, which corresponds to evaluating the product of the measured operators $\hat O_i \hat O_j$ in an infinite temperature state -- thus the correlations will be trivial except for $g=0$, where there is no averaging. The masking effect of such trajectory ensemble average can be mitigated by considering correlation functions which are non-linear in the state -- such as, for example, the equal-time product $\braket{\psi_t | \hat O_i|\psi_t}\times\braket{\psi_t | \hat O_j|\psi_t}$ for the same stochastic wave function $|\psi_t\rangle$ in both quantum mechanical expectation values. We note that the entanglement entropy falls into this class of quantifiers which are non-linear in the state, involving even infinite powers in the case of the von Neumann entropy. 

In this work, we focus on the non-linear connected covariance matrix for the measured operators \begin{eqnarray}\label{eq:covi}
C_{ij} (t) =\overline{\braket{\psi_t | \hat O_i \hat O_j|\psi_t}}- \overline{\braket{\psi_t | \hat O_i|\psi_t}\braket{\psi_t | \hat O_j|\psi_t}}
\end{eqnarray}
in the limit $t\to\infty$. Here, the second part $\sim \overline{\braket{\psi_t | \hat O_i|\psi_t}\braket{\psi_t | \hat O_j|\psi_t}}$ is nonlinear in $\hat\rho^{(c)}$ and therefore does not correspond to an infinite temperature average. It is particularly well suited to distinguish between strong monitoring, where the system  remains close to an eigenstate of the operators $\hat O_j$ (such that $C_{ij}$ is short-ranged or even zero for $i\neq j$), and weak monitoring, where longer-ranged correlations between the measured operators can be established by the Hamiltonian. 
Indeed, such a scenario is realized in our model. For $g\ll 1$, the problem becomes linear in the bosonic formulation. The correlator can be computed analytically, demonstrating an algebraic decay $C_{ij}(t) \sim |i-j|^{-2}$. In the opposite regime $g^{-1}\ll 1$, the pinning of particle density due to the measurements cuts off this correlation function more severely. In the measurement-only limit  $g^{-1} = 0$, the wave function evolves into a product state of onsite occupations $0,1$ of the Dirac fermions with autocorrelations only. In the bosonized language, this leads to a gap opening, giving rise to an effective massive bosonic theory with exponentially decaying correlations. This behavior is backed up by a complementary perturbative calculation for $g^{-1} \ll 1$ of the correlation function in a fermionic lattice model.

The two opposite regimes thus show behaviors which are qualitatively different from each other, pointing at the existence of a phase transition at a finite competition $g_c$.

\textit{Replica approach} -- The ability to capture the qualitative behavior for finite values of $g$ motivates us to study the bosonic non-linear Luttinger model in more detail. To this end, we find it convenient to work in a replica formalism. We introduce a product of replicated, conditioned density matrices $\hat\rho^{(c)}_t\otimes\hat\rho^{(c)}_t$, which allows us to rewrite
\begin{eqnarray}
C_{ij}(t)&=& \overline{\text{Tr}(\hat O_i \hat O_j\hat\rho_t^{(c)})\text{Tr}\hat\rho^{(c)}_t}-\overline{\text{Tr}(\hat O_i\hat\rho^{(c)}_t)\text{Tr}(\hat O_j\hat\rho^{(c)}_t)}\nonumber\\
&=& \frac{1}{2}\text{Tr} \left[(\hat O_i^{(1)}-\hat O_i^{(2)})(\hat O_j^{(1)}-\hat O_j^{(2)})\overline{\hat\rho^{(c)}_t\otimes\hat\rho^{(c)}_t}\right]\label{Eq:Eq3}
\end{eqnarray}
with $\hat O_j^{(1)}=\hat O_j\otimes\mathds{1}$ and $\hat O_j^{(2)}=\mathds{1}\otimes\hat O_j$.
This formulation offers two key advantages: First, it encodes the correlation information in an expectation value linear in the trajectory averaged, replicated state. We then construct a quantum master equation directly for $\hat\rho^{(R_2)}_t \equiv \overline{\hat\rho^{(c)}_t\otimes\hat\rho^{(c)}_t}$. It features a non-linear state dependence, which however becomes unimportant at late times. As a net result, we obtain an evolution linear in the replicated, averaged state $\hat\rho^{(R_2)}_t$. We emphasize that we do not consider quenched disorder in our system, and statistical averages are exclusively reserved for taking the average over random measurement trajectories.

Second, it allows us to introduce new degrees of freedom which capture the physics of the transition more transparently. In the bosonic model, the replicas are acted up on by the measured operators $\hat O^{(l)}=f(\hat \phi_x^{(l)})$ with $l=1,2$ and $f$ either a linear or a nonlinear function of the bosonic operator $\hat \phi_x^{(l)}$, which is associated to density fluctuations. For a linear problem (the problem can be linearized close to $g, g^{-1}=0$), we find that the replicas can be decoupled in terms of center-of-mass (or average) and relative coordinates ${\hat\phi_x^{(a,r)}=\tfrac{1}{\sqrt{2}}}(\hat\phi_x^{(1)}\pm\hat\phi_x^{(2)})$. The center-of-mass coordinates follow an evolution which heats them up indefinitely, while the relative ones obey a Schrödinger-type evolution under an effective non-Hermitian  Hamiltonian, $\ket{\psi^{(r)}_t} = e^{-i\hat H_\text{eff} t}\ket{\psi_0^{(r)}}$. The physical reason for this behavior is that all the replicas experience the same noise realization; therefore, the sum variable (the center-of-mass) collects all the noise-induced fluctuations, while the noise cancels in the difference variable. 

Including the non-linear terms, the decoupling is no longer exact. However, assuming that the center-of-mass degrees of freedom still heat up to infinity irrespectively of the behavior of the relative ones, the center-of-mass degrees of freedom can be traced out. The remaining relative coordinate evolution still takes the above form, and  can be written as a path integral governed by the action of a non-Hermitian  sine-Gordon model,
\begin{eqnarray}
S = \int_X \frac{1}{16\pi}[(\partial_t\phi^{(r)}_X)^2-\eta^2(\partial_x\phi^{(r)}_X)^2]-i\lambda \cos\phi^{(r)}_X],
\end{eqnarray}
where we combined $X=(x,t)$, and with $\eta$ being complex and $\lambda$ real. We derive the flow equations for this model, and find that it exhibits a BKT type phase transition as a function of varying the competition ratio $g$, which enters the coupling constants $\eta(g), \lambda(g)$. This places the measurement-induced phase transition into the BKT universality class. 

This result may be rationalized by the fact that a gapless phase in $(1+1)$ dimensions quite generically is left via the BKT mechanism by the generation of a mass \cite{SonStephanov}. Here the gapless phase is protected by current conservation. Indeed, $\hat \phi$ in the original model describes density fluctuations, resulting in gapless relative modes  $\hat\phi^{(r)}$ at weak monitoring. Upon increasing the measurement strength, \mic{i.e., the rate at which measurement readouts are performed and the amount of information that is extracted per readout,} these modes generate a mass, and thus get pinned.

We also study the generalization from two to $n$ replicas within a multicontour Keldysh approach, representing $Z(n) = \overline{\tr \hat\rho^{(c) n}}$ in terms of a functional integral with $2n$ contours -- two for each copy of the state, as in the usual Keldysh path integral ($n=1$). This is the generating functional for all correlation functions involving $n$ powers of the quantum trajectory projector. The noise average then generates correlations between all the copies of the state. In this framework, and for general Gaussian problems in the long time limit, we find that there is one center-of-mass mode (the symmetric superposition, or Fourier mode $k=0$) which heats up indefinitely, while there are $n-1$ orthogonal ones which do not couple the contours within one copy, and evolve towads a pure state, generalizing the relative coordinate identified above.
We then discuss how the phase transition found at $n=2$ may be generalized to arbitrary $n$ within an Abelian bosonization framework.

\textit{Entanglement entropy in $n$-replica Keldysh approach} -- Finally, we embark on a calculation of the entanglement entropy. To this end, we consider a bipartition of the system into a subsystem $A$ of length $L$ and $\mathds{R}\setminus A$ which is traced out, and decompose the partition function of the subsystem $Z_A(n) = \overline{\tr \hat\rho_A^{(c) n}}$ into a noiseless and a noisy contribution. For free theories, including the Gaussian fixed points of our measurement theory, we show that the noiseless contribution to entanglement entropy exclusively determines the scaling with the bipartition size, while the noisy contribution is independent of the bipartition and amounts at most to a size independent constant. In the gapless regime we find
\begin{eqnarray}
S_{\text{vN}}(L)=\tfrac{1}{3}c(\gamma)\log(L),
\end{eqnarray}
with a $\gamma$-dependent \textit{effective} central charge $c(\gamma)$. It depends on the renormalization group flow of the non-Hermitian sine-Gordon theory. In the non-monitored limit, $c(\gamma \to 0) \to 1$. This reproduces the central charge of a ground state of free massless Dirac fermions.  For weak measurements, $c(\gamma)$ decays monotonously, and drops to zero at the transition point.
In contrast, in the massive regime $g^{-1} \ll 1$, the entanglement entropy saturates to an area law. Only for $\gamma=0$ a thermal initial state produces a volume law entanglement entropy.

\section{Measurement model}\label{sec:thback}

In this section, we construct a model for measurements in many-body systems in the temporal and spatial continuum limit. This model captures the essence of measurement-induced phase transitions, realizing non-commuting operators by means of a kinetic energy Hamiltonian competing with exactly local measurements, and lends itself to a field theory analysis.

\subsection{Temporal continuum limit: Continuous measurements in quantum state diffusion approach}

\emph{Projective vs. continuous measurements} -- Here we describe the framework of continuous measurements that we utilize to model the monitored fermion and boson dynamics. Compared to discrete projective measurements, the continuous measurements are often referred to as weak measurements. Broadly speaking, continuous measurements correspond to the limit where the amount of information extracted per measurement (in the sense of an incomplete readout), and the duration of each measurement, are reduced to zero simultaneously, in a way that their ratio $\gamma$ is kept fixed~\cite{Jacobs_2006, Milburn1993, bookWisemanMilburn}. This comes with the advantage of enabling us to straightforwardly take the continuum limit in time, which in turn is useful for mapping the problem into a field theory in Sec. \ref{sec:RepFielTheory}. In the following we refer to this type of measurement exclusively as 'continuous measurement' or 'monitoring' and reserve the terms 'weak' and 'strong' measurements (or monitoring) to situations where $\gamma$ is small or large compared to the scale of the Hamiltonian. As already pointed out above, continuous measurements are expected to yield the same coarse grained dynamics as their projective counterparts on the time scales of many measurements. Especially they do not exclude the existence of a phase transition~\cite{Schomerus2019, Romito2020}. In fact, we can replace in the competition ratio from the framework of discrete unitary circuits and projective measurements
\begin{eqnarray}
g = \frac{\#\text{projective measurements/time}}{\# \text{unitaries/time}}   \to \frac{\gamma}{J},
\end{eqnarray}
where $\gamma$ is the time scale of the monitoring and $J$ the scale of the Hamiltonian, which can take arbitrary values. Accordingly, we will distinguish \textit{weak monitoring} $\gamma\ll J$ and \textit{strong monitoring} $\gamma\gg J$. Indeed, Ref.~\cite{Schomerus2019} has demonstrated that the existence of a phase transition in random circuits upon increasing $g$ is preserved all along the way from strong projective to weak continuous measurements. 

\emph{Stochastic Schrödinger equation} -- The starting point of our approach is a stochastic Schrödinger equation in the quantum state diffusion framework~\cite{bookWisemanMilburn, Carmichaelbook, Gisin_1992, Milburn1993}. It corresponds to a  monitoring protocol in which the expectation value of an operator $\langle \hat O_l\rangle_t$ is continuously measured, for instance a homodyne detection scheme~\cite{PhysRevLett.120.133601,PhysRevA.98.023852,Milburn1993}. In the quantum state diffusion, the evolution equation $d\ket{\psi_t}=|\psi_{t+dt}\rangle-|\psi_t\rangle$ for the pure state, normalized quantum trajectory wave function $|\psi_t\rangle$ is, 
\begin{align}\label{eq:SSE}
d |\psi_t\rangle &=& -i dt [ \hat H - i\tfrac{\gamma}{2}\sum_i \hat M^2_{i,t}]|\psi_t\rangle + \sum_idW_i \hat{M}_{i,t}|\psi_t\rangle,
\end{align}
for which we define the {\it measurement operators}
\begin{align}
\hat M_{i,t} =& \hat O_i - \langle \hat O_i\rangle_t
\end{align}
for a general set of local measured operators $\hat O_i$, labeled by an index $i$ (e.g. a lattice site index). The dynamical update of the state $\ket{\psi_t}$ for an infinitesimal time interval $dt$ is generated by (i) a deterministic contribution due to a non-Hermitian Hamiltonian $\hat H_{\text{nH}} = \hat H - i\tfrac{\gamma}{2}\sum_i \hat M^2_{i,t}$ involving both the Hamiltonian $\hat H$ and the measurement operators. These contributions are associated with the scale $J$ for the Hamiltonian as anticipated, and  $\gamma$ for the measurements, which is implemented by the measurement operators $\hat M_{i,t}$. In addition, (ii), there is a stochastic contribution: The Wiener process $dW_i$ is a local, real valued Gaussian noise increment, with zero mean $\overline{dW_i}=0$ and variance $\overline{ dW_i d W_j} = \gamma dt \delta_{ij}$~\cite{Milburn1993,Gisin_1992, Jacobs_2006}.

The measurement operators $\hat M_{i,t}$ involve the quantum mechanical expectation value $\langle \hat O_i\rangle_t \equiv \bra{\psi_t} \hat O_i\ket{\psi_t}$ evaluated at the time before the update, describing the measurement feedback. This term makes the quantum trajectory evolution non-linear in the state of the system. It ensures preservation of the norm of the trajectory wave function: It is easily verified that all moments of the norm $\mathcal{N}(n)\equiv\overline{\langle \psi|\psi\rangle^n}$ are conserved, $\partial_t \mathcal{N}(n)=0$ for arbitrary $n$. This is an important condition for describing a physical measurement. For example, the 'raw' quantum state diffusion protocol, obtained by the replacement $\hat M_{i,t} \to \hat O_i$ discarding the feedback term, preserves the norm only on average, $\partial_t \mathcal{N}(n)\sim \gamma n(n-1)\mathcal{N}(n)$, i.e. for $n=1$, and does not qualify for the description of a physical quantum trajectory undergoing measurements.

A useful alternative representation of the stochastic Schrödinger equation utilizes the conditioned projector ${\hat\rho_t^{(c)} = \ket{\psi_t}\bra{\psi_t}}$, which evolves according to (cf. Eq.~\eqref{eq:RepIncrement} and Refs.~\cite{Jacobs_2006,bookJacobs} for details)
\begin{align}\label{eq:rhostoch}
d \hat\rho^{(c)}_t =&  dt(- i [\hat H,\hat\rho^{(c)}_t] - \frac{\gamma}{2}\sum_i [\hat O_i , [ \hat O_i , \hat\rho^{(c)}_t ]])  \\ & + \sum_i dW_i  \{\hat M_{i,t}, \hat\rho^{(c)}_t\}\nonumber ,
\end{align}
and where we used the anti-commutator $\{\hat A,\hat B\}=\hat A\hat B+\hat B\hat A$.

\emph{Connection to projective measurements} -- This becomes particularly transparent when focusing on the stochastic weak measurement evolution alone ($\hat H =0$ for the moment). Consider the long time limit $t\to \infty$ where $dW_i$ acts as a multiplicative noise (multiplied with the state dependent term $\hat M_{i,t}$), and becomes inactive only once the condition 
\begin{eqnarray}
\hat O_i \ket{\psi_t} = \braket{\hat O_i}_t \ket{\psi_t} = O_i^{(\alpha)} \ket{\psi_t}
\end{eqnarray}
is fulfilled -- in other words, once the system has reached an eigenstate of the measurement operator $\hat O_i$, and the instantaneous expectation value $\braket{\hat O_i}_t$ is equal to one of the eigenvalues $O_i^{(\alpha)}$ of $\hat O_i$ (please note that any eigenstate $|\psi_t\rangle$ of $\hat O_i$ fulfills this condition). In this case, also the double commutator in Eq. \eqref{eq:rhostoch} vanishes. Thus, this evolution describes a `continuous collapse' of an initial wavefunction into an eigenstate of the measurement operator, with the stochastic element provided by the different noise realizations. We will refer to these eigenstates of the measurement operators in a quantum optics language as \textit{dark states}, since the measurement evolution stops once the system has reached such state, and the instantaneous expectation values $\langle \hat O_i\rangle_t$ become time-independent.

\emph{Dark state configurational entropy} -- Measurements come with a large configurational entropy (as opposed to the entanglement entropy, building up due to the entangling operations along each pure state trajectory) related to the spectrum of eigenstates of the measurement operators and the their random realization in the measurement process. This is best illustrated by means of a concrete lattice model of $N$ continuously monitored spinless fermions. Consider an entangling tight-binding hopping Hamiltonian
\begin{align}\label{eq:hlatt}
    \hat H=-J\sum_{l} c^\dagger_ic_{i+1}+\text{H.c.},
\end{align}
where $c^\dagger_i,c_i$ are fermionic creation and annihilation operators on a lattice with sites $i=1 , ... , L$, and a continuous measurement implemented by monitoring the local fermion densities 
\begin{align}\label{eq:mlatt}
    \hat O_i=\hat n_i=c^\dagger_i c_i .
\end{align} 
For each of the local fermionic density measurements, there are two dark eigenstates $\ket{\sigma_i}$ with $\sigma_i = 0,1$. A random initial state will ultimately collapse into a pure product state $\ket {\{\sigma_i\} } = \prod_i \ket{\sigma_i}_i$, where the configuration $\{\sigma_i\}$ is random, subject only to the constraint $\sum_i \sigma_i = N$, the total number of fermions in the system. However, for large systems there are exponentially many such pure states, e.g. $\left(\begin{array}{c} 2 N \\ N\end{array}\right)$ at half filling. Of course, this reasoning generalizes to the quantum non-demolition case, where $[\hat H, \hat M_i]=0$ for all $i$; however, the non-commuting hopping Hamiltonian will generate competition between measurement and Hamiltonian evolution, and generically prevent the system from reaching an eigenstate of the measurement operators. Indeed, this model exhibits a measurement-induced phase transition, established in Ref.~\cite{Alberton20} based on numerical simulations. 

The large configurational entropy is seen even more directly upon taking the statistical average over the noise realizations. According to the \^Ito calculus, the noise $dW_i$ at time $t$ is uncorrelated with the state $\hat\rho^{(c)}_t$, such that the noise average of Eq.~\eqref{eq:rhostoch} yields the Lindblad quantum master equation, i.e. the evolution equation for the unconditioned density matrix $\hat\rho_t \equiv \overline{\hat\rho^{(c)}_t}$, 
\begin{eqnarray}\label{eq:lindi}
d \hat\rho_t =  dt(- i [\hat H,\hat\rho_t] - \frac{\gamma}{2}\sum_i [\hat n_i , [ \hat n_i , \hat\rho_t ]]).
\end{eqnarray}
The right hand side indeed describes a completely positive Lindblad operator, for the special case of Hermitian  Lindblad jump operators $\hat n_l$ (generalized measurements using non-Hermitian  operators, and subsequently general Lindblad equations, can be described in the POVM (positive operator valued measurement) framework, cf. \cite{NielsenChuang}). Clearly, while $\hat\rho^{(c)}_t$ describes a pure state, $\hat\rho_t$ represents a mixed state ensemble of pure states. Equation \eqref{eq:lindi} is a deterministic equation which is linear in the state $\hat\rho_t$, as required by the foundations of probability theory for the generator of motion for any complete statistical representation of a system \cite{Primas90}.

In terms of a physical interpretation, the Lindblad equation corresponds to unread (or averaged over) measurements. This averaging over measurement outcomes introduces the extensive configurational entropy, e.g. $S = \log_2(\frac{2N!}{N!^2}) \to 2N$ for a large, half filled system. For $H\neq 0$, the Lindblad equation \eqref{eq:lindi} is solved by a totally mixed, infinite temperature state $\hat\rho \sim \mathbb{1}$ (here $\mathbb{1}$ is the identity in the Hilbert space of fixed total particle number $N$). In fact, this indefinite heating directed towards a stable, totally mixed dynamical fixed point is generic for Hermitian measurement operators competing with a Hamiltonian. This can be illustrated by considering the effect of the Hamiltonian and the Lindblad operators separately. The Hermitian Lindblad operators evolve the system towards a density matrix, which is diagonal in Fock space (or diagonal in the basis of eigenstates of general Hermitian measurement operators $\hat O_l$). The Hamiltonian, due to the non-commutativity $[H,\hat n_l]\neq 0$, induces transitions between different eigenstates and mixes different matrix elements until a totally mixed state $\sim \mathds{1}$ is reached. 

\begin{figure}
  \includegraphics[width=\linewidth]{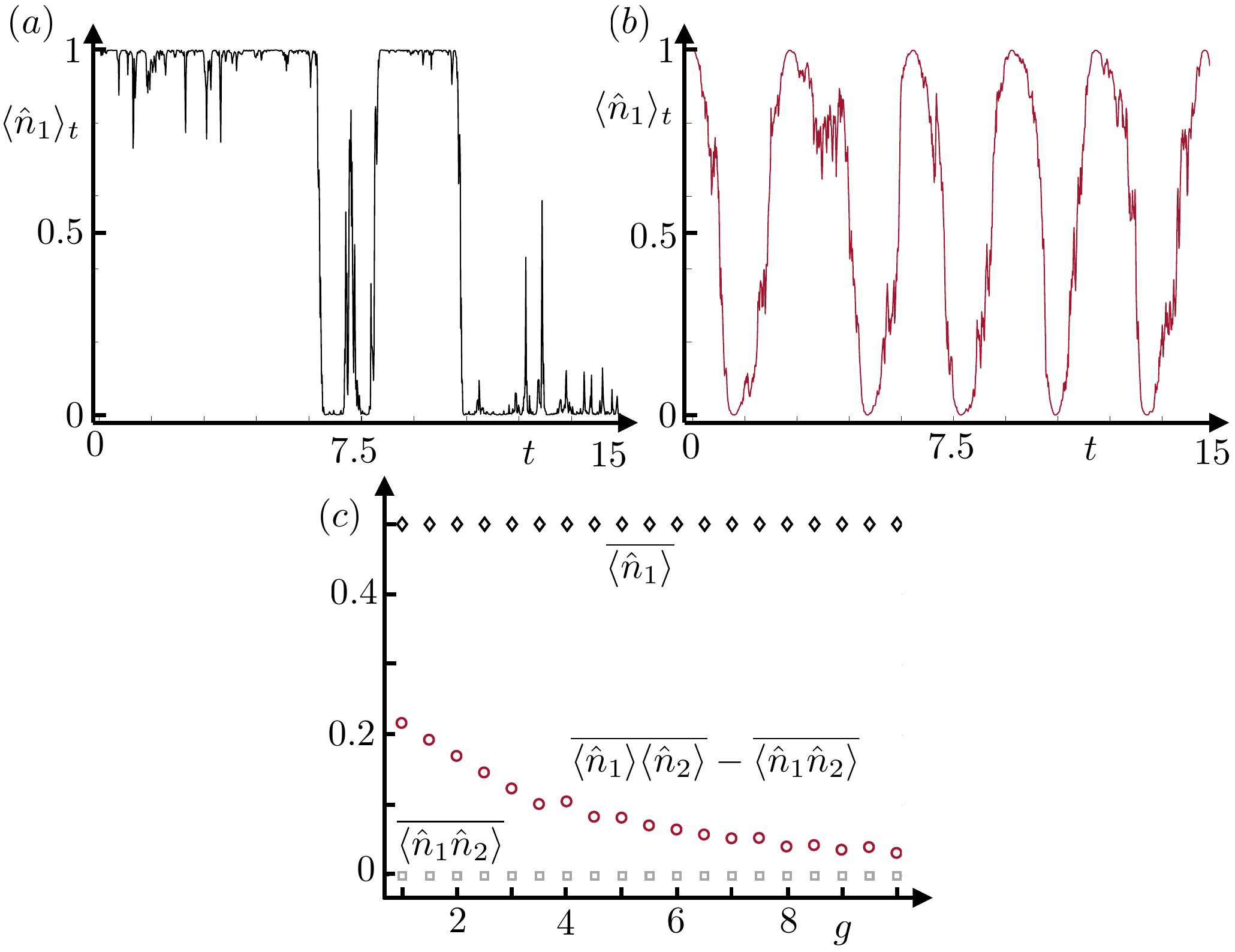}
  \caption{Illustration of competing Hamilton and measurement dynamics in the two-site toy model. (a),(b) Single trajectory dynamics: (a) For strong monitoring $g=10$ the system is pinned to dark states for most of the time, with weak, short-time fluctuations. (b) A dominant Hamilton dynamics ($g=0.5$) leads to depinning, with almost no time spent in the dark states. (c) Trajectory averaged correlation functions: Quantities linear in the state are insensitive to the competition ratio, while the correlation function nonlinear in the state witnesses a monitoring dependence, which develops a non-analytic behavior in the thermodynamic limit.}
  \label{fig-2}
\end{figure}%%%

\emph{Two-site toy model} -- Before delving into the analysis of the many-body problem, we illustrate the competition between Hamiltonian and measurement operators in a minimal toy model, a single fermion on two sites, $L=2, N=1$. The problem is equivalent to a monitored two-level system, with its Hilbert space spanned by the two states $\{\ket{\uparrow} =\ket{0,1}, \ket{\downarrow} =\ket{1,0}\}$, and $\hat H = \sigma_x$, and the operator valued part of $\hat M_{i,t}$ given by $(\mathds{1}_{2\times 2} \mp \sigma_z)/2$ for $i=1,2$, respectively. For $g^{-1} =0$ it continuously collapses into the dark states $\ket{\uparrow}, \ket{\downarrow}$, while for $g=0$, it describes Rabi oscillations. In Fig.~\ref{fig-2}(a,b) we plot the trajectory expectation value of the fermion occupation on site $i=1$. For strong monitoring $g^{-1} \ll 1$, the system is pinned for long times in the dark states, with rare jumps into the other dark state induced by the Hamiltonian, and small fluctuations in the rest periods \cite{Shelving, Shelving2}. Conversely, for weak monitoring $g\ll 1$, the evolution is still closely reminiscent of Rabi oscillations, and there is a vanishing amount of time spent in the dark states. We will be guided by this picture of depinning from the  dark states induced by the Hamiltonian dynamics in our analysis of the many-body problem. 

The toy model also illustrates the importance of a suitable choice of trajectory averaged quantities to detect the  changes in the trajectories upon varying the competition ratio $g$. Fig.~\ref{fig-2}(c) plots averages evaluated after long evolution time. The observables $\overline{\langle \hat n_1\rangle}$ and $\overline{\langle \hat n_1\hat n_2\rangle}$ are linear in the conditional state $\hat \rho_t^{(c)}$: Due to the indefinite heating, they are not sensitive to the value of $g$. In contrast, the trajectory averaged covariance matrix for the occupations 
\begin{align}
C_{12} = \overline{\langle \hat M_{1,t} \hat M_{2,t}\rangle_t} = \overline{\langle \hat n_1 \hat n_2\rangle_t} - \overline{\langle \hat n_1\rangle_t\langle \hat n_2\rangle_t}
\end{align}
is non-linear in the state and tracks a quantitative change as function of $g$, rooted in the non-commuting nature of the Hamilton operator and the measurement operators. In the small Hilbert space considered here, such changes are only of a quantitative nature, i.e., one observes a crossover instead of a sharp transition. In Sec. \ref{SecRic}, we will show that correlators of this type may behave qualitatively differently in the weak and strong monitoring regimes in a many-body system, which provides a clear indication of a phase transition at a finite competition ratio $g_c$.

\subsection{Spatial continuum limit: Measured Dirac fermions and bosonization}\label{sec:fermbos}

\emph{Measured Dirac fermions} -- The above microscopic model of measured fermions offers a measurement-induced phase transition, however in light of the heating for the average measurement a direct low energy reduction to a continuum limit may not be fully justified. Here we take a more conservative approach, and formulate our starting point directly in the spatial continuum in terms of (1+1) dimensional monitored massless Dirac model. The key ingredient shared with the lattice model is a competition between a kinetic Hamiltonian giving rise to delocalization of fermions and creating spatial entanglement, and local, mutually commuting measurements, which localize or pin the fermions, and by themselves drive the system into dark eigenstates of product form. 

The massless Dirac Hamiltonian reads in terms of the spinor $\hat\Psi_x=(\hat\psi_{R,x}, \hat\psi_{L,x})^T$
\begin{align}\label{eq:hfcont}
    \hat H=i v \int_x \hat\Psi^\dagger_x \sigma_z\partial_x\hat\Psi_x,
\end{align}
where $\sigma_z$ is the Pauli matrix. We choose two distinct local measurement operators $\hat O_{1,x}, \hat O_{2,x}$, 
\begin{align}\label{eq:mfcont}
\hat O_{1,x}=\Psi^\dagger_x\Psi_x = \hat J^{(0)}_x, \quad \hat O_{2,x} =\Psi^\dagger_x\sigma_x\Psi_x,
\end{align}
which are both measured independently but with the same rate $\gamma$. The measurement operators commute with each other, but they do not commute with the Hamiltonian, thus realizing the desired competition. They are local, and stabilize product dark states $\ket{\sigma_{a,x}} = \prod_{a,x} \ket{\sigma_{a,x}}_x$ with $a=\pm$, and $\sigma_{a,x}= 0,1$ the occupations in the basis where $\sigma_x$ is diagonal. 

The associated stochastic Schrödinger equation then features a non-Hermitian Hamiltonian,
\begin{align}
    \hat H_{\text{nH}}= \hat H -  i\tfrac{\gamma}{2}\sum_{s=1,2}\int_x \hat M^{2}_{s,x,t},\label{Eq:twomeas}
\end{align}
with $\hat M_{s,x,t} =\hat O_{s,x} - \braket{\hat O_{s,x}}_t$ and uncorrelated noise increments $\overline{dW_{s,x}dW_{s',x'}}=\delta_{s,s'}\delta(x-x')$ for both measurements. In order to characterize the transition we consider the correlation functions 
\begin{align}
    C_{y}=\overline {\langle \hat M_{x,t} \hat M_{x+y,t}\rangle}, \ \hat M_{x,t}=\hat M_{1,x,t}+\hat M_{2,x,t}\label{Eq:Cor}
\end{align} with the symmetric sum of the measurement operators.

We note that this Hamiltonian and the measurement operators would obtain in a naive low energy approach (at half filling), where the continuum limit is performed for the lattice model defined with Eqs.~(\ref{eq:hlatt},\ref{eq:mlatt}), and the continuum fermion field is decomposed as $\hat\psi_x=\hat\psi_{R,x} e^{i2\pi \rho_0x}+\hat\psi_{L,x} e^{-i2\pi\rho_0x}$, where $\rho_0$ here indicates the average fermion density. At half-filling, $\rho_0=0.5$, the sum $\hat M_{x,t}$ in Eq.~\eqref{Eq:Cor} thus corresponds to the local fermion density $\hat\psi^\dagger_x\hat\psi_x$. For the main \mic{part} in this paper, we will not rely on this connection, and consider the continuum model in autonomy. Yet, we will discuss striking phenomenological parallels to lattice models with continuous and stroboscopic measurements in Sec.~\ref{sec:ConnecttoNum}.

\emph{Bosonization} -- The monitored Dirac model lends itself to  bosonization, where the fermion bilinears are replaced by boson fields.
The resulting bosonic model is well suited for a further practical evaluation. The Hamiltonian maps to the Luttinger Liquid Hamiltonian 
\begin{align}\label{SEq10}
\hat H=\frac{v}{2\pi}\int_x [ (\partial_x\hat{\theta}_x)^2+(\partial_x\hat{\phi}_x)^2 ].
\end{align}
It is quadratic in the Hermitian operator $\hat{\phi}_x$ associated to density fluctuations, and its conjugate $\hat{\theta}_x$ connected to phase fluctuations. These operators fulfill the canonical commutation relations $[\partial_x\hat{\theta}_x,\hat{\phi}_{x'}]=-i\pi\delta(x-x')$. The measurement operators transform into 
\begin{align}\label{eq:mbcont}
    \hat O_{1,x}=-\frac{1}{\pi}\partial_x\hat \phi_{x}, \quad \hat O_{2,x}= m\cos(2\hat\phi_{x})
\end{align}
with a 'mass' $m=\mathcal{O}(1)$ that depends on the normal ordering prescription. Clearly, also in bosonized language $\hat O_{s,x}$ do not commute with $\hat H$. The measurement operators both are functions of the operator $\hat \phi_x$ and therefore pairwise commuting $[\hat O_{\alpha,x},\hat O_{\beta,x'}]=0$ for $\alpha,\beta\in\{1,2\}$, and thus stabilize 
the eigenstates of $\hat\phi_x$.
In the measurement-only limit (i.e., for $\hat H=0$) this gives rise to a set of dark states $\hat{M}_{s,x,t}\ket{\Psi_D}=0$, which are eigenstates of the field operator $\hat\phi_x\ket{\Psi_D}=\phi_x\ket{\Psi_D}$ for all positions $x$. The only restriction to the set of real valued eigenvalues $\{\phi_x\}$ is to match the condition for a fixed total particle number, which for periodic boundary conditions reads as $\int_x\cos(2\pi\rho_0x+\phi_x)=0$. This gives rise to an exponentially growing number of dark states with pinned field operator expectation values $\langle \hat\phi_x\rangle_t=\phi_x$, analogous to the lattice fermion dark states with a pinned fermion particle number. 

As soon as the Hamiltonian is switched on ($\hat H\neq0$), a competition arises between the measurement operator $\hat M_{2,x,t}$, which tends to push the system towards one of the measurement dark states, and the conjugate field $\hat \theta_x$ in the Hamiltonian, which favors a depinning of $\hat\phi_x$. We refer to the linear measurement operators $\hat M_{1,x.t}\sim \partial_x(\hat\phi_x-\langle\hat\phi_x\rangle_t)$ as gapless, due to their vanishing in the long wavelength limit (taking $q\rightarrow0$ in momentum space). They turn out to be compatible with the quadratic Hamiltonian, i.e. they do not drive a phase transition. Instead, their presence gives rise to a unique and well-defined stationary state with a scale invariant connected correlation function $\overline{\langle \hat M_{x,t} \hat{M}_{x+y,t}\rangle}\sim |y|^{-2}$. 

What drives the competition between the pinning and depinning of the field operator  $\hat\phi_x$ is the interplay between the local, nonlinear appearance of $\hat\phi_x$ in the measurement operators $\hat M_{2,x}$ and the conjugate fields $\partial_x\hat \theta_x$ in the Hamiltonian. As we detail below in Eq.~\eqref{SEq26}, for strong monitoring, the measurement operators can be linearized around the dark state, yielding a measurement operator $\hat M_{2,x,t}\sim m(\hat\phi_x-\langle\hat\phi_x\rangle_t)$, which does not vanish in the long wavelength limit. It causes exponentially decaying correlations $C_y$, with a correlation length that is comparable to the strong-monitoring correlation length in the fermion lattice model Eq.~\eqref{eq:CorrF}. We therefore denote the linear approximation of $M_{2,x,t}$ as a gapped measurement operator.

\subsection{Limits of weak and strong monitoring: Riccati approach}\label{SecRic}
We proceed by analyzing the connected correlation functions in two limiting cases, where either $M_{2,x,t}$ is irrelevant and can be neglected, or $M_{2,x,t}$ is dominating and obtains the gapped form $\hat M_{2,x,t}\sim m(\hat\phi_x-\langle\hat\phi_x\rangle_t)$. In both cases the measurement operators are linear in the fields $\hat\phi_x$. For linear measurement operators, the correlation function \eqref{Eq:Cor} affords an analytical solution. This allows us to demonstrate that indeed for a gapless measurement operator, the correlations functions obtain a scale-invariant form, while, in contrast, correlations will  not be scale invariant if the measurement operators are gapped.

We identify the limit in which $M_{2,x,t}$ is irrelevant with the limit of weak monitoring, i.e., when the Hamiltonian dominates the dynamics $\gamma\ll v$.  
Then, the pinning of $\hat{\phi}_x$ induced by the nonlinearity in $\hat M_{2,x,t}$ cannot overcome the depinning effect from the Hamiltonian in Eq.~\eqref{Eq:twomeas}. The field operators $\hat{\phi}_x$ will explore an extended region of phase space, and the bounded nonlinear terms are irrelevant and average to zero. We therefore drop $M_{2,x,t}$ from the evolution.
 
In the opposite limit of strong measurements, i.e., $v\ll \gamma$, the state of the system will remain pinned in a measurement dark state $|\Psi_D\rangle$ similar to Fig.~\ref{fig-2}, where this is shown for the fermion density. At random times, it may jump from the vicinity of one dark state to another dark state with a different eigenvalue. These jumps are, however, uncorrelated and exponentially rare in time in the limit of dominant measurements. We take the pinning into account by assuming that the state $|\psi_t\rangle$ remains close to a measurement dark state $|\Psi_D\rangle$ with $\hat\phi_x|\Psi_D\rangle=\phi_x|\Psi_D\rangle$. For instance one may assume $|\psi_t\rangle=|\Psi_D\rangle + |\delta\psi_t\rangle$ with $\langle\delta\psi_t|(\hat\phi_x-\phi_x)^2|\delta\psi_t\rangle/\langle\delta\psi_t|\delta\psi_t\rangle\ll1$. We then shift the operator $\hat\phi_x\rightarrow\phi_{x}+\hat\phi_x$ with respect to the eigenvalue $\phi_{x}$ in the dark state and expand the measurement operators $\hat M_{2,x,t}$ linearly in the small deviation $\hat\phi_x$, yielding 
\begin{align}\label{SEq26}
\hat M_{2,x,t}\sim (\hat{\phi}_x-\langle\hat{\phi}_x\rangle_t).
\end{align}
The more accurate proportionality constant in Eq.~\eqref{SEq26} is $2m\sin(2\phi_{x})$ and depends on the eigenvalues $\phi_x$. We neglect it here (see Appendix~\ref{sec:AppRic} for more details, including the case of inhomogeneously distributed $\phi_x$).

We consider now the dynamics with the quadratic Hamiltonian \eqref{SEq10} and
one dominant measurement operator $\hat M_{x,t}$ which is linear in $\hat\phi_x$, i.e., either $M_{x,t}=\partial_x(\hat\phi_x-\langle\hat\phi_x\rangle_t)$ or $\hat M_{x,t}=\hat\phi_x-\langle\hat\phi_x\rangle_t$, reflecting the two different regimes. This yields the stochastic Schr\"odinger equation
\begin{align}\label{SEq13}
d|\psi_t\rangle = -dt\Big[i \hat H+\frac{\gamma}{2}\int_x \left(\hat M_{x,t}\right)^2\Big]|\psi_t\rangle+\int_x dW_x \hat M_{x,t}|\psi_t\rangle. 
\end{align}
The connected two-point correlation function for linear measurement problems can be solved exactly by methods developed in the context of Kalman filtering  \cite{Kalman61} in classical control theory \cite{bookMeybeck,bookBrunton}, also used in quantum optics \cite{Doherty99,bookWisemanMilburn,bookJacobs}. The problem maps to a Riccati equation, which we solve in the stationary state in both limits (see Appendix~\ref{sec:AppRic} and Ref.~\cite{minoguchi2021continuous} for details).

In the weak monitoring limit, this yields correlation functions in real and momentum space
\begin{align}\label{SEq25b}
C_{y}=\frac{C_0}{|y|^2},\ \ C_{k}=\frac{C_0|k|}{2}.
\end{align}
Here $C_0=\frac{v}{4\gamma}\Big[\Big(\frac{64\gamma^2}{\pi^2v^2}+4\Big)^{\frac{1}{2}}-2\big]^{\frac{1}{2}}$. In the limit $\gamma\rightarrow0$ this reduces to the ground state correlations of a Luttinger Liquid with $C_0=1/\pi$. In the opposite limit $\gamma\gg v$ it assumes the asymptotic form  $C_0=\sqrt{v/2 \gamma\pi}$. 

This result shows that linear measurement operators $\hat M_{x,t}\sim \partial_x\hat \phi_x-\langle \partial_x \hat \phi_x\rangle_t$ stabilize a scale invariant covariance matrix with a $1/|y|^2$ scaling behavior. Precisely this scaling was also detected numerically in the case of measured lattice fermions~\cite{Alberton20}. The only assumption leading to this result in the present framework is that the nonlinear terms in the measurement operators are negligible (or technically, $m =0$). 

In the strong monitoring limit, $M_{x,t}=\hat{\phi}_x-\langle \hat{\phi}_x\rangle_t$ yields
\begin{align}\label{Expcov}
C_{y}\sim \left(\frac{v}{2\pi\gamma }\right)^{|y|}=e^{-|y|/\xi}.
\end{align}
This correlation function decays exponentially with the distance and with a correlation length $\xi^{-1}=\log(2\pi\gamma/v)$. This indicates that the extreme limit $v=0$, which is uncorrelated except for $y=0$, extends smoothly to a short range correlated state at non-zero $v$. To further back up these result beyond the linear approximation performed here, we have computed the correlation length in the related fermionic lattice model defined by Eqs. (\ref{eq:hlatt},\ref{eq:mlatt}) perturbatively in Appendix~\ref{App:strong}. This yields a similar correlation length of $\xi^{-1}\sim\log(\gamma/\nu)$.

To summarize, we obtain a scale invariant correlation function with characteristic $1/|y|^2$ decay at weak monitoring, versus exponentially decaying correlations with a logarithmically growing correlation length at strong monitoring. Strictly speaking, to obtain the first result, we needed to set $\hat M_{2,x,t} =0$, and to perform linearization of $\hat M_{2,x,t}$ in the opposite limit. The viability of this procedure reduces to the question whether or not the nonlinearity in the measurement operator is relevant or irrelevant. This question will be addressed in the next section.

\section{Replica field theory for monitored Luttinger Liquids}\label{sec:RepFielTheory}

In this section, we will provide a theoretical description, which is not restricted to the two limiting cases and also incorporates intermediate monitoring strength. To this end, we develop a field theory description of a two-replica setup, which provides a tool to analyze whether an initial nonlinearity in the measurement operators becomes relevant or irrelevant at large distances. 
The replica approach also grants insights into the relevant degrees of freedom in the measurement evolution, which turn out to be the relative replica fluctuations. 

\subsection{Structure of measurement operator correlation functions}
In order to motivate the replica approach, we revisit the measurement operator correlation function on the lattice $C_{ij}(t)=\overline{\langle \hat M_{i,t} \hat M_{j,t}\rangle_t}$, where again $\hat M_{i,t}=\hat O_i-\langle \hat O_i\rangle_t$ for the measured operator $\hat O_i$. In Eq.~\eqref{Eq:Eq3}, we expressed the correlation function $C_{ij}$ in terms of the conditioned projector $\hat\rho^{(c)}_t$ and the trajectory average of its product $\hat\rho^{(R_2)}_t=\overline{\hat\rho^{(c)}_t\otimes\hat\rho^{(c)}_t}$ as
\begin{align}\label{SEq28}
C_{ij}(t)=\frac{1}{2}\text{Tr} \left[(\hat O_i^{(1)}-\hat O_i^{(2)})(\hat O_j^{(1)}-\hat O_j^{(2)})\hat\rho^{(R_2)}\right].
\end{align}
The density matrix $\hat\rho^{(R_2)}_t$ can no longer be written as a product of pure states and contains inter-replica correlations, and therefore $C_{ij}(t)$ can behave nontrivially.

Equation \eqref{SEq28} shows that (i) the measurement correlation function is the product of two {\it relative} observables $\hat O^{(1)}-\hat O^{(2)}$ (or their fluctuation) in the replicated space and (ii) that this correlation function must be encoded in the {\it linear} statistical average $\hat\rho^{(R_2)}_t$. For this and the following section, we introduce a shorthand notation for expectation values as in Eq.~\eqref{SEq28}. We set 
\begin{align}
    \langle\langle ... \rangle\rangle =\text{Tr}[...\ \hat\rho^{(R_2)}],
\end{align}
which is the expectation value of an operator (or operator product) $...$ with respect to $\hat\rho^{(R_2)}$. It  combines the quantum mechanical average as well as the trajectory average from the previous sections.

\subsection{Two-replica master equation}\label{sec:2RepMaster}
The two-replica density matrix $\hat\rho^{(R_2)}_t$ is no longer a stochastic object but evolves according to a deterministic evolution equation. The infinitesimal increment $d\hat\rho^{(R_2)}_t$ is obtained from taking the statistical average of the product increment
\begin{align}
    d\hat\rho^{(R_2)}_t&=\hat\rho^{(R_2)}_{t+dt}-\hat\rho^{(R_2)}_{t}=\overline{\hat\rho^{(c)}_{t+dt}\otimes\hat\rho^{(c)}_{t+dt}}-\overline{\hat\rho^{(c)}_{t}\otimes\hat\rho^{(c)}_{t}}\nonumber\\
    &=\overline{d\hat\rho^{(c)}_{t}\otimes\hat\rho^{(c)}_{t}}+\overline{\hat\rho^{(c)}_{t}\otimes d\hat\rho^{(c)}_{t}}+\overline{d\hat\rho^{(c)}_{t}\otimes d\hat\rho^{(c)}_{t}}. \label{eq:RepIncrement}
\end{align}
The stochastic increments $\{dW_i\}$ at time-step $t\rightarrow t+dt$ are not correlated with the increments at earlier times. Therefore the trajectory average in Eq.~\eqref{eq:RepIncrement} sets any increments which are only linear in $dW$ to zero and replaces averages over terms quadratic in $dW$ by $\overline{(...) dW_i dW_j}=\overline{(...)}\delta_{ij}\gamma dt$. Here $(...)$ represents any other stochastic variable, such as $\langle \hat O_i\rangle_t$ which does not depend on $dW$ but on stochastic increments at earlier times.

The first and the second term in Eq.~\eqref{eq:RepIncrement} only include the increment of one single density matrix and are thus identical to taking the statistical average over a single-replica master equation for $d\hat\rho^{(c)}$, i.e., 
\begin{align}
    \overline{d\hat\rho^{(c)}_{t}\otimes\hat\rho^{(c)}_{t}}=&dt\overline{\mathcal{L}^{(1)}\left(\hat\rho^{(c)}_{t}\otimes\hat\rho^{(c)}_{t}\right)}+\sum_i \overline{dW_i \left\{\hat M_{i,t}^{(1)},\hat\rho^{(c)}_{t}\otimes\hat\rho^{(c)}_{t}\right\}}\nonumber\\
    =&dt\mathcal{L}^{(1)}\left(\overline{\hat\rho^{(c)}_{t}\otimes\hat\rho^{(c)}_{t}}\right)=dt\mathcal{L}^{(1)}\hat\rho^{(R_2)}_t.
\end{align}
Here, we have defined the intra-replica Liouvillian $\mathcal{L}^{(l)}=i[\hat H^{(l)},\cdot]-\frac{\gamma}{2}\sum_i [\hat O_i, [\hat O_i,\cdot]]$ and measurement operator $\hat M^{(l)}_{i,t}=\hat O^{(l)}-\langle \hat O_i\rangle_t$. The conditioned expectation value $\langle \hat O_i\rangle_t=\text{Tr}\left[\hat O_i \hat\rho^{(c)}_t\right]$ is still a stochastic object, which does, however, not depend on the noise $dW$ at time $t$, as mentioned above. This yields the familiar Lindblad form as in Eq.~\eqref{eq:lindi}, acting on each single replica separately. 

The third term, however, builds up correlations between both replicas and yields an unconventional contribution to the evolution of $\hat\rho^{(R_2)}_t$. The replica-coupling evolution term is
\begin{align}
    \overline{d\hat\rho^{(c)}_{t}\otimes d\hat\rho^{(c)}_{t}}=&\sum_{i,j}\overline{\left\{\hat M_{i,t}^{(1)}, \left\{\hat M_{j,t}^{(2)}, \hat\rho^{(c)}_{t}\otimes \hat\rho^{(c)}_{t}\right\}\right\}dW_i dW_j}\nonumber\\
    =&\gamma dt \sum_i \overline{\left\{\hat M_{i,t}^{(1)}, \left\{\hat M_{i,t}^{(2)}, \hat\rho^{(c)}_{t}\otimes \hat\rho^{(c)}_{t}\right\}\right\}}.
\end{align}
It contains three different classes of couplings, two of which we have to consider with care. The first class are terms composed only of operators, i.e., the ones summarized by
\begin{align}
    \overline{\left\{\hat O_i^{(1)}, \left\{\hat O_i^{(2)}, \hat\rho^{(c)}_{t}\otimes \hat\rho^{(c)}_{t}\right\}\right\}}=\left\{\hat O_i^{(1)}, \left\{\hat O_i^{(2)},  \hat\rho^{(R_2)}_{t}\right\}\right\}.
\end{align}
Here the stochastic average acts only on the density matrix and does not affect the operators at all. The second class of increments is 
\begin{align}
    \overline{\left\{\langle\hat O_i\rangle_t, \left\{ \langle \hat O_i\rangle_t, \hat\rho^{(c)}_{t}\otimes \hat\rho^{(c)}_{t}\right\}\right\}}=4 \overline{\langle\hat O_i\rangle_t^2 \hat\rho^{(c)}_{t}\otimes \hat\rho^{(c)}_{t}}.\label{Eqav1}
\end{align}
Here we face a problem with computing the stochastic average, since the two replica density matrix $\hat \rho^{(R_2)}_t$ does not give access to the individual trajectories of $\langle \hat O_i\rangle_t$. It only allows us to compute the stochastic averages 
$\langle\langle \hat O^{(l)}\rangle\rangle=\overline{\langle\hat O_i\rangle_t}$ or 
$\langle\langle \hat O^{(1)}_j \hat O^{(2)}_i\rangle\rangle=\overline{\langle\hat O_i\rangle_t\langle\hat O_j\rangle_t}$, but this does not include information on the statistical correlations between $\langle\hat O_i\rangle_t$ and $\hat\rho^{c}_t\otimes\hat\rho^{c}_t$, which are required in the average in Eq.~\eqref{Eqav1}. This also holds for the third type of increment, which is
\begin{align}
    \overline{\left\{\hat O_i^{(l)}, \left\{ \langle \hat O_i\rangle_t, \hat\rho^{(c)}_{t}\otimes \hat\rho^{(c)}_{t}\right\}\right\}}=2 \left\{\hat O_i^{(l)},\overline{\langle\hat O_i\rangle_t \hat\rho^{(c)}_{t}\otimes \hat\rho^{(c)}_{t}}\right\}.\label{Eqav2}
\end{align}

This problem indicates the initial point of an infinite coupled hierarchy of replica correlation functions: Indeed, the statistical averages in Eqs.~\eqref{Eqav2} and \eqref{Eqav1} can again be formulated as linear problems in a three- or four-replica framework, respectively (e.g., $\overline{\langle\hat O_i\rangle_t \hat\rho^{(c)}_{t}\otimes \hat\rho^{(c)}_{t}}=\text{Tr}_{(3)}[\hat O^{(3)}_i \hat \rho^{(R_3)}_t]$ can be written as a partial trace over the third replica in a three-replica density matrix $\hat\rho^{(R_3)}_t$). This illustrates that once the realm of a single replica description of the problem is left, an infinite hierarchy of replica equations emerges. In order to find a solution for $\hat\rho^{(R_2)}$, this hierarchy has to be truncated at finite order. 

Here we perform the truncation on the level of the two-replica density matrix, which amounts to treating the stochastic correlations between $\langle\hat O_i\rangle_t$ and $\hat\rho^{(c)}_{t}\otimes \hat\rho^{(c)}_{t}$ in a mean-field type decoupling by setting 
\begin{align}
     \overline{ \langle \hat O_i \rangle_t \hat\rho^{(c)}_{t}\otimes \hat\rho^{(c)}_{t} } &\equiv \langle\langle \hat O^{(1,2)}_i\rangle\rangle_t \hat \rho^{(R_2)}_t,\label{MF1}\\
     \overline{ \langle \hat O_i \rangle_t^2 \hat\rho^{(c)}_{t}\otimes \hat\rho^{(c)}_{t} } &\equiv \langle\langle \hat O^{(1)}_i\hat O^{(2)}_i\rangle\rangle_t \hat \rho^{(R_2)}_t.\label{MF2}
\end{align}

We will see later that if the operators $\hat O_l$ are linear functions of boson or fermion fields (as it is the case for the operators $\hat M_{1,x,t}$ in the bosonized framework), then the mean-field decoupling in Eqs.~\eqref{MF1}, \eqref{MF2} has no effect on the relative correlation function $C_{ij}$, i.e., the mean field approximation is exact for these correlators. The measurement expectation values then only show up in a center-of-mass coordinate $\hat O_i^{(1)}+\hat O_i^{(2)}$, which completely decouples from relative coordinates $\hat O^{(1)}_i-\hat O^{(2)}_i$ and effectively heats up to infinite temperature. This heating of the center-of-mass  coordinate remains operational when the linear operator $\hat M_{1,x,t}$ and the non-linear operator $\hat M_{2,x,t}$ are measured simultaneously; in turn this justifies the mean-field approximation also in this case. The Hermitian nature of the measurement operators, and the subsequent heating, simplifies the problem significantly here. For a general set of only nonlinear and non-Hermitian measurement operators, including nonlinear feedback operations, the mean-field decoupling is, however, not immediately justified.

After performing the statistical average including the mean-field decoupling, one obtains a deterministic evolution equation for $\hat\rho^{(R_2)}$, the two-replica master equation
\begin{align}
    \partial_t \hat\rho^{(R_2)}_t=&\mathcal{L}^{(1)}\hat\rho^{(R_2)}_t+\mathcal{L}^{(2)}\hat\rho^{(R_2)}_t-8\gamma \hat\rho^{(R_2)} \tilde C(t)\label{eq:RepCoup}\\
    &+2\gamma  \sum_l\left\{\hat O^{(2)}_{i}-\langle\langle O^{(2)}_{i}\rangle\rangle, \left\{\hat O^{(1)}_{i}-\langle\langle \hat O^{(1)}_{i}\rangle\rangle, \hat\rho^{(R_2)}_t\right\}\right\}.\nonumber
\end{align}
Here, the normalization function is $\tilde{C}_t=\sum_i \left(\langle\langle \hat O^{(1)}_i\hat O^{(2)}_i\rangle\rangle_t-\langle\langle\hat O^{(1)}_i\rangle\rangle_t\langle\langle\hat O^{(2)}_i\rangle\rangle_t\right)$. The first line is the regular master equation for two uncoupled replicas evolving under the Liouvillians $\mathcal{L}^{(l)}$. If no additional terms were present, the Liouvillians would heat up each replica towards an infinite temperature state.

The second line in Eq.~\eqref{eq:RepCoup} is what distinguishes the two-replica setup from a single replica, which would only heat up to infinite temperature. It leads to the build-up of inter-replica correlations, and transforms the replica master equation into a non-Lindblad form. This equation is still non-linear in the state due to the expectation values $\langle \hat O_i\rangle_t$, contained in the operators $\hat M_{i,t}$ and in $\tilde{C}_t$. However, the equation is deterministic, and the expectation values are no longer stochastic variables. In the stationary state, they can be replaced by a number $\langle \hat O_i\rangle_t= o_i$.  We will analyze the consequences of this evolution equation in the following section. 

\subsection{Two-replica master equation in bosonized framework}\label{sec:2bosonReplica}
We now formulate the bosonized measurement setup introduced in Sec.~\ref{sec:fermbos} in terms of the replica master equation \eqref{eq:RepCoup}. This will allow us to establish the connection between the generator of dynamics, and the correlation functions.  To this end, we extend the bosonized Hamiltonian and the measurement operators to replica space by introducing
\begin{align}\label{SEq32}
\hat H^{(l)}&=\frac{v}{2\pi}\int_x (\partial_x\hat\phi_x^{(l)})^2+(\partial_x\hat\theta_x^{(l)})^2,\\
\hat M_{1,x,t}^{(l)}&=-\tfrac{1}{\pi}\partial_x (\hat\phi_x^{(l)}-\langle\hat\phi_x^{(l)}\rangle_t), \\
\hat M_{2,x,t}^{(l)}&=2m(\cos(2\hat\phi_x^{(l)})-\langle \cos(2\hat\phi_x^{(l)})\rangle_t),
\end{align}
where again the label $l$ indicates how the operator acts in replica space (see the definition above Eq.~\eqref{SEq28}). The two-replica Hamiltonian is $\hat H=\hat H^{(1)}+\hat H^{(2)}$. Due to independent noise increments for $\hat M_{1,x,t}$ and $\hat M_{2,x,t}$, the two different measurements are uncorrelated and contribute independently to the replica master equation. This can be implemented in the two-replica master equation by replacing the index $i\rightarrow(s,x)$ with $s=1,2$ and $x$ the continuous position.

To gain a better understanding of the two-replica dynamics, we again start by considering only measurements of the linear operator $\hat M_{1,x,t}$ without measuring $\hat M_{2,x,t}$. Apart from the constant terms, this yields the two-replica  master equation
\begin{align}\label{SEq33}
\partial_t\hat\rho^{(R_2)}=&
i\left[\hat\rho^{(R_2)}, \hat H\right]-\frac{\gamma}{2\pi}\sum_{l=1,2}\int_x
\left[\partial_x\hat\phi^{(l)}_x,\left[\partial_x\hat\phi^{(l)}_x,\hat\rho^{(R_2)}\right]\right]  \\
&+\frac{\gamma}{\pi}
\int_x \left\{\partial_x\hat\phi^{(2)}_x-\langle\langle\partial_x\hat\phi_x^{(2)}\rangle\rangle,\left\{\partial_x\hat\phi^{(1)}_x-\langle\langle\partial_x\hat\phi_x^{(1)}\rangle\rangle,\hat\rho^{(R_2)}\right\}\right\}.\nonumber
\end{align}

The master equation is quadratic in the field operators, but couples operators with different replica index $l$. The terms can be decoupled by performing a coordinate transformation into a replica center-of-mass field $\hat\phi^{(a)}_x, \hat\theta^{(a)}_x$ and a relative field $\hat\phi^{(r)}_x, \hat\theta^{(r)}_x$ (describing inter-replica fluctuations) according to the unitary transformation ${\hat\phi^{(a,r)}_x=(\hat\phi^{(1)}_x\pm\hat\phi^{(2)}_x)/\sqrt{2}}$, ${\hat\theta^{(a,r)}_x=(\hat\theta^{(1)}_x\pm\hat\theta^{(2)}_x)/\sqrt{2}}$. It preserves the bosonic commutation relations between the operators $[\partial_x\hat\theta^{(l)}_x,\hat\phi^{(l')}_y]=-i\delta_{l,l'}\delta(x-y)$ for $l,l'=a,r$. 

The quadratic Hamiltonian $\hat H$ remains a sum  $\hat H=\hat H^{(r)}+\hat H^{(a)}$ with $\hat H^{(l)}$ as defined in Eq.~\eqref{SEq32}, but with center-of-mass and relative operators. The measurement part of the evolution, however, is transformed such that the relative and center-of-mass decouple in the new basis. The master equation for $\hat\rho^{(R_2)}$ is therefore separable, which suggests a product ansatz $\hat\rho^{(R_2)}=\hat\rho^{(r)}\otimes\hat\rho^{(a)}$. Inserting this ansatz into Eq.~\eqref{SEq33} yields
\begin{align}\label{SEq35}
\partial_t\hat\rho^{(r)}=&i[\hat\rho^{(r)},\hat H^{(r)}]-\frac{\gamma}{\pi^2} \int_x\left\{ (\partial_x\hat\phi^{(r)}_x)^2,\hat\rho^{(r)}\right\},\\
\partial_t\hat\rho^{(a)}=&i[\hat\rho^{(a)},\hat H^{(a)}]\label{SEq36}\\&+\frac{2\gamma}{\pi^2}\int_x\left(\partial_x\hat\phi^{(a)}-\langle\langle\partial_x\hat\phi^{(a)}\rangle\rangle\right)\hat\rho^{(a)}\left(\partial_x\hat\phi^{(a)}-\langle\langle\partial_x\hat\phi^{(a)}\rangle\rangle\right).\nonumber
\end{align}

This decomposition is a remarkable result, which gives the two replica master equation a new interpretation. The density matrix $\hat\rho^{(r)}$ of the fluctuations evolves according to a non-Hermitian Schrödinger equation with an effective Hamiltonian \begin{align}
\hat H_{\text{eff}}^{(r)}=\frac{v}{2\pi}\int_x (\partial_x\hat\theta_x^{(r)})^2+(1-i\frac{2\gamma}{v\pi})(\partial_x\hat\phi_x^{(r)})^2,\label{eq:EffHamQuad1}
\end{align}
but without feedback from a contour-coupling term (i.e., no coupling between the left and the right side of $\hat\rho^{(R_2)}$). The density matrix of the relative degrees is thus evolving towards a stationary state $\hat\rho^{(r)}_t=e^{-i\hat H^{(r)}_{\text{eff}}t}\hat\rho^{(r)}_0e^{i\hat H^{(r)}_{\text{eff}}t}\rightarrow|\psi^{(r)}_D\rangle\langle\psi^{(r)}_D|$ which approaches a dark state of the effective Hamiltonian $\hat H_{\text{eff}}^{(r)}|\psi^{(r)}_D\rangle=0$. In Appendix~\ref{app:darkstate}, we demonstrate that $H_{\text{eff}}^{(r)}$ indeed has one unique dark state, which is reached from any initial state during the time evolution. We show that the correlation functions of the replica field operators in the dark state, $\langle\psi^{(r)}_D|\partial_x\hat\phi^{(r)}_x\partial_y\hat\phi^{(r)}_y|\psi^{(r)}_D\rangle$ are identical to the correlation functions \eqref{SEq25b} obtained from the Riccati approach, which provides the exact solution for the covariance matrix.  

This observation is independent of the mean-field decoupling of the statistical fluctuations in Eqs.~\eqref{MF1}, \eqref{MF2}. The reason is that due to the replica symmetry (i.e. invariance of the theory under exchange $\hat \phi^{(1)} \leftrightarrow \hat\phi^{(2)}$), independently of the noise realization we have $\langle \hat \phi^{(1)}\rangle=\langle \hat \phi^{(2)}\rangle$. This implies $\sqrt{2}\langle \hat \phi^{(r)}\rangle=\langle \hat \phi^{(1)}-\hat \phi^{(2)}\rangle=0$. Therefore, the neglected noise correlations are only affecting $\hat\rho^{(a)}$, but not the density matrix of the fluctuations $\hat\rho^{(r)}$. The dark state therefore describes the exact steady state of the replica fluctuations. 

We now turn to the measurement correlation function Eq.~\eqref{SEq28}, here under the assumption that the nonlinear terms are irrelevant,  $\hat M_{2,x,t}=0$. The relative  measurement operator is $\hat M^{(1)}_{x,t}-\hat M^{(2)}_{x,t} =\frac{\sqrt{2}}{\pi}\,\partial_x\hat\phi^{(r)}_x$, yielding the measurement correlation function $C_{y}=\frac{1}{\pi^2}\langle\psi^{(r)}_D|\partial_x\hat\phi^{(r)}\partial_x\hat\phi_{x+y}^{(r)}|\psi^{(r)}_D\rangle$, the expectation value of $\partial_x\hat\phi^{(r)}\partial_x\hat\phi_{x+y}^{(r)}$ in the dark state. The scale invariance of the effective Hamiltonian $\hat H_{\text{eff}}^{(r)}$ implies that also the measurement correlation functions in this limit become scale invariant and algebraically decaying, confirming the observation from Sec. \ref{SecRic}.

We can also infer what happens in the limit when $\hat M_{2,x,t}$ is measured with a large mass $\gamma m\gg v$, where the linearization $\hat M_{2,x,t}\rightarrow \hat\phi^{(l)}_x-\langle\hat\phi^{(l)}_x \rangle_t$ is justified. Using the same arguments as for the case $m=0$, this yields a separable evolution equation for $l=r,a$ with a modified effective Hamiltonian in Eq.~\eqref{SEq35} $\hat H^{(r)}_{\text{eff}}\rightarrow  \hat H^{(r)}_{\text{mass}}$ and 
\begin{align}
\hat H_{\text{mass}}^{(r)}= \hat H^{(r)}_{\text{eff}}-i\gamma\int_x (\hat\phi_x^{(r)})^2.\label{eq:EffHamQuad2}
\end{align}
In $\hat H_{\text{mass}}^{(r)}$ the additional dissipative mass term $\sim \gamma$ introduces a length scale. The measurement correlations are therefore still described by a pure quantum state but now correlations functions in the state $|\tilde\psi^{(r)}_D\rangle$ are exponentially decaying in space. 

The effective cooling and the evolution towards a single dark state in the relative degrees of freedom $\hat\theta^{(r)}_x, \hat \phi^{(r)}_x$ have to be contrasted with the behavior of the  center-of-mass fields $\hat\theta^{(a)}_x, \hat \phi^{(a)}_x$ described by Eq.~\eqref{SEq36}. Their evolution does not follow an effective Schrödinger equation but instead displays enhanced statistical fluctuations. 
In order to determine the fate of the local correlation functions for the center-of-mass degrees of freedom, we consider their evolution equation
\begin{align}\label{eq:growth}\partial_t\langle\langle \hat\phi_x^{(a)}\hat\phi_x^{(a)}\rangle\rangle=&\text{Tr}\left[(\hat\phi_x^{(a)})^2 \partial_t\hat\rho^{(a)}\right]=-i\langle\langle [\hat H^{(a)},\hat\phi_x^{(a)}\hat\phi_x^{(a)}]\rangle\rangle\nonumber\\&+\tfrac{2\gamma}{\pi^2}\int_{x'}\langle\langle (\partial_{x}\hat\phi^{(a)}_{x'}-\langle\langle\partial_x\hat\phi^{(a)}_{x'}\rangle\rangle)^2(\hat\phi_x^{(a)})^2\rangle\rangle.
\end{align}
Due to the quadratic evolution equation for $\hat\rho^{(a)}_t$, higher order correlation functions such as in second line of Eq.~\eqref{eq:growth} can be decoupled via Wick's theorem. This  yields a nonlinear, strictly positive growth of $\langle\langle \hat\phi_x^{(a)}\hat\phi_x^{(a)}\rangle\rangle$ in time. The commutator with the Hamiltonian yields a term proportional to $\langle\langle \hat\phi_x^{(a)}\hat\theta_x^{(a)}\rangle\rangle$, which, however, has a similar non-linear evolution equation (just like $\langle\langle \hat\theta_x^{(a)}\hat\theta_x^{(a)}\rangle\rangle$). 
In consequence this yields an asymptotic value $\langle\langle \hat\phi^{(a)}_x\hat\phi^{(a)}_x\rangle\rangle\rightarrow\infty$, which is consistent with an infinite temperature state in the center-of-mass degrees of freedom. 

\subsection{Nonlinear master equation for the replica fluctuations}
 Now, we will focus on the general case in which the measurement operator $\hat M_{2,x,t}$ cannot be neglected and is a nonlinear function of $\hat\phi^{(l)}_x$. We will use the insights from the previous discussion to obtain the master equation for the replica fluctuations. We demonstrate that also in this case the time evolution of the relative degrees of freedom is described by an effective Hamiltonian. It is a non-Hermitian analogue of the sine-Gordon Hamiltonian.

We start with the full two-replica master equation in the basis of relative and center-of-mass degrees of freedom
\begin{widetext}
\begin{align}
    \partial_t\hat\rho^{(R_2)}=&-i[\hat H^{(r)}_{\text{eff}}\hat\rho^{(R_2)}-\hat\rho^{(R_2)}(\hat H^{(r)}_{\text{eff}})^\dagger]-i[\hat H^{(a)},\hat\rho^{(R_2)}]+\frac{2\gamma}{\pi^2}\int_x\left(\partial_x\hat\phi^{(a)}-\langle\langle\partial_x\hat\phi^{(a)}_x\rangle\rangle\right)\hat\rho^{(R_2)}\left(\partial_x\hat\phi^{(a)}_x-\langle\langle\partial_x\hat\phi^{(a)}_x\rangle\rangle\right)\label{eq:fullMaster}\\
    &-\gamma m^2\int_x \sum_{\sigma=\pm}\left[\cos\left(\sqrt{2}(\hat\phi_x^{(a)}+\sigma\hat\phi_x^{(r)})\right),\left[\cos\left(\sqrt{2}(\hat\phi_x^{(a)}+\sigma\hat\phi_x^{(r)})\right),\hat\rho^{(R_2)}\right]\right]\nonumber\\&+2\gamma m^2\left\{\cos\left(\sqrt{2}(\hat\phi_x^{(a)}+\hat\phi_x^{(r)})\right)-\langle\langle\cos\left(\sqrt{2}(\hat\phi_x^{(a)}+\hat\phi_x^{(r)})\right)\rangle\rangle,\left\{\cos\left(\sqrt{2}(\hat\phi_x^{(a)}-\hat\phi_x^{(r)})\right)-\langle\langle\cos\left(\sqrt{2}(\hat\phi_x^{(a)}-\hat\phi_x^{(r)})\right)\rangle\rangle,\hat\rho^{(R_2)}\right\}\right\}.\nonumber
\end{align}
\end{widetext}
This equation may look overwhelming at first sight, but we will now show how its complexity can be drastically reduced by making use of the insights from the previous section. We will eliminate the coupling of the relative degrees of freedom to the center-of-mass modes, to obtain an effective master equation for $\hat\rho^{(r)}$.  

In order to do this, we trace out the center-of-mass degrees of freedom in Eq.~\eqref{eq:fullMaster}, i.e., we set $\partial_t \hat\rho^{(r)}\equiv \text{tr}_{(a)}\partial_t\hat\rho^{(R_2)}$, where $\text{tr}_{(a)}$ denotes the trace over the center-of-mass modes. We also set $\langle\langle ... \rangle\rangle_a= \text{tr}_{(a)}(...\rho^{(R_2)})$.  In order to integrate out the center-of-mass modes, we have to consider that they couple to the relative modes nonlinearly. However, we have seen above that the quadratic part of Eq.~\eqref{eq:fullMaster} is separable, i.e., $\rho^{(R_2)}=\rho^{(r)}\otimes\rho^{(a)}$, and pushes the correlation functions towards $\langle\langle \hat\phi^{a}_x\hat\phi^{a}_x\rangle\rangle_a\rightarrow\infty$, corresponding to a density matrix $\hat\rho^{(a)}$ in an infinite temperature state. 

The bounded vertex operators $\exp(i\sqrt{2}\hat\phi^{(a)}_x)$ alone do not push the center-of-mass mode away from an infinite temperature state. They rather become irrelevant in this limit, which follows from $\langle\langle \exp(i\sqrt{2}\hat\phi^{(a)}_x)\rangle\rangle_a=\exp(-\langle\langle\hat\phi^{(a)}_x\hat\phi^{(a)}_x\rangle\rangle_a)\rightarrow0$, where we assumed that we can apply Wick's theorem. This also indicates that the coupling between the relative and the center-of-mass modes will not modify the correlation functions of the latter. While in general the precise form of $\hat\rho^{(R_2)}$ may be unknown, this suggests very weak correlations between the two sets of modes. We therefore approximate the density matrix by a product  $\hat\rho^{(R_2)}\sim \hat\rho^{(r)}\otimes \hat\rho^{(a)}$. The the only assumption we need regarding $\hat\rho^{(a)}$ is that it is in a Gaussian state with $\langle\langle \hat\phi^{(a)}_x\hat\phi^{(a)}_x\rangle\rangle_a\rightarrow\infty$~\footnote{It is also possible to perturbatively incorporate correlations between the center-of-mass and the relative degrees of freedom beyond a product state assumption by applying time-dependent perturbation theory in the Mori-Zwanzig formalism \cite{Zwanzig}. However, these corrections are always less relevant than the first order correction used here.}. 

In this case, tracing out the center-of-mass modes from Eq.~\eqref{eq:fullMaster} can be performed analytically. First, the dependence on the expectation values $\langle\langle \cos\left(\sqrt{2}(\hat\phi^{(a)}_x\pm\hat\phi^{(r)}_x)\right)\rangle\rangle=0$ vanishes. Second, only non-linear vertex operators $\sim\exp(i...)$, which are independent of $\hat\phi^{(a)}_x$, persist upon taking the trace with respect to the average degrees of freedom. Thus only the terms in the third line of Eq.~\eqref{eq:fullMaster} yield a non-trivial contribution to the evolution equation of $\rho^{(r)}$. We illustrate this by tracing out the center-of-mass mode in the cross term
\begin{widetext}
\begin{align}
    \langle\langle \cos\left(\sqrt{2}(\hat\phi^{(a)}_x+\hat\phi^{(r)}_x)\right)\cos\left(\sqrt{2}(\hat\phi^{(a)}_x-\hat\phi^{(r)}_x)\right)\rangle\rangle_a=\frac{1}{2}\langle\langle \cos\left(\sqrt{8}\hat\phi^{(a)}_x\right)+\cos\left(\sqrt{8}\hat\phi^{(r)}_x\right)\rangle\rangle_a=\frac{1}{2}\cos\left(\sqrt{8}\hat\phi^{(r)}_x\right). 
\end{align}
\end{widetext}

From this procedure, we obtain the evolution equation for the relative replica fluctuations. In the following we will only focus on this degree of freedom and drop the label $(r)$~\footnote{The trace has to be evaluated with care. Due to the noise average there is no straightforward cyclic permutation rule for operators in the absolute and relative basis.}. After keeping the most relevant nonlinear terms only, the master equation for the fluctuations is
\begin{align}\label{SEq37}
\partial_t\hat\rho=&i(\hat\rho \hat H^\dagger_{sg}-\hat H_{sg}\hat\rho),
\end{align}
where $\lambda=\gamma m^2$, and the non-Hermitian sine-Gordon Hamiltonian reads
\begin{align}\label{eq:effsineGordon}
    \hat H_{sg}=&\frac{v}{2\pi}\int_x \left[(\partial_x\hat\theta_x)^2+\left(1-\tfrac{2i\gamma}{v\pi}\right)(\partial_x\hat\phi_x)^2\right]\\
    &-i\lambda\int_x(\cos(\sqrt{8}\hat\phi_x)-1).\nonumber
\end{align}
Again we encounter a non-Hermitian Hamiltonian evolution for the relative degrees of freedom. 

On large length- and long time-scales, the irrelevance of higher order perturbative corrections (scaling $\sim \cos(\zeta\sqrt{8}\hat\phi)$ at $\zeta$-th order) again leads to an effective cooling of the replica fluctuations towards the dark state of the non-Hermitian Hamiltonian $\hat H_{sg}$. Depending on the strength of the nonlinearity and the imaginary part in the kinetic energy term, this dark state is scale invariant at large distances, or has built in a length-scale yielding either algebraically or exponentially decaying correlations. For a Hermitian sine-Gordon Hamiltonian, a renormalization group treatment predicts a phase transition between a scale invariant ground state and a gapped ground state according to the Berezinskii-Kosterlitz-Thouless paradigm~ \cite{ZinnJustin,Amit_1980,Kosterlitz_1973}, and we may conjecture that a similar scenario applies to the dark state of the non-Hermitian Hamiltonian. We will examine the nature of the transition in terms of a renormalization group analysis of $\hat H_{sg}$ in the following section. 

We conclude this section with an additional remark. The effective Hamiltonian $\hat H_{sg}$ is not PT symmetric~\footnote{Both $(\partial_x\hat\phi_x)^2$ and $\cos(\sqrt{8}\hat\phi_x)$ are PT symmetric and therefore multiplying them with an imaginary unit generates PT non-symmetric terms.} and therefore its eigenvalues are generally complex. However, the imaginary part of the eigenvalues is always negative, which guarantees dynamical stability of the time-evolution. An evolution with a non-Hermitian Hamiltonian with complex eigenvalues, however, does not preserve the norm of the state. This is possible because there is no constraint that ensures the norm of $\hat\rho^{(r)}$ and $\hat\rho^{(a)}$ to be conserved individually. The norm of $\hat\rho^{(R_2)}$ is preserved as argued in Sec.~\ref{sec:thback}.

\section{Renormalization group approach to the non-Hermitian sine-Gordon theory}\label{sec:RG}
The sine-Gordon Hamiltonian in Eq.~\eqref{eq:effsineGordon} describes the long-time and large-distance behavior of the replica fluctuations. In order to determine whether or not the $\cos$-nonlinearity is relevant or irrelevant at large distances, i.e., whether it generates a mass term $\sim m^2\hat\phi_x^2$ or vanishes and restores a scale-invariant action, we perform a renormalization group analysis. While the renormalization group (RG) flow of the sine-Gordon model is well established in thermal equilibrium (i.e., for a Hermitian Hamiltonian), only few non-Hermitian situations have been considered previously \cite{FENDLEY_1993, Ashida_2017,Sarkar2021}. 

A convenient starting point for a renormalization group treatment is the non-Hermitian sine-Gordon action $S_{sg}$. It is derived by writing the propagator ${U(t,0)=\langle \{\phi_{x,f}\}| e^{-i\hat H_{sg}(t_f-t_i)}|\{\phi_{x,i}\}\rangle}$ in terms of a path integral
\begin{align}\label{eq:prop}
   U(t,0) = \int_{\phi_{x,0}=\phi_{x,i}}^{\phi_{x,t}=\phi_{x,f}}\mathcal{D}[\{\phi_{x,t} \}] e^{iS(\{\phi_{x,t}\})}
\end{align}
over the real fields $\{\phi_{x,t}\}$ by the conventional Trotterization procedure, e.g., see Appendix \ref{App:path_integral}. 

The real-time action for the non-Hermitian sine-Gordon Hamiltonian \eqref{eq:effsineGordon} can be brought into the Lagrangian form (with $X=(x,t)$)
\begin{align}
    S=\int_X\left\{ \frac{K}{16\pi}\left[\frac{1}{\eta}(\partial_t\phi_X)^2-\eta(\partial_x\phi_X)^2\right]-i\lambda\cos(\phi_X)\right\}.\label{eq:sineGordonField}
\end{align}
Here, $K, \lambda$ are the flow parameters for the renormalization group treatment with the microscopic value of $K=\eta$ and $\eta^2=1-\frac{2i\gamma}{v\pi}$. The parameter $\eta$ can been eliminated from the action \eqref{eq:sineGordonField} by a complex Wick rotation $(x,t)\rightarrow(\eta^{\frac{1}{2}}x,i\eta^{-\frac{1}{2}}t)$. This yields an imaginary time path integral, which is well-defined for the here realized case $\text{Im}(\eta)<0$, i.e., when there is no dynamical instability.

\begin{figure}
    \includegraphics[width=0.49\textwidth]{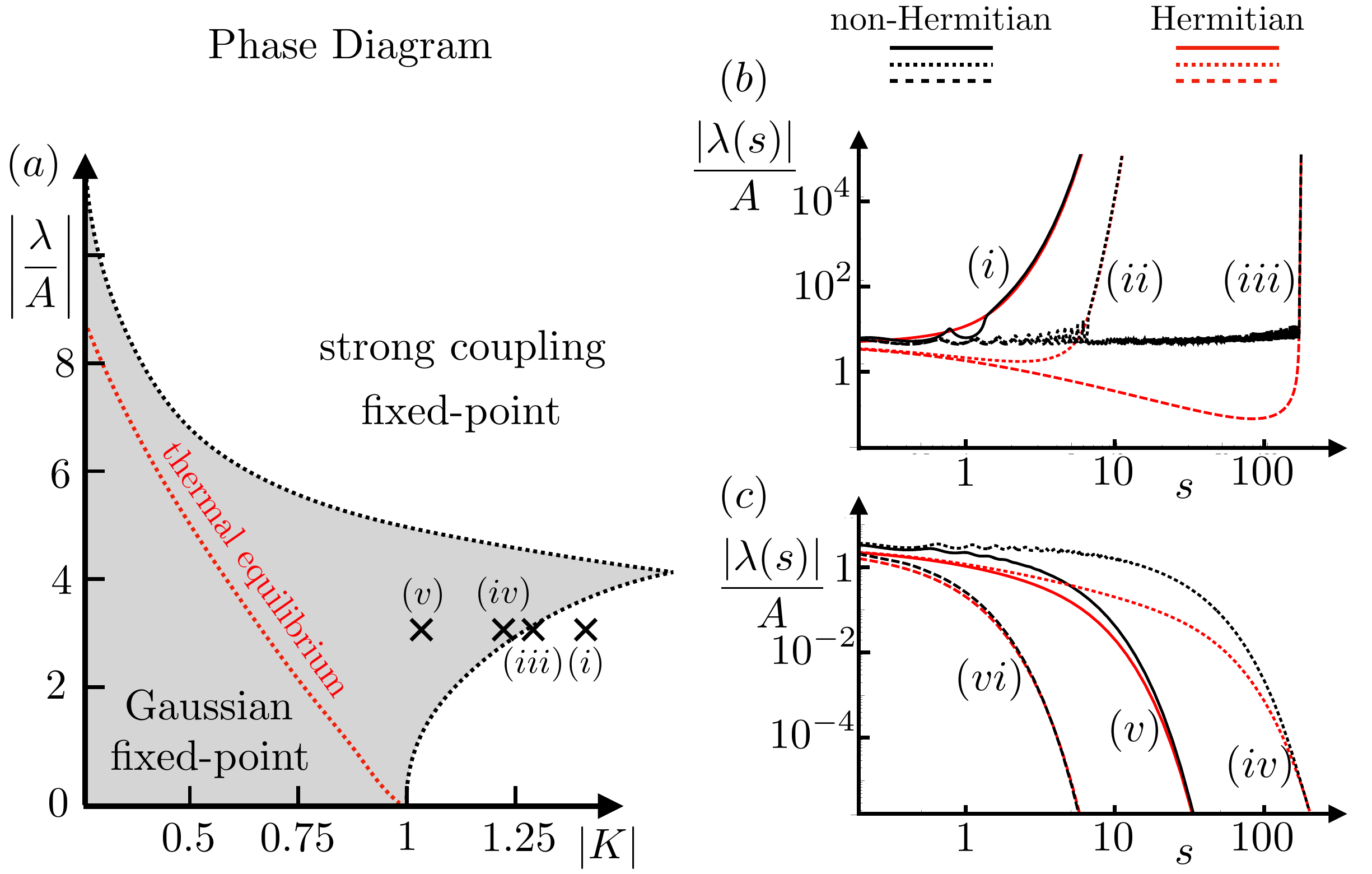}
    \caption{(a) Phase diagram of the non-Hermitian sine-Gordon field theory, obtained from numerical integration of the flow equations~\eqref{eq:Lren}, \eqref{eq:Kren}. In the grey region the theory flows towards the Gaussian fixed point located at the origin. The phase boundary is signalled by the black dotted line. For comparison we also show the transition line for the Hermitian (thermal equilibrium) theory. Figures (b) and (c) display the renormalization group flow towards the strong coupling fixed point (b) and the Gaussian fixed point (c) for the non-Hermitian sine-Gordon theory and the corresponding Hermitian sine-Gordon theory (with comparably larger starting value of $K$). The initial parameters of the curves (i) and (iii)-(v) are displayed in (a), the remaining curves correspond to $|K|=1.3$ for (ii) and $|K|=3/4$ in (vi). All curves start at $\lambda/A=3$.}
    \label{fig:KTflow}
\end{figure}
We already note that the $\cos$-nonlinearity here becomes irrelevant in the limit $\gamma\rightarrow 0$ (corresponding to $K,\eta\rightarrow1$), i.e., in the limit of zero measurement. This is due to the small prefactor of the Gaussian part $\sim \frac{1}{16\pi}$, which yields strong fluctuations of the field $\phi_X$. This prefactor arises from rescaling the fields with the large factor $\phi_X\to \phi_X/\sqrt{8}$ to eliminate the numerical prefactor of $\phi_X$ in the $\cos(\sqrt{8}\phi_X)$-term in the sine-Gordon Hamiltonian Eq.~\eqref{eq:effsineGordon}. It guarantees that the Gaussian fixed point is robust against an infinitesimal measurement rate. 

The renormalization group flow equations for the couplings $K, \lambda$ are derived in the conventional way, but here we consider the parameters $K, \lambda, \eta$ to be arbitrary complex numbers. The fields $\phi_X=\phi_X^{(<)}+\phi_X^{(>)}$ are decomposed into long-distance ($\phi_X^{(>)}$) and short-distance ($\phi_X^{(<)}$) fluctuations, where the long-distance fluctuations correspond to momenta $|k|<\Lambda/\xi$ and the short distance fluctuations correspond to momenta $\Lambda/\xi<|k|<\Lambda$. Here $\Lambda$ is a short-distance cutoff ($\Lambda=\pi$ in units of the lattice spacing) and $\xi=e^{s}$, where $s$ is the rescaling parameter controlling the renormalization group flow. The renormalization group flow is then obtained by integrating out the short-distance modes perturbatively in $\lambda$ and rescaling $x\rightarrow x/\xi$. This yields the flow equations
\begin{align}
    \partial_s \lambda=&\Big(2-\frac{2}{K}\Big)\lambda,\label{eq:Lren}\\
    \partial_s K=&-\lambda^2 A,\label{eq:Kren}
\end{align}
where $A$ is a positive number of order $\mathcal{O}(1)$ determined by the propagator of the Gaussian theory \cite{Amit_1980,Kadanoff77} (see App.~\ref{App:SineGordonRG}).   

The perturbative flow equations for the non-Hermitian sine-Gordon model in Eq.~\eqref{eq:effsineGordon} take the same form as those in the BKT scenario, but are more general due to the generally complex nature of the couplings $K, \lambda$. They include thermal equilibrium and some established non-Hermitian theories in particular limits, which can be immediately reproduced: If $K$ is real and $\lambda$ is purely imaginary, i.e., $\lambda= i|\lambda|$,  then the nonlinearity $\lambda$ is generally relevant for an initial $K>1$, and $K$ is monotonously growing under coarse graining, reproducing the conventional KT-flow diagram known from thermal equilibrium \cite{Amit_1980,Kadanoff77}. On the other hand, if $K$ and $\lambda$ are both real (corresponding to a purely imaginary nonlinearity but real Gaussian part), then $K$ is continuously decreasing, such that the only asymptotic fixed point of the renormalization group flow is the Gaussian one~\cite{FENDLEY_1993}.

In our situation, where the parameter of the Gaussian theory $K$ is generally complex, both $K, \lambda$ will not only experience a flow of their magnitude, but also of their complex phase. In this case, the magnitude of the parameter $K$ in Eq.~\eqref{eq:Kren} initially experiences both periods of growth and periods of shrinking during the RG flow, depending on the complex phase of $\lambda$. Since $\lambda$ is real initially, one starts with an RG flow that reduces $K$, shifting the critical point at which $\lambda$ becomes relevant to larger values of $K$ compared to the corresponding equilibrium RG flow.  Figure~\ref{fig:KTflow}(a) displays the phase diagram obtained from solving Eqs.~\eqref{eq:Lren}, \eqref{eq:Kren} numerically for the initial condition $K=\eta$.

The analysis shows that, while the initial, short distance behavior of the RG is affected by the complex valued nature of the coefficients, the asymptotic flow channels in on the one familiar from the Hermitian case. In consequence, the phase boundary is deformed compared to the latter, but the two familiar phases and their large distance behavior is reproduced, cf. Fig.~\ref{fig:KTflow}. In more detail, after an initial period of rapid phase oscillations, $\lambda$ either flows to zero, or the complex phases $\varphi_K, \varphi_\lambda$ of $K=|K|e^{i\varphi_K}, \lambda=|\lambda|^{i\varphi_\lambda}$ get pinned with respect to each other, fulfilling the relation $2\varphi_\lambda-\varphi_K=\pm\pi$. This happens typically at $s=\mathcal{O}(1)$ and from then on, the flow equations \eqref{eq:Lren}, \eqref{eq:Kren} describe the familiar thermal KT flow. See Fig.~\ref{fig:KTflow}(b) for examples of RG flow curves towards the strong coupling fixed point and Fig.~\ref{fig:KTflow}(c) for examples flowing towards the Gaussian fixed point.

\section{n-replica Keldysh construction}\label{sec:nkeldy}
Above we have studied two replicas, identifying center-of-mass and relative coordinates as useful degrees of freedom to capture the transition: In particular, for Gaussian problems, these coordinates decouple exactly. The center-of-mass coordinate undergoes indefinite heating, the relative coordinate is governed by a non-Hermitian  Hamiltonian. In this section, we address the question how this decomposition generalizes to $n$ replicas. Extending the theory to $n$ replicas enables us to compute higher order correlation functions, which are of $n$-th order in the conditioned density matrix $\rho^{(c)}$. In particular, it gives access to the computation of Rényi entropies and the von Neumann entanglement entropy, which we will discuss in this section. The construction is done within a replicated Keldysh functional integral formalism \cite{ALEINER2016378,Naoto,Shenker}, and yields two key results: First, we show that the transition established previously in the $2$-replica setup extends to the $n$ replica scenario. Second, we find that for Gaussian problems there continues to be an exact decoupling into one mode analogous to the center-of-mass coordinate, and $n-1$ modes analogous to the relative one; non-linear problems can then be treated on similar grounds as detailed above for the two-replica case.  

\subsection{General construction}
\begin{figure}
  \includegraphics[width=\linewidth]{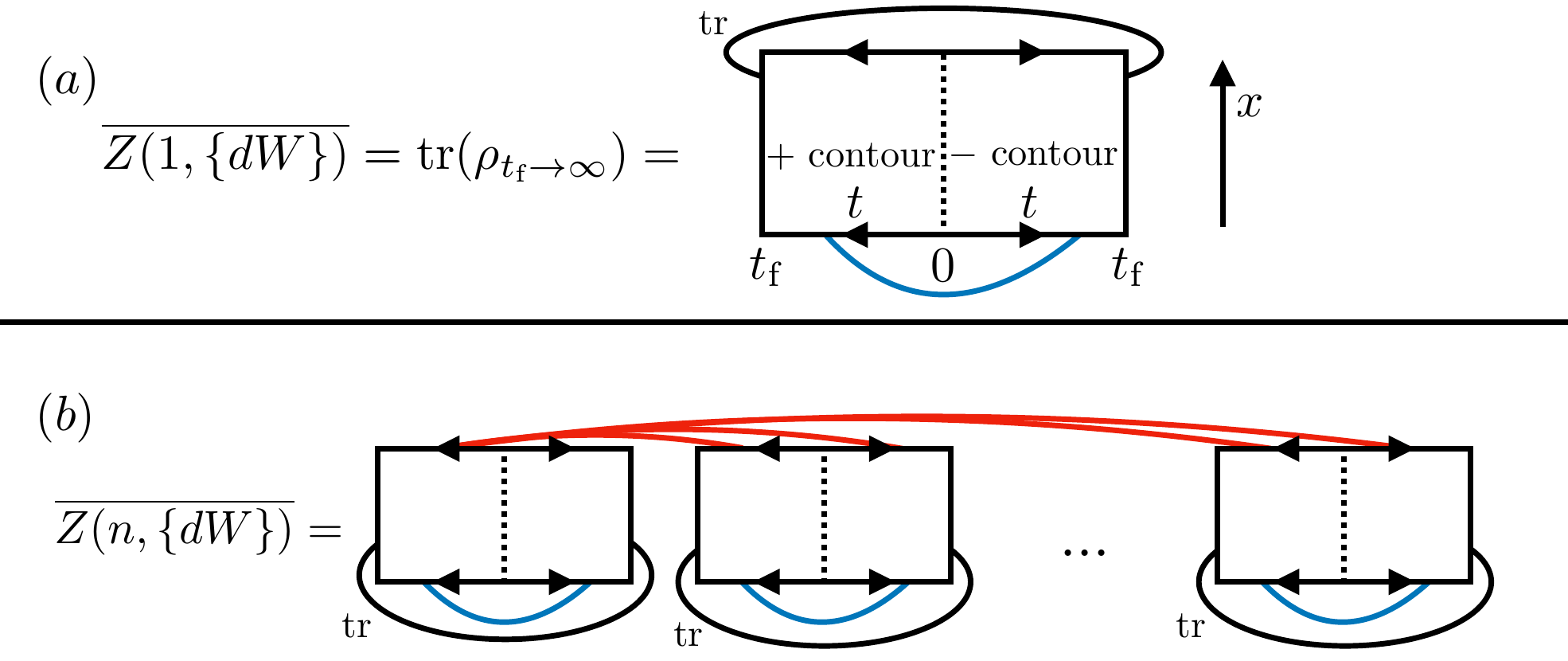}
  \caption{(a) Keldysh representation of a single replica. Time evolution of the noise averaged density matrix is captured by two Keldysh contours running horizontally; the spatial dimension is represented by the vertical lines. The trace operation is represented by a black line connecting final times $t_f$, for a selected position $x$. The noise average (lower blue line) yields the usual path integral representation of a Lindblad equation, with characteristic contour coupling terms.  (b) Keldysh representation of $n$ replicas. In addition to the coupling of contours within each replica, the average over the stochastic term couples all replicas (upper red lines). The trace operation acts on each replica individually, and is indicated by black lines.
  }
  \label{fig-MultiKeldysh}
\end{figure}

Starting point is again Eq. \eqref{eq:rhostoch} for the conditioned trajectory projector. We are then interested in  the update (see Eq.~\eqref{eq:RepIncrement})
\begin{eqnarray}\label{eq:drhon}
d \hat\rho^{(R_n)} = \overline{d \Big[\otimes_{l=1}^n \hat\rho^{(c)}\Big]}
\end{eqnarray}
of the trajectory average over $n$ identical copies of the system.
We can iterate this update in time, and represent the process in terms of a path integral by inserting coherent state resolutions of identity after each time step in the usual way. The resulting path integral is still conditioned on the noise realization, we finally take the noise  average, yielding the $n$-replica partition function $Z(n) = \overline{Z(n,\{dW\})}$. This bears similarities to the Keldysh double- or multi-contour constructions, which have been introduced to study out-of-time ordered correlation functions (OTOCs), with the major difference that the noise average introduces additional contour couplings compared to a Hermitian or purely Lindbladian evolution \cite{ALEINER2016378,Ansari16,Naoto,Shenker}. The procedure is illustrated in Fig.~\ref{fig-MultiKeldysh}. 

We consider again the general scenario of a continuously monitored quantum system, whose wave function evolution is described by the stochastic Schrödinger equation \eqref{eq:SSE}. The present construction is general, and encompasses fermions or bosons alike; for the sake of concreteness and notation, we first focus on a (1+1)-dimensional fermion system in spatial continuum, with a normal ordered Hamiltonian $\hat H=\hat H[\hat\psi^\dagger_{x}, \hat\psi_{ x}]$ 
and local, Hermitian measurement operators $\hat M_{s,x,t}=\hat O_{s,x}-\langle\hat O_{s,x}\rangle_t$, where $\hat O_{s,x} = \hat O_{s,x}[\hat\psi^\dagger_{x}, \hat\psi_{x}]$. Here $x$ is the position and the index $s$ distinguishes potentially different types of local measurements.
We leave implicit internal degrees of freedom of the fermion field, like $L,R$ for the Dirac fermions considered above. 

We illustrate the construction of the $n$-replica path integral step by step. For the sake of clarity, we will drop the spatial and measurement indices $x,s$ for now, set $\gamma=1$, and restore these quantities at the end. The evolution operator $\hat V_{dt}$ evolving the state $|\psi_{t}\rangle\rightarrow |\psi_{t+dt}\rangle=\hat V_{dt}|\psi_t\rangle$ then is
\begin{align}
    \hat V_{dt}=\exp\left[-(i \hat H+\hat M_t^2)dt+dW\hat M_t\right], 
\end{align}
which when expanded up to second order (since $dW^2=dt$) yields the stochastic Schr\"odinger equation for the increment $d|\psi_t\rangle=|\psi_{t+dt}\rangle-|\psi_t\rangle$. 

The single replica partition function for the conditioned projector $Z(1,\{dW\})\overset{t\rightarrow\infty}{=}\text{Tr}(\hat\rho^{(c)}_t)\overset{t\rightarrow\infty}{=}\text{Tr}(|\psi_t\rangle\langle\psi_t|)$ then requires two time strings or contours, corresponding to acting $\hat V_{dt}$ from the left ($+$ contour) and $\hat V_{dt}^\dagger$ from the right ($-$ contour) onto $\hat\rho^{(c)}_t$ at each time step. 
This can be expressed via a two-contour path integral over the (Grassmann valued, for fermions) fields $\psi_{\sigma}, \bar\psi_{\sigma}$, which carry a contour index $\sigma=\pm$.

%%%
The $n$-replica partition function $Z(n,\{dW\})=\text{Tr}[\otimes_{n=1}^n \hat\rho^{(c)}_t]$ is obtained from the product over $n$ independent trajectory projectors $\hat\rho^{(c)}$, yielding the partition function $Z(n,\{dW\})=[Z(1,\{dW\})]^n$. In order to express this via a single path integral, we add a replica index $l$ to the fields $\psi_{\sigma}^{(l)}, \bar\psi_{\sigma}^{(l)}$, accounting for each individual copy $\hat\rho^{(c)}$. The path integral expression for the stochastic $n$-replica partition function then is
\begin{align}
    Z(n,\{dW\})=\int \mathcal{D}[\Psi] \exp\left[i(S_{n,H}[\Psi]+S_{n,dW}[\Psi])\right]. \label{eq:nReppf}
\end{align}
The action is composed of a Hamiltonian part $S_{n,H}[\Psi]$ and a measurement part $S_{n,dW}[\Psi]$,
\begin{align}
    S_{n,H}[\Psi]=&\sum_{\sigma=\pm}\sum_{l=1}^n \sigma \int_t \left(\bar\psi_{\sigma,t}^{(l)} i\partial_t \psi_{\sigma,t}^{(l)}-H[\bar\psi_{\sigma,t}^{(l)},\psi_{\sigma,t}^{(l)}]\right),\label{unitaction}\\
    S_{n,dW}[\Psi]=&i\sum_{\sigma=\pm}\sum_{l=1}^n\int_t\left( [M_{\sigma,t}^{(l)}]^2-dW_tM_{\sigma,t}^{(l)}\right).\label{measact}
\end{align}
Here we have bestowed $dW_t$ a temporal index, since increments at different times are uncorrelated and the averaging prescription in the operator formalism is replaced by $\overline{dW_t dW_{t'}}=\delta(t-t')$. For fermions, the functional integral has anti-periodic boundary conditions in time, which we do not make explicit in the notation here.

The partition function $Z(n,\{dW\})$ is the product of $2n$ independent path integrals ($n$ forward and $n$ backward contours). They share in common that all contours couple to the same noise increment $dW_t$, which acts as a contour-independent source term. Performing the noise average via integration over the Gaussian distributed increments $dW_t$, the partition function $Z(n)=\overline{Z(n,\{dW\})}$ results, which is no longer of product structure. The corresponding path integral is
\begin{align}\label{eq:totact}
Z(n) = \int \mathcal D \Psi \ \exp\left[i (S_{n,H}[\Psi]+S_{n,M}[\Psi])\right],
\end{align}
where the unitary part $S_{n,H}[\Psi]$~\eqref{unitaction} remains unchanged by the noise average. The measurement action $S_{n,M}[\Psi]$ will now be discussed for the cases $n=1$ and $n>1$ separately.

For the single replica, taking the average over the stochastic increment couples the $(\pm)$-contours, as illustrated in Fig.~\ref{fig-MultiKeldysh}(a), and yields
\begin{align}
    S_{1,M}[\Psi]=&i\int_t M_{+,t}^2+M_{-,t}^2-\frac{1}{2}(M_{+,t}+M_{-,t})^2\nonumber\\
    =&\frac{i}{2}\int_t (O_{+,t}-O_{-,t})^2.
\end{align}
It therefore removes any state-dependent term ($\sim \langle \hat O_{s,x}\rangle_t$), and makes the noise averaged single replica evolution linear in the state. 
This is then equivalent to the path integral representation of a Lindblad equation. In particular, for the Hermitian  Lindblad operators considered here, the stationary state described by it is at infinite temperature.

Next we consider the case $n>1$, which is illustrated in Fig.~\ref{fig-MultiKeldysh}(b). In addition to intra-replica contour couplings, the noise average produces additional inter-replica couplings (red lines in Fig.~\ref{fig-MultiKeldysh}(b)). Here, the measurement expectation values generally do not drop out, and we have to perform the average in the mean-field decoupling approximation outlined in Sec.~\ref{sec:2RepMaster}. In the path integral description this amounts to the approximation that $M_{\sigma,t}^{(l)}$ is independent of the history of $dW_{t}$. Importantly, for the same reason as in the two-replica operator formulation, the thus obtained action functional features linear, noise averaged trajectory expectation values, such as $\overline{\langle \hat O_{s,x}\rangle_t}$. These are now averages obtained from the path integral and, focusing on the long time limit, we replace them directly by their stationary values, e.g. $\overline{\langle \hat O_{s,x}\rangle_t} \to o_s$. The action obtained from this average is 
\begin{align}
    S_{n,M}[\Psi]=&i\int_t \sum_{l=1}^n\left([M_{+,t}^{(l)}]^2+[M_{+,t}^{(l)}]^2\right)-\frac{1}{2}\left(\sum_{l=1}^nM_{+,t}^{(l)}+M_{-,t}^{(l)}\right)^2.
\end{align}

Restoring the spatial and measurement indices in  $\hat O_{s,x}, \hat\psi^\dagger_x, \hat\psi_x$ and considering arbitrary $\gamma$, the measurement-induced action is
\begin{widetext}
\begin{eqnarray}\label{eq:MeasIndAction}
S_{n,M}[\Psi] &=& 
-i\gamma \int_X\left(\sum_{l=1}^n\sum_s \left[(O^{(l)}_{s,+,X})^2+(O^{(l)}_{s,-,X})^2\right]-2(\Omega_{s,+,X}+\Omega_{s,-,X}) o_s-\tfrac{1}{2}\left[\Omega_{s,+,X}+\Omega_{s,-,X}-2n\cdot o_s\right]^2\right),
\end{eqnarray}
\end{widetext}
with $\Omega_{s,\pm,X}=\sum_{l=1}^n O^{(l)}_{s,\pm, X}$. 
Eq.~\eqref{eq:MeasIndAction}  yields an interesting structure. While the first term is of the form of an effective non-Hermitian Hamiltonian acting on each replica individually, the second term features inter-contour couplings only between two collective degrees of freedom $\Omega_{s,\pm,X}$. For $n>1$ this potentially enables degrees of freedom orthogonal to $\Omega_{s,\pm,X}$ (see Sec. \ref{sec:bosmeas} below), which are free of contour coupling terms and only evolve according to an effective Hamiltonian.

\subsection{Bosonization of the multi-replica theory and decoupling of the Gaussian theory}\label{sec:bosmeas}
We apply this setup to the continuum measurement model presented in Sec.~\ref{sec:fermbos}. In the fermionic formulation defined by Eqs.~(\ref{eq:hfcont},\ref{eq:mfcont}), we first have the massless Dirac action associated to unitary evolution 
\begin{align}
    S_{n,H}[\Psi]=\sum_{l=1}^n \sum_{\sigma=\pm}\sigma\int_X \bar\Psi^{(l)}_{\sigma,X}i(\partial_t-v \sigma_z\partial_x)\Psi^{(l)}_{\sigma,X}.
\end{align}
We consider two independent sets of measurement operators
\begin{align}
 O^{(l)}_{1,\sigma,X}=\bar\Psi^{(l)}_{\sigma,X} \Psi^{(l)}_{\sigma,X}, \quad  O^{(l)}_{2,\sigma,X}=\bar\Psi^{(l)}_{\sigma,X}\sigma_x \Psi^{(l)}_{\sigma,X},
\end{align}
each measured with rate $\gamma$.
In the bosonized formulation, Eqs.~~(\ref{SEq10},\ref{eq:mbcont}), the action is defined with the expressions
\begin{align}
    S_{n,H}[\phi]=&-\frac{1}{2\pi}\sum_{l=1}^n \sum_{\sigma=\pm}\sigma\int_X \phi^{(l)}_{\sigma,X}\partial^2\phi^{(l)}_{\sigma,X}, \nonumber\\
      \hat O_{1,\sigma, X}[\phi]=&-\frac{1}{\pi}\partial_x \phi^{(l)}_{\sigma,X}, \quad \hat O_{2,\sigma, X}=m \cos(2\phi^{(l)}_{\sigma,X}),\label{eq:NRepBos}
\end{align}
where $\partial^2 \equiv \partial_t^2 -\partial_x^2$.

Before we discuss the general, nonlinear theory, we now focus on the Gaussian bosonic setting, relevant to the cases of weak and strong monitoring in the above model. It is defined with a quadratic Hamiltonian and measurement operators that are linear functions of Hermitian (bosonic) field operators, i.e.,  $O^{(l)}_{\sigma,X}=D\phi^{(l)}_{\sigma,X}$, where $D=m\pi, i\partial_x$ can be either a mass or a derivative. We then find a decoupling analogous to the one in the 2-replica case discussed in Sec.~\ref{sec:2bosonReplica}. 

\begin{figure}\includegraphics[width=\linewidth]{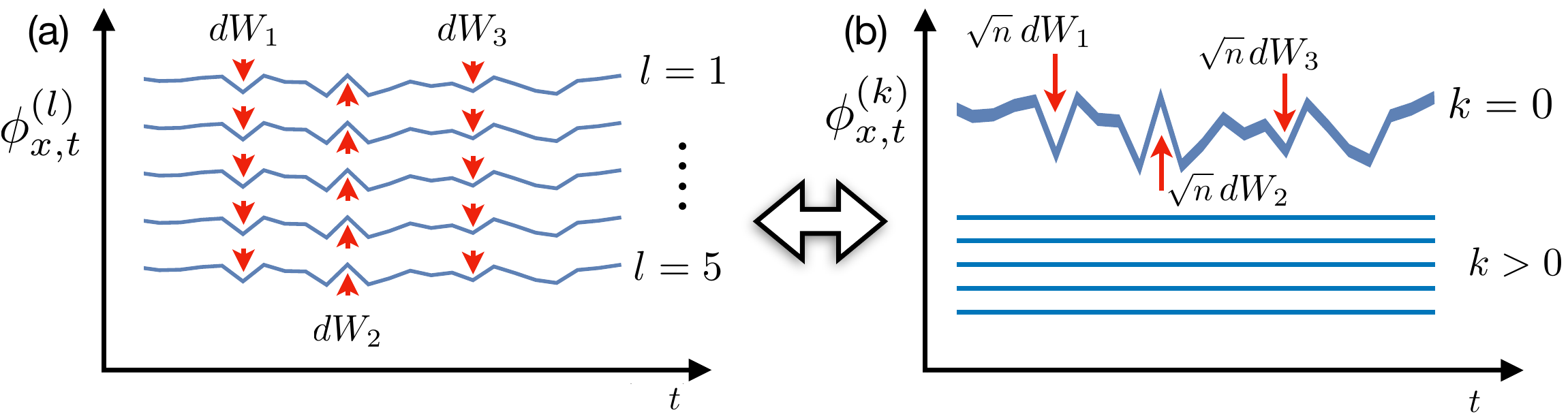}\caption{Illustration of the coupling of the fields $\phi_{X}$ to the measurement noise at a fixed position $x$ and for different times $t$: (a) In the replica basis $\{\phi^{(l)}_{\sigma,X}\}$, each field couples to the the same (spatiotemporal) measurement noise $dW_t$ (the individual fields are given an offset with respect to each other for better visibility). (b) In the Fourier basis $\{\phi_{\sigma,X}^{(k)}\}$, defined in Eq.~\eqref{eq:FourierRep}, the $k=0$ or center-of-mass mode couples to an enhanced noise $\sqrt{n}dW_t$ while all the relative modes $(k>0)$ decouple from $dW_t$ and are noise-free.}
  \label{Fig-noise}
\end{figure}%%%

In order to motivate this decoupling on an intuitive level, we start from the action $S=S_{n,H}+S_{n,dW}$ in Eq.~\eqref{eq:nReppf}, for which the noise has not yet been integrated out. In the bosonized framework, the action of the Hamiltonian part $S_{n,H}$ is the described by Eq.~\eqref{eq:NRepBos}, and the measurement part $S_{n,dW}$ by Eq.~\eqref{measact} with the boson measurement operators $\hat O_{\alpha,\sigma,X}$ being linearized. One can then derive the equation of motion for the field $\phi_{+,X}^{(l)}$ of the $l$-th replica from the saddle point condition $\frac{\delta S}{\delta\phi_{+,X}^{(l)}}$. This yields 
\begin{eqnarray}
\partial_t^2\phi^{(l)}_{+,X}=(\partial_x^2-\frac{\gamma i}{\pi v}D^2)\phi^{(l)}_{+,X}+dW_t.
\end{eqnarray}
This equation is linear in both the fields $\phi_{+,X}^{(l)}$ and the noise $dW_t$, and each replica $l$ experiences the same noise realization (see Fig.~\ref{Fig-noise}). Therefore, it is possible to introduce generalized a center-of-mass modes and relative modes, where the first absorbs all the noise, while it cancels for the latter. To this end, we Fourier expand for each $l$ in the equation of motion
\begin{eqnarray}\label{eq:FourierRep}
\phi^{(l)}_{\sigma,X}=\frac{1}{\sqrt{n}}\sum_{k=0}^{n-1}  e^{-i\frac{2\pi k l}{n}}\phi^{(k)}_{\sigma,X},
\end{eqnarray}
and use $dW_t = \frac{1}{\sqrt{n}}\sum_{k=0}^{n-1}  e^{-i\frac{2\pi k l}{n}}\delta_{k,0}(\sqrt{n} dW_t)$, to compare coefficients over $k$. Indeed, all relative modes $k\neq 0$ are then noiseless, while the center-of-mass mode $k=0$ is affected by a noise of strength $\sqrt{n} dW_t$,
\begin{eqnarray}
\partial_t^2\phi^{(k)}_{+,X}=(\partial_x^2-\frac{\gamma i}{\pi v}D^2)\phi^{(k)}_{+,X}+\sqrt{n}dW_t \delta_{k,0}.
\end{eqnarray}

In order to proceed, the noise average is performed, which then yields the averaged $n$-replica action 
$S_n[\phi]=S_{n,H}+S_{n,M}$, where $S_{n,H}$ is again provided by Eq.~\eqref{eq:NRepBos} and the averaged measurement action $S_{n,M}$ corresponds to Eq.~\eqref{eq:MeasIndAction} with the corresponding boson measurement operators, i.e.,  in terms of bosonic fields $\phi= (\phi^{(0)} , ... , \phi^{(n-1)})^T, \phi_X^{(k)} = ( \phi_{+,X}^{(k)}, \phi_{-,X}^{(k)})$, and again linearized. This yields the sum
\begin{align}\label{eq:linact}
S_n[\phi] = \sum_{k=0}^{n-1}S^{(k)}[\phi^{(k)}].\end{align}
The above action has one  center-of-mass or Fourier $k=0$ mode $\phi^{(k=0)}=\frac{1}{\sqrt{n}}\sum_l\phi^{(l)}$, which is described by
\begin{align}
S^{(0)}[\phi^{(0)}]=-\frac{1}{2\pi}\int_X\Big\{& \sum_{\sigma=\pm}[\phi_{\sigma,X}^{(0)}(\sigma \partial^2-\frac{\gamma i}{\pi v}D^2)\phi_{\sigma,X}^{(0)}]\\&+\frac{n\gamma i}{\pi v}\phi_{c,X}^{(0)}D^2\phi_{c,X}^{(0)}\Big\}\nonumber,
\end{align}
where  $\phi_{c,X}^{(l)}=\frac{\phi_{+,X}^{(0)}+\phi_{-,X}^{(0)}}{\sqrt{2}}$.
Due to the intra-replica contour coupling term (last contribution in Eq.~\eqref{eq:MeasIndAction}) there is heating to an infinite temperature state. This can be inferred from the Green's function of the average mode in Fourier space ($D=m\pi,q$)
\begin{align}\label{eq:g0}
    (G^{(0)}_{q,\omega})^{-1}=\sigma_z(\omega^2-q^2)+\mathds{1}_{2\times 2}\frac{i\gamma(n-2)}{2v\pi}D^2+i\frac{\gamma n}{2v\pi}D^2\sigma_x.
\end{align}
It has four poles at frequencies 
\begin{align}
    \omega_{\pm,\pm}=\pm\sqrt{q^2\pm D^2\frac{\gamma}{\pi v}\sqrt{n-1}}.
\end{align}
For $n>1$, this implies that at least two of the poles lie on the real axis \footnote{This behavior persists under introducing a regularization: The latter would enter Eq.~\eqref{eq:g0} as an additional matrix $i\epsilon \omega \sigma_z$, the effect of which is overwritten by any non-zero measurement rate $\gamma$, and does not shift the poles away from the real axis.}. They cause all the matrix elements of the equal-time correlation function for a given momentum mode $q$ to diverge
\begin{align}
    G^{(0)}(q,t=0)=\int_\omega G^{(0)}(q,\omega)\rightarrow\infty .
\end{align}
This indicates unbounded fluctuations of $\phi^{(0)}_X$, characteristic of the infinite temperature state, and in line with the findings discussed around Eq.~\eqref{eq:growth}. 

In addition, the action $S_n[\phi]$ features $n-1$ orthogonal relative modes $\phi^{(k)}, k = 1 , ... , n-1$. These do not involve intra-replica contour coupling in their action ($k>0$)
\begin{align}
S^{(k)}[\phi^{(k)}]=-\frac{1}{2\pi}\int_X \sum_{\sigma=\pm}\left[ \phi_{\sigma,X}^{(k)}(\sigma\partial^2-\frac{\gamma i}{\pi v}D^2)\tilde\phi_{\sigma,X}^{(k)}\right],\label{eq:freebosonact}
\end{align}
where we have introduced the shortcut $\tilde\phi^{(k)}_{\sigma,X}=\phi^{(n-k)}_{\sigma,X}$.
 
Each set of relative fields $\phi^{(k>0)}_X$ evolves according to a non-Hermitian action, yielding a Keldysh Green's function
\begin{eqnarray}
iG^{K (k >0)}(Q) = \braket{\phi_{c,Q}^{(k>0)}\tilde\phi_{c,-Q}^{(k>0)}} = \frac{-i \tilde\gamma  D^2}{(q^2-\omega^2)^2+ D^4\tilde\gamma^2}.\label{eq:relKeldGreen}
\end{eqnarray}
The corresponding equal-time correlation functions
\begin{eqnarray}
G^{K (k >0)}(q,t=0)=2\text{Im}\Big[\big(i\tilde \gamma  D^2-q^2\big)^{-\frac{1}{2}}\Big]
\end{eqnarray}
show the same scaling behavior as for a Luttinger Liquid in the ground state, i.e., $G^{K (k >0)}(q,t=0)\sim |q|$ for $D\sim q$, but with a different, $\gamma$-dependent amplitude.

As a consequence of the decoupling, the $n$-replica partition function for linear measurement operators $O_X^{(l)}\sim D\phi_X^{(l)}$ and a quadratic Hamiltonian factorizes into $n$ independent products. This remains true when coupling the replica fields $\phi_X$ linearly to local sources $h_X$. For later purposes we introduce such coupling to source terms via $\int_X h_X^T\phi_X$, with $h= (h^{(0)} , ... , h^{(n-1)})^T, h_X^{(k)} = ( h_{c,X}^{(k)}, h_{q,X}^{(k)})^T$. The sourced $n$-replica partition function then obtains by Gaussian integration and reads
\begin{eqnarray}\label{eq:znh}
Z(n)[h] &=& \prod_{k=0}^{n-1} Z^{(k)}[h], \\\nonumber
Z^{(k)}[h]&=&\braket{e^{i \int_X h_X^T\phi_X}}= Z^{(k)}[0] \exp\left[-\tfrac{1}{2}\hspace{-2mm}\int_{X,X'}\hspace{-2mm}h^{(k)}_{\sigma,X}\langle\phi^{(k)}_{\sigma,X}\tilde\phi^{(k)}_{\sigma' ,X'}\rangle \tilde h^{(k)}_{\sigma'  X'}\right].
\end{eqnarray}
This allows us to efficiently compute $n$-replica correlation functions of arbitrary power in the fields $\phi_X$, including the $n$-th order Rényi entropies, which we will discuss in the following section.

Based on the bosonic formulation, and our understanding of the Gaussian theory, now we are in the position to discuss the measurement-induced phase transition imposed by a nonlinear measurement operator $O_{2,\sigma,X}$. For $n=2$ replicas, this problem is equivalent to the boson model that we analyzed in Secs.~\ref{SecRic}-\ref{sec:RG}. 
For $n>2$, we can proceed in the same manner and bring the Gaussian part of the action, i.e., the sum $S_{n,H} +S^{(0)}_{n,M}$, into a replica diagonal form by applying the replica-Fourier transform described in Eq.~\eqref{eq:FourierRep}. The 'hot' or infinite-temperature mode $\phi^{(k=0)}_{\sigma,X}$ can then again be integrated out, yielding the remaining action for the $n-1$ equivalent relative modes with $\phi^{(k>0)}_{\sigma,X}$. These modes are coupled by the $\cos$-nonlinearities, which contain their sum as arguments, e.g., the term $\cos(\sqrt{2}\hat\phi^{(1)}_X)\rightarrow\cos(\sum_{k>0} a_k^{(1)}\hat\phi_X^{(k)})$ after integrating out the $k=0$ mode and with the Fourier coefficients $a^{(l)}_k=\sqrt{\frac{2}{n}}e^{i2\pi lk/n}\in\mathds{C}$. The Gaussian part of the action for $k>0$, however, does not couple different replica sectors $k, k'$ and furthermore is identically the same for each $k>0$. Therefore the common BKT perturbative renormalization group scheme boils down to the individual but identical, multiplicative renormalization of the factors $e^{\pm i\phi^{(k)}_{\sigma, X}}$ for each $k>0$. While deriving the set of flow equations for $n>2$ may be tedious, the details outlined above ensure that in the Gaussian sector no coupling terms between different replica indices $k, k'$ can be generated. This in turn ensures that the Gaussian fixed point for $n>2$ is of the same form as for $n=2$, describing $n-1$ decoupled free boson theories.

\subsection{Entanglement transition}\label{sec:EntEnt}
In this section, we discuss how the measurement-induced phase transition modifies the structure of the entanglement entropy. On both sides of the transition, the effective theory at long wavelengths is a theory of free bosons. The entanglement entropy of the bosons can thus be determined from conventional approaches for free theories, e.g., from the boson covariance matrix~\cite{Vidal2003,Peschel_2003,Cramer2006}, which is discussed in Appendix~\ref{app:entro}. Here, however, one is mainly interested in the {\it fermion} entanglement entropy, i.e., in the entanglement entropy of the original, microscopic fermion degrees of freedom.

In order to capture the entanglement entropy of the free Dirac fermions~\footnote{The compactification radius for the Dirac fermions subject to measurements is the same as for noninteracting fermions, i.e., $K=1$.}, we implement the approach outlined in Refs.~\cite{Casini_2005, Casini_2009}. Starting with the definition of the Rényi entropies, this approach then predicts that the {\it fermion} entropies are exclusively determined from the {\it boson } correlation functions of the relative replica modes $k>0$. The structural difference between the $n-1$ replica fluctuations modes at the weak and strong measurement fixed points then already give a hint on the expected scaling behavior of the entanglement entropy. In the weakly and strongly monitored limits, it exhibits subextensive logarithmic growth or area law saturation, respectively, confirming a phase transition in the entanglement entropy. We use RG arguments to interpolate between these regimes. We also determine the {\it effective} central charge in the weak monitoring regime and find that it behaves non-universally. Our approach then confirms an algebraic decrease of the effective central charge $c(\gamma)\sim \gamma^{-\kappa}$ as a function of the measurement strength $\gamma$ in the weak monitoring regime, and a sharp drop of $c(\gamma)$ at the measurement-induced transition, both of which was reported for monitored lattice fermions in Refs.~\cite{Alberton20, chen2020}.

\textit{General construction} -- 
To compute the entanglement entropy in the measurement problem, we leverage the techniques developed in Ref.~\cite{Casini_2005} for free relativistic fermions in $(1+1)$ dimensions to our replica field theory (Rényi entropies for quantum heat engines have been obtained in a similar multicontour formalism in \cite{Ansari16}). More precisely, we first derive a general expression for the $n$-th Rényi entropy and for the von Neumann entanglement entropy for the monitored fermions. Then we show that the fermion entropies are directly related to the correlation functions of the bosonized model.

For a general bipartition of the one-dimensional system into two disjoint subsystems $A=[x_0,x_0+L]$ and $B=\mathds{R}\setminus A$, the $n$-th R\'enyi entropy for the measured system is defined as 
\begin{align}\label{eq:entro}
S_n=\tfrac{1}{1-n}\ \overline{\log Z_A(n, \{dW\})},\  Z_A(n,\{dW\}) \equiv \text{tr}[ (\hat\rho_A^{(c)})^n].\ \ \
\end{align}
Here, $Z_A(n,\{dW\})$ is the reduced partition function for the subset $A$, which is obtained by first performing a partial trace over $B$ to obtain $\hat\rho^{(c)}_A=\text{tr}_B\hat\rho^{(c)}$, and then taking the trace of the $n$-replica matrix product $(\hat\rho_A^{(c)})^n$.

The reduced partition function $Z_A(n, \{dW\})$ can computed from the real-time path integral for the $n$-replica partition function Eq.~\eqref{eq:nReppf}. In order to do so, the partial trace and the matrix multiplication in Eq.~\eqref{eq:entro} have to be implemented by a set of additional boundary conditions, which we discuss in detail in Appendix~\ref{sec:EntApp} (see also Refs.~\cite{Casini_2005, Casini_2009}). The key requirement here is that the boundary conditions can be imposed via a gauge potential, which only modifies derivatives of the fields. After bosonization, this yields the reduced partition function 
\begin{eqnarray}\label{eq:ZA}
Z_{A}(n, \{dW\})=\prod_k \langle  \exp\left(-\sqrt{2}i\frac{k}{n}(\phi^{(k)}_{c,x_0}-\phi^{(k)}_{c,x_0+L})\right)\rangle.
\end{eqnarray}
Here $\phi^{(k)}_{c,X}$ is the replica Fourier transform of $\phi^{(l)}_{c,X}=\frac{1}{\sqrt{2}}(\phi^{(l)}_{+,X}+\phi^{(l)}_{-,X})$, defined in Eq.~\eqref{eq:FourierRep}, and the average is performed in the steady state, i.e., with respect to the bosonic measurement action discussed in Sec.~\ref{sec:bosmeas}. Crucially, the center-of-mass mode with $k=0$ only yields a trivial contribution to Eq.~\eqref{eq:ZA}. The particular values of the noise $\sim dW$ therefore do not modify the reduced $n$-replica partition function, i.e., $Z_{A}(n, \{dW\})=Z_{A}(n, 0)$. This decoupling of the reduced partition function, and therefore of the Rényi entropies, from the noise is a consequence of the decoupling of all the $k>0$ modes from the noise. It should therefore be a general property of Gaussian theories, and it is also found for the boson entropies in Appendix~\ref{app:entro}. Our interpretation of this decoupling is that each individual wave function, while showing random fluctuations
induced by the local measurements on the basis of local observables, has the same global entanglement pattern.

The result in Eq.~\eqref{eq:ZA} may be viewed as the bosonic field $\phi^{(k)}_{c,X}$ being coupled to two point-like, opposite charges located at the beginning $x_0$ and the end $x_0+L$ of the non-traced-out interval $A$. A similar relation has been found in Refs.~\cite{bao2021symmetry, zhang2021syk} for monitored Majorana modes, for which the full charges are replaced by half-charges~\footnote{In the framework of a classical XY model, the charges appear as (half-) vortices, matching the terminology in the references.}.   %It amounts to source fields $h^{(k)}_{q,X} = 2\frac{k}{n}[\delta(x-x_0) -\delta(x-x_0-L) ]\delta(t-t_f)$, yielding time-local correlations according to Eq.~\eqref{eq:znh}, which are independent of $t_f$ in the stationary state. 
In the limit of either weak or strong monitoring, the bosonic theory can be linearized and we can use Eqs. \eqref{eq:znh}-\eqref{eq:ZA} to get explicit results for the Rényi entropies. Apart from constant contributions $\sim Z^{(k)}[0]$, we obtain
\begin{eqnarray}\label{eq:entrolutt}
S_n=\frac{1}{1-n}\sum_{k=1}^{n-1}\frac{2k^2}{n^2} \langle\phi_{c,x_0}^{(k)}\phi_{c,x_0}^{(k)}-\phi_{c,x_0}^{(k)}\phi^{(k)}_{c,x_0+L}\rangle.
\end{eqnarray}
The correlation functions of the replica Fourier modes $k>0$ are all independent of $k$ and identical. The summation over $k$ can therefore be performed, and we can infer the von Neumann entanglement entropy $S=\lim_{n\rightarrow1}S_n$, yielding
\begin{eqnarray}\label{eq:entrolutt1a}
S=\frac{2}{3} \langle\phi_{c,x_0}^{(k>0)}\phi^{(k>0)}_{c,x_0+L}\rangle.
\end{eqnarray}

\textit{Weak monitoring entropy, $\gamma\ll v$} -- 
If the nonlinearity is irrelevant, the renormalization group flow approaches a Gaussian fixed point at which $\lambda=0$. The theory is then characterized by a single quantity, i.e., the fixed point value of $K$. The correlation function at this fixed point is 
\begin{eqnarray}
\langle \phi^{(k>0)}_{c;q}\phi^{(k>0)}_{c;-q}\rangle= \text{Re}\Big[\frac{\pi}{K|q|}\Big] \Rightarrow \langle \phi^{(k>0)}_{c,x_0}\phi^{(k>0)}_{c,x_0+L}\rangle=2c(\gamma)\log(L).\nonumber
\end{eqnarray}
In this case, Eq.~\eqref{eq:entrolutt1a} predicts the von Neumann entanglement entropy and the effective central charge
\begin{eqnarray}
S=\frac{1}{3}c(\gamma)\log(L), \quad c(\gamma ) =\text{Re}(1/K).\label{eq:centralcharge}
\end{eqnarray}

For $\lambda=0$ but $\gamma>0$, the system is already initialized at a Gaussian fixed point with $K=\sqrt{1-i\frac{2\gamma}{\pi v}}$. Then the prefactor of the $\log$-scaling behaves $\sim 1/\sqrt{\gamma}$ for sufficiently large $\gamma$. A suppression of the effective central charge with the measurement strength has been observed to be a general feature of monitored free fermions, including away from the critical point~\cite{Alberton20,chen2020}.

\begin{figure}\includegraphics[width=\linewidth]{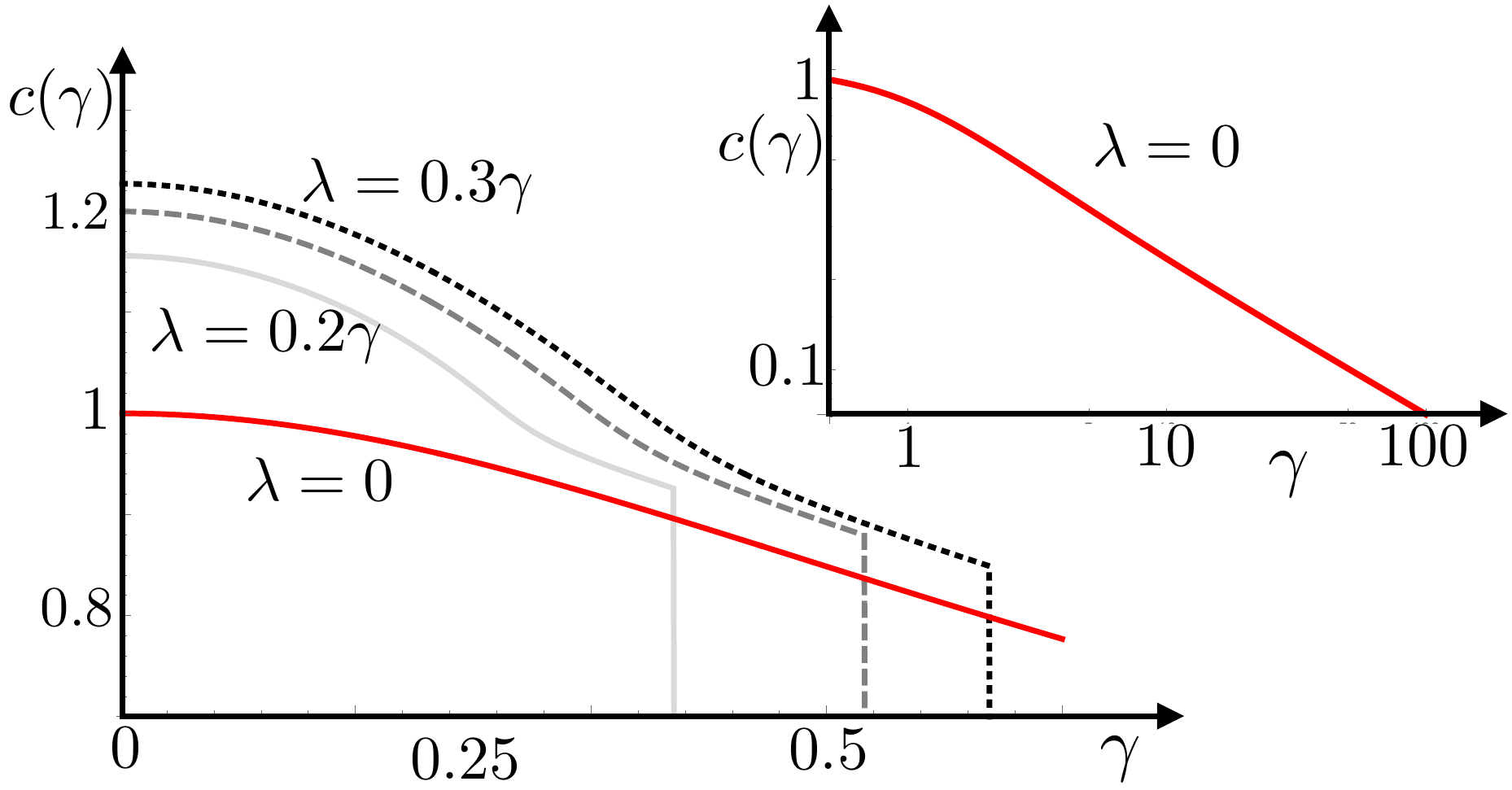}\caption{Prefactor of the logarithmic scaling function (effective central charge) $c(\gamma)$ obtained from integrating the RG equation \eqref{eq:Kren} for the parameter $K$. For $\lambda=0$, we observe the scaling $c(\gamma)\sim\gamma^{-0.5}$ (red line and inset). For $\lambda\neq0$, a transition into the area law phase happens at a critical measurement strength $\gamma_c(\lambda)$, where $c(\gamma)$ drops to zero. In the weak measurement regime, $c(\gamma)$ increases with $\lambda$. The lines correspond to $\lambda=0.2\gamma$ (light gray, bold), $\lambda=0.25\gamma$ (dark gray, dashed) and $\lambda=0.3\gamma$ (black, dotted).}
  \label{fig-CC}
\end{figure}

For the general case $\gamma>0, \lambda>0$, the nonlinear terms lead to a renormalization of the Gaussian part of the theory. In particular, $K$ receives renormalization corrections, and should be evaluated at the fixed point for a description of the asymptotic long distance physics. Generally, we observe a growth of the parameter $K$ during the renormalization group flow. Inserting the fixed point values of $K$ into the above formula then yields a slightly enhanced suppression of the boson correlation function and of the effective central charge, followed by a sudden jump to $c(\gamma)\rightarrow0$ at the transition, which is illustrated in Fig.~\ref{fig-CC}.

%%%

\textit{Strong measurement entropy, $\gamma\gg v$} -- 
For strong measurements, the problem is gapped. Repeating the calculation from above but with non-zero mass $m$ yields, for a bipartition of the system with subsystem sizes $L\gg m^{-1}$,
\begin{align}
S_{\text{vN}}(L)=\frac{c(\gamma)}{3}\log\left(\frac{L}{\sqrt{1+m^2L^2}}\right) \approx
\frac{c(\gamma)}{3}\log m^{-1} .\label{eq:arealawent}
\end{align}
The entanglement entropy saturates and does no longer depend on $L$. The overall prefactor of the entropy decays again monotonously with $\gamma$. Here, however, the inverse mass $m^{-1}$ is a measurement-induced correlation length. It is also $\gamma$-dependent, approaching the short-distance cutoff ($=1$ on the lattice) of the theory in the limit $\gamma\rightarrow\infty$.

\textit{Zero monitoring, $\gamma=0$} --
We have observed that in the limit $\gamma\rightarrow0$ the entropy approaches the behavior of a Luttinger Liquid theory in the ground state with $c(0)=1$. However, this has to be taken with care, since it involves two non-commuting limits. Setting $\gamma=0$ first, i.e., on the level of the microscopic theory, then the state of the system after finite time always depends on the initial state of the system. However, when evaluating the entanglement entropy for $\gamma>0$ first and then taking the limit $\gamma\rightarrow0$, any initial state dependence drops out. This is the consequence of the ‘cooling’ effect for the relative modes with $k\neq 0$, which govern the entanglement entropy, cf. Eq.~\eqref{eq:entrolutt}.

For instance, for a thermal state at temperature $T$, the initial condition in the Keldysh framework is usually implemented via an infinitesimal contribution to the quantum-quantum sector~\cite{Kamenev}
\begin{align}
    S^{qq}=\int_P 2i\epsilon\omega\coth(\frac{\omega}{2T})\phi_{q,-P}\phi_{q,P},
\end{align}
with $\epsilon=0^+$ infinitesimal, and $\phi_{q,P}=\frac{\phi_{+,P}-\phi_{-,P}}{\sqrt{2}}$. This initial contribution appears in the quadratic sector and is identical for each replica, such that it is directly carried over to the relative and center-of-mass coordinates. For any non-vanishing measurement rate $\gamma>0$ the infinitesimal contribution $\sim \epsilon\omega\coth(\frac{\omega}{2T})$, however, is overwritten by a finite imaginary contribution $\sim \gamma$ in Eq.~\eqref{eq:relKeldGreen}. Only at the isolated point $\gamma=0$, the initial state matters, and yields the correlation functions of a Luttinger Liquid at finite temperature $\langle \phi^{(r)}_{\sigma,x,t}\phi^{(r)}_{\sigma,y,t}\rangle\sim |x-y|$ on distances $|x-y|>v/T$. For an initial non-zero temperature (or energy density), we thus recover the expected volume law of the entanglement entropy. This is in line with numerical results, which show that the volume law is only asymptotically stable in the absence of monitoring \cite{Cao2019}, and is replaced by a logarithmic scaling at any non-zero monitoring rate \cite{Alberton20}.

\textit{Discussion} -- The entanglement entropy for the fermions in our setup reflects both the scaling behavior with system size as well as an monotonously decreasing effective central charge, both of which was numerically observed for monitored lattice fermions~\cite{Alberton20,chen2020}. The scaling behavior also includes the isolated point of zero monitoring.

As already anticipated above, there exists an alternative entanglement measure in this setup, namely the entanglement entropy of the {\it boson} degrees of freedom, which is discussed in Appendix~\ref{app:entro}. %The fermions are related to the boson degrees of freedom via a nonlinear (and nonlocal) transformation and per se there is no particular reason why both degrees of freedom should inherit the same entanglement structure. 
In equilibrium, the ground state entanglement entropy of the fermions and the bosons coincides~\cite{Calabrese_2009, Calabrese_2009}. Here, indeed we find that both display the same system-size scaling behavior of the entanglement entropy, i.e., volume, logarithmic and area law in the corresponding phases. However, the numerical prefactors of the entanglement entropy, e.g., the effective central charge $c(\gamma)$, do not coincide for fermions and bosons.

The comparison with related numerical works~\cite{Alberton20,bao2021symmetry,mueller2021measurementinduced,minato2021fate} confirms the picture drawn from the analytical computation of the fermion entanglement above. The missing of an equivalence between fermionic and bosonic entropy may well be related to a so far not well understood dichotomy between the signatures of conformality in the entanglement entropy and the correlation functions, and the $\gamma$ dependence of the effective central charge. We believe that this is an interesting observation, which may trigger a future study of conformal invariance in measurement-induced dynamics.

We would like to add a word of caution. The approach used here to compute the entanglement entropy for the fermion degrees of freedom is based on Refs.~\cite{Casini_2005}. It is known to work for free fermions, but to fall short for interacting Dirac fermions, i.e., for Luttinger Liquids, for which the boson fields $\phi_x, \theta_x$ are compact variables with a compactification radius different from one~\cite{Casini_2009}. In our setup, free fermions are subject to continuous measurements of the fermion density, and the combined dynamics is therefore Gaussian. The bosonization procedure is thus the same as for noninteracting fermions, which hints at the extension of Refs.~\cite{Casini_2005, Casini_2009} to the measurement setup to be justified. As we already stated above, this is confirmed by the matching of Eqs.~(\ref{eq:centralcharge}, \ref{eq:arealawent}) with numerical simulations of lattice fermions (or Majorana modes)~\cite{Alberton20,chen2020,bao2021symmetry}. It also correctly predicts an algebraic growth of the entanglement entropy (and the power-law exponent) for free fermion models with long-range hoppings~\cite{mueller2021measurementinduced}.

\subsection{Further implications and predictions}\label{sec:ConnecttoNum}
\textit{BKT essential scaling} -- The replica field theory introduced in this work predicts a measurement-induced BKT transition from a critical phase phase with scale invariant correlations and a logarithmic entanglement growth into an area law phase with exponentially decaying correlations. Key signatures of this phase transition depend on how the critical point $\gamma=\gamma_c$ is approached. (i) When approached from the critical phase, we report an effective central charge $c(\gamma)\sim \gamma^{-\kappa}$, which decays algebraically with the measurement strength, and then sharply drops to zero at the transition point (in the thermodynamic limit), see Fig.~\ref{fig-CC}. (ii) When approached from the area law phase, we predict a non-zero correlation length $\xi\sim \exp(\alpha/\sqrt{\gamma-\gamma_c})$ which displays the BKT essential scaling.

These predictions are based on a microscopic model of $U(1)$-symmetric Dirac fermions, subject to measurements of local fermion densities. From the viewpoint of symmetries and universality, we expect the effective long-wavelength theory, i.e., the boson replica field theory in Eqs.~\eqref{SEq37}, \eqref{eq:effsineGordon}, to apply to any Gaussian evolution protocol, which (i) preserves the $U(1)$-symmetry and (ii) features a competition between localizing measurements and a delocalizing Hamiltonian. 
Indeed, the outlined characteristic signatures have been observed in a variety of setups, which match conditions (i) and (ii). These include lattice fermions subject to continuous measurements~\cite{Alberton20, minato2021fate}, random circuit models with projective measurements \cite{bao2021symmetry}, measurement-only dynamics~\cite{ippoliti2020} and models with stroboscopic non-unitary dynamics~\cite{chen2020,jian2020criticality}.  In Ref.~\cite{bao2021symmetry} a $U(1)$-symmetry even is absent on the microscopic level and only emerges on the multi-replica level. For the case of lattice fermions, the BKT transition has been clearly identified via finite size scaling of the effective central charge $c(\gamma)$~\cite{Alberton20, minato2021fate}. For systems with finite size $L<\infty$, $c(\gamma)$ does not drop to zero in the area law phase but decays $\sim \log(L)^{-2}$ due to the essential BKT scaling of the correlation length.

Our formalism can also be readily applied to measurement-only dynamics, studied for instance in Ref.~\cite{ippoliti2020}. In our present setup we consider a competition between unitary dynamics, which delocalizes the fermions and non-unitary measurements, which pin the particles to their positions. In the case of measurement-only dynamics both sources of the competition, i.e. pinning {\it and} delocalization are caused by a non-unitary dynamics. All terms in the effective Hamiltonian \eqref{eq:effsineGordon} then become purely imaginary, yielding $\hat H_{\text{sg}}^\dagger=\hat H_{\text{sg}}$. The measurement dark state is then described by the ground state of the hermitian Hamiltonian $\hat H=-i \hat H_{\text{sg}}$. In this case, the dark state transition is equivalent to the ground state phase transition in the conventional sine-Gordon model. 

\textit{Measurements in long-range and pairing models} -- 
Natural and promising extensions of the short-ranged and $U(1)$-symmetric model studied here are setups in which one (or both) of the above conditions are relaxed, while staying in the manifold of Gaussian states. Both cases share in common that they yield a nonlinearity in the fermion phase $\hat \theta$, e.g., long-range hopping $\hat c^\dagger_{l}\hat c_{l+m}+\text{h.c.}\sim \cos(\hat \theta_l-\hat \theta_{m+l})$ for $m>1$~\cite{mueller2021measurementinduced}, and pairing $\hat c^\dagger_l\hat c^\dagger_{l+1}+\text{h.c.}\sim\cos(2\hat \theta_l)$.
In thermal equilibrium, a term $\sim \cos(2\hat\theta)$ is always relevant. In a measurement-induced dynamics, however, it couples multiplicatively to the absolute coordinate, i.e., an infinite temperature state. When averaging over the center-of-mass mode  $\langle \cos(2\hat \theta)\rangle_{k=0}=0$ the first order terms then vanish, and the leading order contribution arises in second order in the expansion, cf. Eq.~\eqref{eq:fullMaster}.

The replica field theory then predicts the following two scenarios: (i) For the case of long-range hopping, the second order terms describe a non-Hermitian hopping process with a squared hopping amplitude $t_{l,m}\rightarrow t_{l,m}^2$. This enables a new dynamical phase with an algebraic scaling of the entanglement entropy~\cite{mueller2021measurementinduced, minato2021fate}. (ii) For the case of fermion pairing, the second order expansion yields a term $\sim iU(\hat c^\dagger_{l}\hat c_{l+m}+\text{h.c.})^2$. It is again particle number conserving and has the form of an attractive interaction. This then again yields a sine-Gordon Hamiltonian as in Eq.~\eqref{eq:effsineGordon}, but with an effective, reduced measurement rate $\sim i(\gamma-U)(\partial_x\hat \phi)^2$. Consequently, one should observe qualitatively the same measurement-induced dynamics as in the model studied here, but with a critical measurement strength shifted to larger values of $\gamma$. Precisely this scenario was observed for a monitored Ising model in Ref.~\cite{turkeshi2021measurementinduced}.

\textit{Relaxation dynamics and purification} -- The replica field theory allows us to draw a clear and quantitative picture of the relaxation dynamics towards the steady state: The decoupling of the center-of-mass from the relative modes gives rise to the non-Hermitian evolution equation \eqref{SEq37}. For {\it any} initial state $\hat\rho$, this describes the evolution into a pure state $\hat\rho\rightarrow |\psi_D\rangle\langle\psi_D|$, which is the unique dark state of $\hat H_{\text{sg}}$. The dynamical critical exponent imprinted by the Hamiltonian is always $z=1$, as for a gapless, relativistic problem. The qualitatively different structure of the Hamiltonian $\hat H_{\text{sg}}$ in both phases, however, gives rise to qualitatively different relaxation dynamics: (i) in the area law phase, the Hamiltonian has a dissipative gap $i\tau^{-1}\sim\gamma$ \eqref{eq:EffHamQuad2}, which leads to an exponentially fast relaxation into the dark state with a global rate $\sim\tau^{-1}$. When approaching the transition into the critical phase, the gap vanishes and the relaxation time $\tau$ has to diverge. Due to $z=1$, the near critical relaxation time  $\tau\sim \xi$ is proportional to the correlation length $\xi$, and therefore features the same universal scaling behavior, i.e., the essential BKT scaling.
(ii) In the critical phase, the system also approaches a dark state, but the Hamiltonian is gapless. This naturally yields an algebraic relaxation $\sim t^{-1}$ into the dark state.

The relaxation behavior can be applied readily to the concept of a measurement-induced {\it purification transition}~\cite{Gullans2019}. For an initial mixed state with $\hat\rho^2\neq\hat\rho$, the measurement-induced evolution towards a dark state leads to a purification of $\hat\rho$. This phenomenon has been observed numerically also for monitored random circuit models in Ref.~\cite{Gullans2019}. There, a measurement-induced phase transition has been characterized by different purification dynamics on both sides of the transition. Here, we observe a transition between two purifying phases: (i) in the area law phase, the state purifies within a finite time $\tau$, which is set by the BKT correlation length. (ii) in the critical regime, the evolution is scale invariant, and the system purifies algebraically in time $\sim t^{-1}$.

\section{Conclusions and Outlook}
\label{sec:conclusio}

In this work, we have constructed a general field theory approach to measurement-induced phase transitions, and applied it to demonstrate a BKT type phase transition in a concrete model of measured massless Dirac fermions in $(1+1)$ dimensions. 

In general measurement problems, the measured wave function evolves in a pure state, however in a combined deterministic and random way. The pure state character enables transitions similar to quantum phase transitions, resulting from the competition of non-commuting operators. The randomness of the state in each realization of the quantum trajectory however requires statistical analysis to assess such transitions. Averages linear in the quantum trajectory projector are equivalent to expectation values in an infinite temperature state density matrix, at least for Hermitian measurement operators. We thus introduce the generating functionals for the $n$th moment of the state, $Z(n) = \overline{\tr [\hat\rho^{(c)}]^n}$, worked out for both the operatorial (for $n=2$) and a Keldysh functional integral formulation. This allows us to identify the relevant degrees of freedom for the description of measurement-induced phase transitions. A particularly strong structure emerges for free theories, which often form the basis for more sophisticated analysis: There, we find an exact decoupling of the generating functional into one 'hot' center-of-mass mode,  which indeed heats up indefinitely, and $n-1$ 'cold' relative modes whose dynamics can be recast in terms of the pure state evolution generated by an effective Hamiltonian, describing a kind of cooling. Beyond free theories, this framework allows us to define and compute correlation functions which are non-linear in the state. It also provides a formula for the computation of entanglement entropies.

A key advantage of this approach is that it is formulated in terms of (replicated) microscopic degrees of freedom. We exemplify the strengths of this formalism in the bosonized version of the fermion model, leveraging powerful bosonization and RG techniques to the many-body measurement problem: The effective Hamiltonian for relative modes is given by a quantum sine-Gordon model with complex coefficients, and the non-linearity is found irrelevant or relevant for weak or strong monitoring, respectively. Conceptually, the flow towards a Gaussian theory -- either a gapless, or a gapped scalar boson -- established in this way leads to an effective decoupling into 'hot' and 'cold' modes at long wavelength. Practically, this allows us to connect the microscopic physics to macroscopic phases, establishing a gapless and a gapped phase, with a logarithmic scaling and an area law saturation of the Rényi entropy. The gapless phase asymptotically exhibits an emergent conformal invariance. The critical point is in the BKT universality class.

We expect the toolbox developed here to provide further insight into the nature of measurement-induced phase transitions. For example, building on the striking parallels of the phenomenology found here with spinless fermionic lattice models suggests that the transition from a gapless CFT phase to a gapped one in (1+1) dimensions will persist in the presence of interactions between lattice fermions, at least assuming the validity of a naive connection between lattice fermions and the interacting Luttinger liquid: In that case, an interacting fermion Hamiltonian would simply manifest in a renormalized  Luttinger parameter, and none of the qualitative conclusions of the present work would be changed. More generally, our approach should help to construct further models showing such transitions, including in higher dimensions. In particular, an important goal is to identify a volume-to-area law transition: A possible mechanism for such a transition is an incomplete decoupling into 'hot' and 'cold' modes at long wavelength, leaving the cooling of the 'cold' modes imperfect, in turn enhancing entanglement growth. In turn, connecting such a possible mechanism to the presence or absence of integrability of the dynamics would be intriguing~\cite{Lunt2020,doggen2021generalized}.

Conceptually, it will be interesting to explore the relation of the damping dynamics under the non-Hermitian Hamiltonian to pure states to the purification scenario developed in Refs.~\cite{Gullans2019,Lunt2020}. Finally, our analysis clarifies that, while entanglement entropies provide a hallmark signature of measurement-induced transitions, they are not at the root of it, and correlation functions in the $n$th moment do provide similar information on the existence of the transition. The picture of a depinning transition from the eigenstates of the measurement operators developed here sparks the hope that the transition could be witnessed in experiment by suitable statistical analysis of quantum trajectories in highly controlled quantum systems undergoing monitoring. 

\acknowledgements 
We thank O.~Alberton, T.~Bintener, T.~Botzung, P.~Calabrese, A.~Daley, M.~Gullans, D.~Huse, K.~von Keyserlingk, M.~Knap, J.~Knolle, B.~Ladewig, M.~McGinley, Y.~Malo, M.~Müller, T.~Müller, A.~Nahum, F.~Pollmann, A.~Rosch  and S.~Roy for fruitful discussions. We acknowledge support from the Deutsche Forschungsgemeinschaft (DFG, German Research Foundation) under Germany's Excellence Strategy Cluster of Excellence Matter and Light for Quantum Computing (ML4Q) EXC 2004/1 390534769, and by the DFG Collaborative Research Center (CRC) 183 Project No. 277101999 - project B02. S.D. acknowledges support by the European Research Council (ERC) under the Horizon 2020 research and innovation program, Grant Agreement No. 647434 (DOQS), Y.M. by the Austrian Science Fund (FWF) through Grant No. P32299 (PHONED) and M.B. by the DFG SPP 1929 GiRyd.

\appendix
\section{Correlation functions in the strong measurement limit}\label{App:strong}

Here we demonstrate the emergence of a gap in the strong measurement limit via the exponential decay of the  connected measurement correlation function $C_{ij}=\overline{\langle \hat M_{i,t} \hat M_j(t)\rangle}$. To this end, we work on a lattice, utilizing the microscopic fermion model defined in Eqs.~(\ref{eq:hlatt},\ref{eq:mlatt}), a modification which should not affect the long range decay properties that we are interested in. For strong measurements $0<J\ll \gamma$, we can determine the solution for $C_{ij}$ perturbatively in $J/\gamma$. Its explicit time evolution before performing the trajectory average is obtained by applying Eq.~\eqref{eq:SSE} to the definition of $C_{ij}$. It yields
\begin{align}
d C_{ij}=&-dt\left[4\gamma\sum_k C_{ik}C_{kj}+\langle [i\hat H,\hat M_{i,t}\hat M_{j,t}]\rangle \right]\label{eq:PertC1}\\
&+\sum_k dW_k \left(\langle c^\dagger_i c_k\rangle\langle c^\dagger_kc_j\rangle\langle c^\dagger_j c_i\rangle+\langle c^\dagger_i c_j\rangle\langle c^\dagger_j c_k\rangle\langle c^\dagger_kc_i\rangle\right).\nonumber 
\end{align}
The second row of this equation is obtained by applying Wick's theorem to the higher order correlation functions occurring in the equation of motion. It contains the random Wiener increments and is proportional to a product of three non-local fermion bilinears. Each individual expectation value in the second row vanishes in the steady state if $J=0$, and therefore can at most be $O(J/\gamma)$. This yields a subleading scaling of the entire product $\sim dW J^3/\gamma^3$. 

In order to proceed with the perturbative expansion, it is convenient to introduce the correlation function $B_{ij}\equiv \langle \hat M_{i,t} [i\hat H,\hat M_j(t)]\rangle$. It evolves according to 
\begin{align}\label{eq:PertC2}
d B_{ij}=& dtJ \langle \hat M_{i,t} \Big(\hat M_{j+1,t}-2\hat M_{j,t}+\hat M_{j-1,t}\Big)\rangle\\&-dt\gamma \left[B_{ij}+\sum_lC_{il} B_{lj} \right] +\text{nonlocal terms}.\nonumber 
\end{align}
The combination of Eqs.~\eqref{eq:PertC1} and \eqref{eq:PertC2} provides important information. First, any nonlocal function, which cannot be expressed exclusively via fermion densities, will be exponentially suppressed in its time evolution, and its stationary state expectation value is at most of order $O(J/\gamma)$. This is contrasted by the evolution of the correlation function $C_{ij}$ in Eq.~\eqref{eq:PertC1}, which is dominated by the product $C_{ik}C_{kj}$ and yields a slow $\sim t^{-1}$ decay over time. Second, in both equations, the remaining stochastic terms are products of three nonlocal operator expectation values, and therefore at least of order $O(J^3/\gamma^3)$. They therefore will be neglected in a perturbative treatment. 

Without the remaining stochastic increments, Eqs.~\eqref{eq:PertC1} and \eqref{eq:PertC2} yield the stationary state solution
\begin{align}
C_{ij}=\frac{1}{2\gamma}\Big[\Big(\gamma^2+2J^2T\Big)^{\frac{1}{2}}\Big]_{ij}-\frac{\delta_{ij}}{2}.
\end{align}
Here $T_{ij}=-\delta_{i,j+1}-\delta_{i,j-1}+2\delta_{ij}$ is the discrete lattice Laplacian. The square root is a non-local function, but for small $J/\gamma$ it can be expanded in a Taylor series. The leading order contribution connecting two sites of separation $|i-j|$ is the $|i-j|$-th term of the expansion, yielding 
\begin{align}\label{eq:CorrF}
C_{ij}=& \frac{(-1)^{|i-j+1|}(2|i-j|-3)!!}{2^{|i-j+1|} |i-j|!}\left(\frac{2J^2}{\gamma^2}\right)^{|i-j|}\nonumber\\&\sim |i-j|^{3/2} e^{-|i-j|/\xi}.
\end{align}
This describes exponentially decaying correlations with a perturbative correlation length $\xi$ given by  $1/\xi=2\log \frac{\gamma}{J\sqrt{2}}$.

\section{Riccati equation for linearly monitored bosons}\label{sec:AppRic}
For a quadratic Hamiltonian and linear measurement operators, the steady state covariance functions $C^{AB}_{ij}$ of the form $C^{AB}_{ij}=\frac{1}{2}\overline{\langle \{\hat A_i-\langle \hat A_i\rangle, \hat B_j-\langle\hat B_j\rangle\}\rangle}$ for a set of bosonic operators  $\hat A_i, \hat B_j$ can be determined analytically. This is a consequence of three key properties of their evolution equation: (i) the Hamiltonian part of the evolution equation is a linear function of the  $C^{AB}_{ij}$, (ii) the Lindblad-type contribution to the evolution equation yields double commutators with the measurement operators $\sim [\hat M^Q_k,[\hat M^Q_k, \hat A_i \hat B_j]]$, which due to bosonic commutation relations are either proportional to the identity $\sim \mathds{1}$ or vanish, (iii) importantly, for all  stochastic terms $\sim dW_i \langle ...\rangle$ that occur in the evolution, the quantum mechanical expectation values $\langle ... \rangle=0$ vanish independently of the  stochastic increments $dW_i$. This is true for any Gaussian state (including initial product states) $|\psi_{t}\rangle$.  Consequently, the stochastic terms $\sim dW_i$ drop out, and the equation for the covariance matrix becomes deterministic.

Here, we demonstrate how one can solve for the stationary state covariance functions in such a setting explicitly. As a first step, we intermittently re-discretize the linear bosonic problem, taking the continuum limit at the end -- this will not affect the long-distance properties we are interested in. We thus start from a general set of Hermitian, bosonic operators $Q_i$, $P_i$ on a lattice with the sites $i$. Their spectrum is real and continuous, and they form a pair of conjugate variables, $[Q_k, P_j] = i\delta_{k,j}$. We will then later identify these operators with the discretized counterparts $\hat\phi_i, \hat\theta_i$ from the bosonized measurement setup introduced in Sec.~\ref{sec:fermbos}. The correspondence will be either $(Q_i, P_i)=(\hat\phi_i, \frac{\hat\theta_{i+1}-\hat{\theta}_{i-1}}{2})$ or $(Q_i, P_i)=(\frac{\hat\phi_{i+1}-\hat{\phi}_{i-1}}{2}, \hat\theta_i)$ depending on whether we consider weak or strong monitoring. 

We assume a quadratic Hamiltonian, which is expressed by the Hermitian matrices $V$, $W$ via
\begin{align}\label{SEq12}
\hat H=\sum_{i,j}\left(\hat Q_i V_{ij}\hat Q_j+\hat P_i W_{ij}\hat P_j\right).
\end{align}
The measurement operator shall be linear in $\hat Q_i$, i.e., $\hat M^Q_i=\hat Q_i-\langle \hat Q_i\rangle_t$ yielding the stochastic Schr\"odinger equation
\begin{align}\label{SEq13lin}
d|\psi_t\rangle = -dt\Big[i \hat H+\frac{\gamma}{2}\sum_i \left(\hat M^Q_i\right)^2\Big]|\psi_t\rangle+\sum_i dW_i \hat M^Q_i|\psi_t\rangle. 
\end{align}

We are interested in the covariance functions  ${C^{AB}_{ij}}\equiv \frac{1}{2}\overline{\langle \{\hat A_i-\langle \hat A_i\rangle_t, \hat B_j-\langle \hat B_j\rangle_t\}\rangle}$, where the operators $\hat A_i, \hat B_i\in\{ \hat Q_i,\hat P_i\}$. 
Starting from Eq.~\eqref{SEq13lin}, the evolution equation for the operator expectation value $\langle\hat O\rangle$ for a general operator $\hat O$ according to the \^Ito product rule~\cite{bookWisemanMilburn} is
\begin{equation} \label{eq:dO}
\begin{split}
d \langle \psi_t \vert \hat O \vert \psi_t \rangle 
= &  
(d\langle \psi_t \vert)\hat O \vert \psi_t \rangle 
+ \langle \psi_t \vert \hat O \vert d\vert \psi_t \rangle  \\
& + 
(d\langle \psi_t \vert)\hat O (d\vert \psi_t \rangle). 
\end{split}
\end{equation}

Combining Eq.~\eqref{SEq13lin} and Eq.~\eqref{eq:dO} we obtain
\begin{align} \label{eq:dO_Heisenberg}
d\langle \hat O \rangle  = 
&  i \langle [\hat O,\hat H]\rangle d t 
- \frac{\gamma}{2}dt \sum_k  \langle [\hat Q_k,[\hat Q_k,\hat O]] \rangle\\
& + \sum_j d W_{k} \left[\langle \left\{\hat Q_k, \hat O\right\}\rangle - 2 \langle \hat Q_k \rangle \langle\hat O\rangle\right]. \nonumber
\end{align}
In the limit where (i) the measurement $\hat Q_k$ is linear and (ii) the Hamiltonian $\hat H$ and the operator
$\hat O$ is at most a quadratic function of $\hat P_k, \hat Q_k$, we stay withing the manifold of states fully characterized by the covariance matrix \cite{Habib04}. 

For the covariance functions we're interested in, the second line in Eq.~\eqref{eq:dO_Heisenberg} vanishes, which we argue below. Inserting then the operator $\hat O=\frac{1}{2}\{\hat A_i-\langle\hat A_i\rangle, \hat B_j-\langle \hat B_j\rangle\}$ into the first line in Eq.~\eqref{eq:dO_Heisenberg} yields the Riccati equation
\begin{align}\label{SEq18}
\frac{d}{dt} C^{AB}_{ij}=&\frac{i}{2}  \langle [\hat H, \{A_i-\langle A_i\rangle, B_j-\langle B_j\rangle\}]\rangle\\
&+\gamma \left(\delta_{ij}\delta_{B,P}\delta_{A,P}-4 \sum_k C^{AQ}_{ik}C^{QB}_{kj}\right),\label{SEq19}
\end{align}
where $\delta_{i,j}$ is a Kronecker-$\delta$ in the operator index and the product $\delta_{B,P}\delta_{A,P}$ equals $1$ if the operators $A=B=P$ and zero otherwise.
Matrix Riccati equations \cite{bookSerafini,Kalman61,Doherty99} of this type have a unique and well defined steady state.
The quadratic form of the Hamiltonian ensures that the unitary part Eq.~\eqref{SEq18} remains linear in $C^{AB}$, while the nonlinear part ensures convergence of the evolution towards a well-defined stationary state. This gives rise to three types of correlation functions, $C^{QQ}, C^{PP}$ and $C^{QP}$ ($C^{PQ}=(C^{QP})^\dagger$). Their stationary values can be solved from $\frac{d}{dt}C^{AB}=0$ for each combination individually.

One of the key advantages of this setting which renders the conditioned dynamics solvable, is that by dropping the second line in \eqref{eq:dO_Heisenberg}, the equations of motion for $C^{AB}$ become deterministic. Dropping this line is, however, not an approximation. It turns out that for the choice of $\hat O=\frac{1}{2}\{\hat A_i-\langle\hat A_i\rangle, \hat B_j-\langle \hat B_j\rangle\}$ the second line is identical to zero, independently of the value of the increments $dW_k$.
This can be seen by focusing on a single term $k$ of the sum, and $\hat o=(\hat A_i-\langle\hat A_i\rangle) (\hat B_j-\langle \hat B_j)$. This yields  
\begin{align}\label{eq:H_super}
\langle \left\{\hat Q_k, \hat o\right\}\rangle - 2 \langle \hat Q_k \rangle \langle\hat o\rangle=& \langle\{\hat Q_k, \hat A_i\hat B_j\}\rangle-\langle\{\hat Q_k, \hat A_i\}\rangle\langle\hat B_j \rangle\\ &-\langle\{\hat Q_k, \hat B_j\}\rangle\langle\hat A_i \rangle-2\langle \hat Q_k\rangle \langle\hat A_i\hat B_j\rangle. \nonumber
\end{align}
The RHS of this equation is identical to zero if the cubic expectation value is evaluated in a Gaussian state, i.e., in a state where Wick's theorem for the decoupling of higher order expectation values can be applied. Since the stochastic Schr\"odinger equation \eqref{SEq13lin} is quadratic in the bosonic operators, this applies to any initial Gaussian state, which we assume here. Equations \eqref{SEq18}, \eqref{SEq19} are therefore the exact evolution equations for the covariance matrix for any initial state of this type. 

Inserting the Hamiltonian \eqref{SEq12} into the equation of motion and solving for $\frac{d}{dt}C_{ij}^{AB}=0$, we find for the cases $A=B=P$ and $A=B=Q$ the following equations
\begin{align}
0 = &  - V C^{QP} - C^{PQ}V - 4 \gamma C^{PQ}C^{QP} + \gamma \mathbb{1},  \label{eq:CQP}\\
0 = & \,\,\, W C^{PQ} + C^{QP}W - 4\gamma C^{QQ}C^{QQ}, \label{eq:CQQ}.
\end{align}
The two equations can be solved straightforwardly, yielding 
\begin{align}
C^{QP} = & 
\frac{1}{2}\biggl[ \mathbb{1} + \frac{1}{4\gamma^2}V^2\biggr]^{\frac{1}{2}}
- \frac{1}{4\gamma}V ,\\
C^{QQ}=&\left\{\frac{W}{8\gamma^2}\left[\left(4\gamma^2\mathds{1}+V^2\right)^{\frac{1}{2}}-V\right]\right\}^{\frac{1}{2}},\label{QQSol}
\end{align}
 which depend only on the matrices $V, W$ defining the Hamiltonian. Next we discuss the implications for the weak and strong measurement limits.

At weak monitoring, we identify $\hat Q_l=\frac{\hat \phi_{l+1}-\hat \phi_{l-1}}{2}$, which is the lattice analogue of $\partial_x\hat \phi_x$. A direct comparison shows that the lattice equivalent of the Hamiltonian \eqref{SEq10} is described by \eqref{SEq12} with matrices $V_{ij}=\delta_{ij}\frac{\pi v}{2}$ and $W_{ij}=\frac{v}{2\pi}\left(2\delta_{ij}-\delta_{i,j+1}-\delta_{i,j-1}\right)$, i.e. a diagonal matrix and a lattice Laplacian. Inserting this expression into the correlation function \eqref{QQSol} and taking the continuum limit yields Eqs.~\eqref{SEq25b} from the main text.

For strong monitoring we only consider the gapped measurement operator
\begin{align}\label{SEq26a}
\hat M_{2,x,t}=-\tilde m_{x}(\hat{\phi}_x-\langle\hat{\phi}_x\rangle_t),
\end{align}
where we inserted the mass $\tilde m_{x}=2m\sin(2\phi_{x})$, which depends on the eigenvalues $\phi_x$ of $\hat\phi_x$ in the corresponding dark state. The corresponding lattice measurement operator is $\hat Q_l=-\tilde{m}_l^2\hat \phi_l$ with $\tilde{m}_l=2m\sin(2\phi_{l})$ and its conjugate $\hat P_l=\frac{\hat\theta_{l+1}-\hat\theta_{l}}{2}$. Compared to the weak monitoring limit, this requires the exchange of $V\leftrightarrow W$, yielding $V=\frac{v}{2\pi}T$ and $W_{ij}=\delta_{ij}\frac{\pi v}{2}$ with the familiar lattice Laplacian $T_{ij}=2\delta_{ij}-\delta_{i,j+1}-\delta_{i,j-1}$. 

To treat the general case of spatially inhomogeneous masses $\tilde m_l$, we have to modify the measurement part of the Riccati equation \eqref{SEq19} according to
\begin{align}\label{SEq18b}
\frac{d}{dt} C^{AB}_{ij}=&\frac{i}{2}  \langle [\hat H, \{A_i-\langle A_i\rangle, B_j-\langle B_j\rangle\}]\rangle\\
&+\gamma \left(\tilde m_i^2\delta_{ij}\delta_{B,P}\delta_{A,P}-4 \sum_k \tilde m_k^2C^{AQ}_{ik}C^{QB}_{kj}\right).\label{SEq19b}
\end{align}
Solving this equation along the lines of the homogeneous Riccati equation yields the solution
\begin{align}\label{SEq27}
C=\frac{v}{4\gamma}\Big[\Big(T^2+\frac{4\pi^2\gamma^2}{ v^2}D^2\Big)^{\frac{1}{2}}-T\Big]^{\frac{1}{2}}.
\end{align}
Here $D$ is a diagonal matrix $D_{l,s}=\delta_{l,s}\tilde m_l^4$ containing the masses $\tilde m_l$. 
For the case of homogeneously distributed masses $\tilde{m}_l=\tilde m$, the square root can be expanded in a Taylor series in $T$ in the limit $\gamma\gg v$. This yields the covariance matrix in Eq.~\eqref{Expcov} in the main text, which decays exponentially with the distance with a correlation length $\xi^{-1}=\log(2\pi\gamma \tilde m^2/v)$. 

We can also convince ourselves that the more realistic scenario including inhomogeneously distributed masses $\tilde m_{l}$ does not alter this picture qualitatively. In Eq.~\eqref{SEq27}, a spatially dependent mass enters multiplicatively. The algebraic equations can then be solved numerically and one finds that this modifies the correlation length approximately such that that $\tilde m^2$ is replaced by the lattice average $\overline{\tilde{m}^2}=\frac{1}{N}\sum_{i=1}^N m_i^2$. This still leads to an exponential suppression of the covariance matrix with the distance. In the numerical confirmation we used a random, uniform distribution of the masses $\tilde m_l$.

\section{Review of constructing the sine-Gordon path integral}\label{App:path_integral}

Here we briefly review the construction of the bosonized path integral for the Hamiltonian \eqref{eq:effsineGordon}. For clarity, we use a hat on quantum operators and no hat for real fields. For the field operator $\hat\phi_x$ and its conjugate $\hat\Theta_x=\partial_x\hat\theta_x/\pi$, the kinematics follow from the commutation relation 
\eq{Real1}{
[\hat\Theta_y,\hat\phi_x]=-i\delta(x-y) \Rightarrow \langle \phi_x |\Theta_x\rangle =e^{i\Theta_x\phi_x},
}
where $\hat\phi_x |\phi_x\rangle=\phi_x|\phi_x\rangle$ and $\hat\Theta_x|\Theta_x\rangle=\Theta_x|\Theta_x\rangle$, and the Hamiltonian
\eq{Real2}{
\hat H=\frac{v}{2\pi}\int_x\left[ (\pi\hat\Theta_x)^2+\eta^2(\partial_x\hat\phi)^2+i\lambda(\cos(\sqrt{8}\hat\phi)-1)\right].
}
We perform the usual Trotterization and the  $l$-th trotterized time step is computed by using the completeness relations $\mathds{1}=\int d\phi_x |\phi_x\rangle\langle\phi_x|=\int d\Theta_x |\Theta_x\rangle\langle\Theta_x|$. This yields for an infinitesimal  step
\eq{SEq41}{
e^{iS_{l,x}}&=& \langle \phi_{l+1,x}|e^{-i\hat H_{sg}\delta t}|\phi_{l,x}\rangle\\
&=&\int d\Theta_x \langle \phi_{l+1,x}|\Theta_x\rangle\langle \Theta_x| (\mathds{1}-i\delta t \hat H)|\phi_{l,x}\rangle\nonumber\\
&=&\int d\Theta_x e^{i\Theta_x(\phi_{l+1,x}-\phi_{l,x})-i\delta t\frac{v}{2\pi} (\pi^2\Theta_x^2+\eta^2(\partial_x\phi_x)^2+i\lambda(\cos(\sqrt{8}\phi_x)-1))}\nonumber\\
&=&e^{\frac{i\delta t }{16\pi}\left(\frac{(\phi_{l+1,x}-\phi_{l,x})^2}{\delta t^2}-
\eta^2(\partial_x\phi_x)^2-i16\pi\lambda\cos(\phi_x).
\right)}\cdot\text{const.}
}
In the last step, we performed a rescaling of the field $\phi_x\rightarrow \phi_x/\sqrt{8}$ to eliminate the prefactor in the cosine, and a rescaling of time $v\delta t\rightarrow \delta t$, and factored out an overall normalization constant. 

Multiplying the Trotter increments and taking the limit of infinitesimal time steps $\delta t\rightarrow dt$ yields the non-Hermitian action ($X=(x,t)$)
\eq{SEq42}{
S=\int_X\frac{1}{16\pi}\left[(\partial_t\phi_X)^2-\eta^2(\partial_x\phi_X)^2\right]-i\lambda \cos\phi_X. \label{SEq43}
}
\section{Renormalization group equations for the sine-Gordon model}\label{App:SineGordonRG}
In order to derive the renormalization group equations for the couplings $K, \lambda$ we decompose the fields into short and long distance modes, $\phi_X=\phi_X^{(<)} +\phi_X^{(>)}$. The long distance modes correspond to momenta $|k|\le \Lambda/\xi$ for a short-distance cutoff $\Lambda=\pi$ (in units of the lattice spacing) and a dimensionless parameter $\xi=e^s$. The short-distance modes correspond to the small interval $\Lambda/\xi<|k|\le\Lambda$. 
In each RG step, the short-distance modes are integrated out perturbatively, and then momentum and position are rescaled as $k\rightarrow k\xi, x\rightarrow x/\xi$.

The RG equations for the sine-Gordon model are most easily derived in (2+0)-dimensions, which is why we start by performing a generalized Wick rotation $(x,t)\rightarrow(\eta^{\frac{1}{2}}x,i\eta^{-\frac{1}{2}}t)$ towards imaginary time action. This yields the (2+0)-dimensional action
\begin{align}
S=\int \frac{K}{16\pi} (\nabla\phi_X)^2+i\lambda\cos(\phi_X).
\end{align} 
The free Green's function of the theory in momentum space then is
\eq{SEq44}{
G_{\Omega,k}=\langle \phi_{\Omega,k}\phi_{-\Omega,-k}\rangle=\frac{16\pi}{K(\Omega^2+k^2)},
}
where $\Omega$ denotes the imaginary time frequency.

Integrating out the short-distance modes perturbatively yield the renormalized action 
\eq{SEq45}{
S^{(>)}=-\log(\langle e^{-S}\rangle_{<}) =S_0^{(>)}-\log\langle e^{-\delta S}\rangle_<,
}
where we have denoted $\delta S=\int_X\frac{i\lambda}{16\pi}\cos\phi_X$ and $S_0=S-\delta S$. Up to second order in $\lambda$ one finds
\begin{align}\label{SEq46}
S^{(>)}=S_0^{(>)}+\langle\delta S\rangle_<-\frac{1}{2}\left(\langle\delta S^2\rangle_<-\langle\delta S\rangle_<^2\right).
\end{align}
The linear term yields the correction
\begin{align}
\langle \delta S\rangle_<=i\lambda\int_X \langle\cos(\phi_X^{(<)}+\phi_X^{(>)})\rangle_<=i\lambda\int_X \cos(\phi_X^{(>)})e^{-\frac{1}{2}\langle(\phi_X^{(<)})^2\rangle_s}.\nonumber
\end{align}
After rescaling space and time $(x,\tau)\rightarrow (x,\tau)/\xi$ this leads to
\eq{SEq49}{
\lambda\rightarrow\lambda \xi^{2-\frac{2}{K}}.
}
Equation \eqref{SEq49} shows that up to first order in $\lambda$, the renormalization group flow predicts that, due to $K=1-i\frac{\gamma}{\pi v}$, in the limit of weak monitoring $\gamma\ll v$ the coupling is marginal and the second order correction decides of whether $\lambda$ is growing or not under the RG flow.

The second order correction in $\lambda$ is a bit more subtle but still manageable. It yields
\begin{align}
\langle \delta S^2\rangle_s-\langle \delta S\rangle_s^2
=&-\frac{\lambda^2\xi^{-\frac{4}{K}}}{2}\int_{X,Y}\Big[\cos(\phi_X^{(>)}-\phi_Y^{(>)})\Big(e^{\langle\phi_X^{(<)}\phi_Y^{(<)}\rangle_<}-1\Big)\nonumber\\
&+\cos(\phi_X^{(>)}+\phi_Y^{(>)})\left(e^{-\langle\phi_X^{(<)}\phi_Y^{(<)}\rangle_<}-1\right)\Big].\label{SEq50}
\end{align}
The average $\langle \phi_X^{(<)}\phi_Y^{(<)}\rangle_<$ being performed over the short-distance modes only ensures that it remains local and that the differences in Eq.~\eqref{SEq50} are only non-zero for $X\approx Y$. For the second term, this yields $\cos(\phi_X^{(>)}+\phi_Y^{(>)})\rightarrow \cos(2\phi_X^{(>)})$, which is less relevant than the $\cos(\phi_X^{(>)})$ term (due to the additional exponential suppression in each RG step) and is commonly neglected. The derivative term is expanded around $\delta X=Y-X$, which yields $\cos(\phi^{(>)}_X-\phi_{Y}^{(>)})\approx 1-\frac{1}{2}(\delta X\nabla\phi_X^{(>)})^2$. The Green's function in Eq.~\eqref{SEq44} is even and symmetric in $\Omega,k$, and thus cross averages $x\cdot \tau$ vanish. We thus finally obtain
\eq{SEq51}{
\langle \delta S^2\rangle_s-\langle \delta S\rangle_s^2&=&-\frac{(\lambda\xi^{-\frac{2}{K}})^2}{2}\int_{X} A(\xi)(\nabla\phi_X)^2
}
with
\eq{SEq52}{
A(\xi)=\int_{x,\tau} x^2 \left(e^{G_{x,\tau}^{(<)}}-1\right).
}

To leading order, the new function $A(\xi)$ can be seen as a constant, typically of order $\mathcal{O}(1)$.
Performing the substitution $\xi=e^{s}$ with $0<s\ll1$, we find the flow equations from the main text.

\section{Dark state wave function and correlations}\label{app:darkstate}
Here we analytically compute the dark state $|\psi_D\rangle$ of the non-Hermitian Hamiltonian \eqref{eq:EffHamQuad1} in the main text and show that $|\psi_D\rangle$ is unique and can be reached from any initial state. The non-Hermitian Hamiltonian in momentum space is
\begin{align}
H=\frac{1}{2\pi}\int_q q^2\left(\hat\theta_q\hat\theta_{-q}+(1-i\gamma)\hat\phi_q\hat\phi_{-q}\right),
\end{align}
where we have set $v=1$ and we have rescaled $\gamma\rightarrow \gamma\pi/2$ for brevity. We are interested in the dark state $|\psi_D\rangle$ of this Hamiltonian, i.e. in the state $H|\psi_D\rangle=0$. Therefore, we introduce boson operators $b^\dagger_q,  b_q$ with $[b_q, b^\dagger_k]=\delta_{q,k}$ via
\begin{align}
\hat\phi_q&=-i \sqrt{\frac{\pi}{2|q|}}\text{sgn}(q) (b^\dagger_q+b_{-q}), \\
\hat\theta_q&=i\sqrt{\frac{\pi}{2|q|}}(b^\dagger_q-b_{-q}).
\end{align}
Inserting them into the Hamiltonian yields
\begin{align}\label{off-diag}
H=\frac{1}{4}\int_q |q| \left(2b^\dagger_qb_q+2b_{-q}b^\dagger_{-q}-i\gamma(b^\dagger_{q}+b_{-q})(b^\dagger_{-q}-b_{q})\right).
\end{align}
If the Hamiltonian were Hermitian, one would now diagonalize the problem by a canonical transformation, preserving bosonic commutation relations. For the present non-Hermitian  Hamiltonian, a canonical transformation can be found that brings it into a tri-diagonal form, which still allows us to extract the dark state properties. Here, this requires a particularly simple Bogoliubov transformation, which is independent of the momentum mode $q$, i.e. we can reduce the problem to tri-diagonalizing the Hamiltonian
\begin{align}
h=\underbrace{(b^\dagger_q, b_{-q})}_{=B^\dagger_q}\underbrace{\left(\begin{array}{cc} 2-i\gamma& -i\gamma\\ -i\gamma & 2-i\gamma\end{array}\right)}_{=M}\underbrace{\left(\begin{array}{c}b_q \\ b^\dagger_{-q}\end{array}\right)}_{=B_q}=B^\dagger_q M B_q. 
\end{align}
Any transformation of the form $B_q=V C_q$ through some matrix $V$ and onto a new set of bosonic operators $C_q=(c_q, c^\dagger_{-q})^T$ has to respect bosonic commutation relations 
\begin{align}
1=B^\dagger_q \sigma_z B_q=C^\dagger_q V^\dagger \sigma_z VC_q\overset{!}{=}C^\dagger_q  \sigma_zC_q \Rightarrow V^{-1}=\sigma_z V^\dagger\sigma_z.\nonumber
\end{align}
The RHS of this equation yields a representation of the matrix $V$ and its inverse
\begin{align}
\left(\begin{array}{c}b_q\\ b^\dagger_{-q}\end{array}\right)=\left(\begin{array}{cc}\alpha^* & -\beta \\ -\beta^*& \alpha \end{array}\right) \left(\begin{array}{c}c_q\\ c^\dagger_{-q}\end{array}\right) \Leftrightarrow \left(\begin{array}{c}c_q\\ c^\dagger_{-q}\end{array}\right)=\left(\begin{array}{cc}\alpha & \beta \\ \beta^*& \alpha^* \end{array}\right)\left(\begin{array}{c}b_q\\ b^\dagger_{-q}\end{array}\right),
\end{align}
with the constraint $|\alpha|^2-|\beta|^2=1$. 
We identify the matrix elements
\begin{align}
\alpha&=\frac{2(1+\sqrt{1-i\gamma})-i\gamma}{\sqrt{|2(1+\sqrt{1-i\gamma})-i\gamma|^2-\gamma^2}},\nonumber \\
\beta&=\frac{-i\gamma}{\sqrt{|2(1+\sqrt{1-i\gamma})-i\gamma|^2-\gamma^2}},
\end{align}
which yield the Hamiltonian in the basis of the new boson operators $c_q, c^\dagger_q$
\begin{align}
h=\epsilon (c^\dagger_q c_q+c^\dagger_{-q}c_{-q})+\eta c_{q}c_{-q}.\label{eq:Jordnorm}
\end{align}
Here, $\epsilon=2\sqrt{1-i\gamma}$ and $\eta=\frac{-4i\gamma}{2+\sqrt{1-i\gamma}-\sqrt{1+i\gamma}}$ are complex numbers. Although the canonical transformation does not diagonalize the non-Hermitian Hamiltonian, Eq.~\eqref{eq:Jordnorm} has one unique dark state $|\psi_D\rangle$, which is the vacuum in the basis of the $c_{q}, c_{-q}$ states, $c_{q}|\psi_D\rangle=0, \ \  \forall  \ q$. 
The diagonal terms $\sim \epsilon c^\dagger_q c_q$ ensure that any other state is exponentially suppressed over time due to the imaginary part of $\epsilon$, $\text{Im}(\epsilon)<0$ and the term $\sim \eta$ ensures that the vacuum is reached from any other state, even if the initial state has no overlap with the vacuum. The condition $c_{q}|\psi_D\rangle=0$ allows us to obtain correlation functions in the dark state $|\psi_D\rangle=|0\rangle$
\begin{align}
\langle 0|b^\dagger_q b_q|0\rangle&=|\beta|^2=\langle 0|b_q b^\dagger_q|0\rangle-1,\\
\langle 0|b_{-q} b_q|0\rangle&=-\alpha^*\beta=\langle 0|b_{q}^\dagger b_{-q}^\dagger|0\rangle^*.
\end{align}
With this result, we can determine the correlation functions of the Luttinger Liquid field operators (remember that $[\hat\phi_q, \hat\theta_{-q}]=\frac{\pi}{q}$), 
\begin{align}
\langle 0|\hat \phi_q \hat\phi_{-q}|0\rangle&=\frac{\pi}{2|q|}\left(1+2|\beta|^2-\alpha^*\beta-\beta^*\alpha\right)\nonumber\\
&=\frac{\pi}{2|q|\gamma}\sqrt{2\sqrt{\gamma^2+1}-2}, \label{Corlut}\\
\frac{1}{2}\langle 0|\{\hat \phi_q ,\hat\theta_{-q}\}|0\rangle&=\frac{\pi}{2q}(\alpha^*\beta-\beta^*\alpha)=-i\frac{\pi}{2q}\frac{\gamma}{1+\sqrt{1+\gamma^2}},\nonumber\\
\langle 0|\hat \theta_q \hat\theta_{-q}|0\rangle&=\frac{\pi}{2|q|}\frac{2\sqrt{1+\gamma^2}\text{Re}(\sqrt{1-i\gamma})}{1+\sqrt{1+\gamma^2}}.
\end{align}
The result for the correlation function in Eq.~\eqref{Corlut} is identical to the correlation function of the $\hat\phi$ field operators in the Riccati approach~\eqref{SEq25b} (after rescaling of $\gamma$ and multiplication $\hat\phi_q\rightarrow iq/\pi \hat\phi_q$, which was measured in the Ricatti equations). One can easily show, that the remaining correlation functions are also identical in both approaches. Since the Riccati results are obtained without any approximation, this demonstrates that the dark state of the non-Hermitian Hamiltonian indeed describes the steady state of the replica fluctuations. 

\section{Fermion Entanglement Entropy in the Keldysh setting}\label{sec:EntApp}
\begin{figure}
  \includegraphics[width=\linewidth]{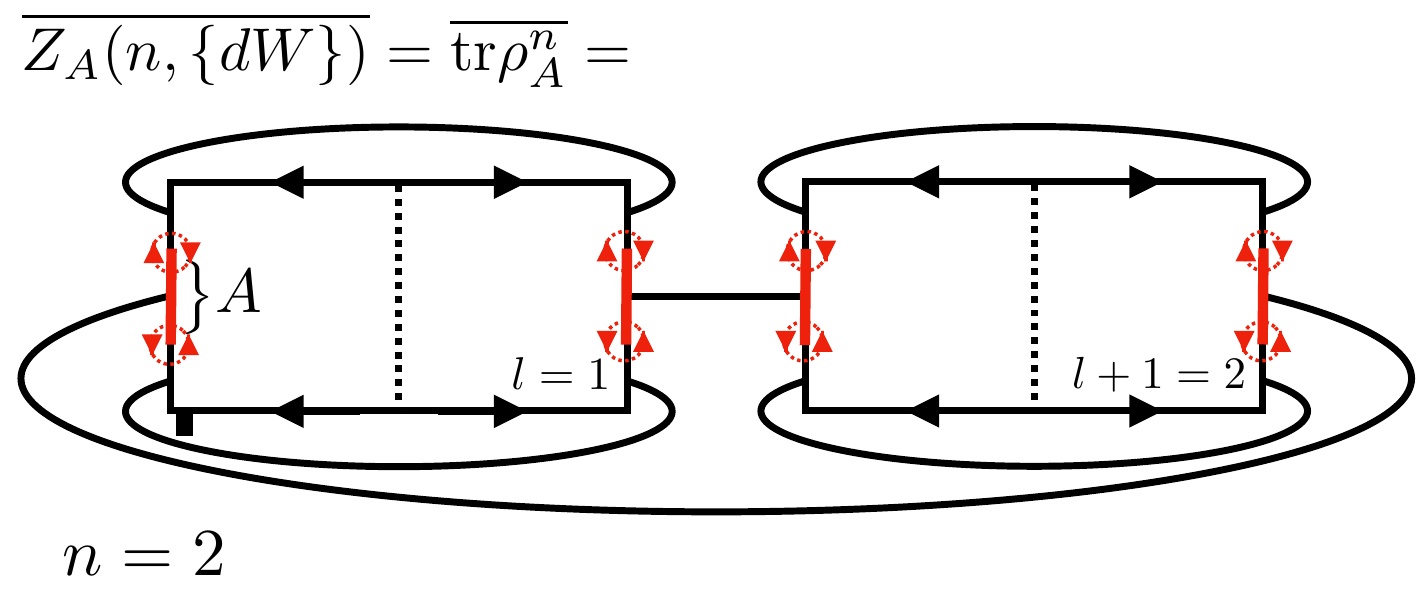}
  \caption{Graphical representation of the reduced partition sum. Over the interval A, consecutive replicas are coupled. For the rest of the system $\mathds{R}\setminus A$, replicas are decoupled from each other. The singular gauge $A_\mu^{(k)}$ appears as charges of opposite sign at the boundaries of $A$.
  }
  \label{fig-boundary}
\end{figure}%%%

To compute the Rényi entropies in practice, we start from the fermionic formulation, where the partial trace is implemented on the $n$-replica functional integral as twisted boundary conditions on the replicas:
\begin{align}
Z_A(n,\{dW\}) =& \int \mathcal{D} [\Psi] T T_A e^{iS},\label{eq:contTrafo0}\\
T \Psi^{(l)}_{+,x,t_f} =&-\Psi^{(l)}_{-,x,t_f},\label{eq:contTrafo1}\\
T_A \Psi^{(l)}_{\sigma,x,t_f}=&\left\{\begin{array}{c} \Psi^{(l)}_{\sigma,x,t_f}\,\, \forall x \notin A\\\Psi^{(l+\sigma )}_{\sigma,x,t_f}\,\, \forall x \in A\end{array}\right. ,\label{eq:contTrafo2}
\end{align}
where $l$ is defined $\mod (n)$, and $S$ is the fermion action before performing the noise average. The boundary conditions are illustrated  graphically in Fig.~\ref{fig-boundary}. In the common functional integral representation of a partition function $Z(1,\{dW\})$ above, we evaluate the density matrix at time $t_f$ and perform the trace over coherent states, which for fermions imposes the antiperiodic boundary condition $\Psi^{(1)}_{+,x,t}=-\Psi^{(1)}_{-,x,t}$ (Fig.~\ref{fig-MultiKeldysh}(a)). This is implemented by applying the operator $T$ defined in Eq.~\eqref{eq:contTrafo1}. For the $n$-replica partition function $Z(n,\{dW\})$, as constructed in Eq.~\eqref{eq:nReppf}, each replica is traced out independently, and the antiperiodic boundary condition \eqref{eq:contTrafo1} is implemented by $T$ acting on each replica individually (Fig.~\ref{fig-MultiKeldysh}(b)). For the partially traced $n$-replica partition function $Z_A(n,  \{ dW\})$, we want to implement the same condition for $x\notin A$. For $x\in A$, however, we need $\Psi^{(l)}_{+,x,t}=-\Psi^{(l+1)}_{-,x,t}$. The second condition is implemented by acting the operator $T_A$ on the action, which shifts the fields by one replica index if $x\in A$ and acts as the identity if $x\notin A$. In order to make this transformation contour-symmetric, we also implement it on the minus contour. Here, however, the inverse contour hopping $(l)\rightarrow(l-1)$ has to be applied.

Following the steps of Ref.~\cite{Casini_2005}, we can Fourier transform the replica space fermion fields according to $\Psi^{(k)}_{\sigma,X}=\sum_{l=1}^n e^{i\frac{2\pi k l}{n}}\Psi^{(l)}_{\sigma,X}$, which makes them eigenstates of the operator $T_A$ with $T_A\Psi^{(k)}_{\sigma,X}=e^{i\sigma \frac{2\pi k}{n}}\Psi^{(k)}_{\sigma,X}$ if $x\in A$. Local fermion bilinears are invariant under the action of $T_A$, i.e., $T_A \bar\Psi^{(k)}_{\sigma, X}\Psi^{(k)}_{\sigma,X}=\bar\Psi^{(k)}_{\sigma, X}\Psi^{(k)}_{\sigma,X}$ since the phases of $\bar\Psi^{(k)}_{\sigma, X}$ and $\Psi^{(k)}_{\sigma,X}$ cancel each other. In the vicinity of the points $(x_0, t_f)$ and $(x_0+L, t_f)$, nonlocal products, such as $\bar\Psi^{(k)}_{\sigma, X}\partial_{x,t}\Psi^{(k)}_{\sigma,X}$, however, are generally not invariant under $T_A$.

The action of $T_A$ on the fermion fields can be implemented by a singular gauge transformation on the fields~\cite{Casini_2005}
\begin{eqnarray}
T_A\Psi^{(k)}_{\sigma,X}= \exp\left(i\sigma\int_{X_0}^X dX'^\mu A^{(k)}_\mu(X') \right)\Psi^{(k)}_{\sigma,X},
\end{eqnarray}
where $\mu$ indicates position $x$ ($\mu=0$) or time $t$ ($\mu=1$). The gauge field obeys the boundary conditions 
\begin{align}
    \oint_{C(x_0,t_f)}dX^\mu A^{(k)}_\mu(X)=- \oint_{C(x_0+L,t_f)}dX^\mu A^{(k)}_\mu(X)=-\frac{2\pi k}{n},
\end{align}
where the integrals are performed along a curve $C(x_0,t_f), C(x_0+L,t_f)$ that encloses the points $(x_0, t_f)$ or $(x_0+L, t_f)$.

This eliminates the operator $T_A$ from the average in \eqref{eq:contTrafo0}, trading it for a shift in the partial derivatives ${\partial_\mu\Psi^{(k)}_{\sigma,X}\rightarrow (\partial_\mu+i\sigma A^k_\mu) \Psi^{(k)}_{\sigma,X}}$. This corresponds to a 'quantum' gauge transformation, defined by acting on the Keldysh fields $\sigma = \pm$ in the opposite way. 

Inserting the gauge field into the action yields the expression for the partially traced $n$-replica Keldysh partition function \eqref{eq:contTrafo0}. The only modification compared to the action~\eqref{eq:nReppf} is thus an additional fermion bilinear coupling to the gauge field $\sim A^{(k)}_\mu$; in particular, the noise averaging is not altered by this modification. We thus obtain the obtain the result
\begin{eqnarray}
Z_{A}(n)=\langle \prod_k \exp\left(i\int dX A^{(k)}_\mu J^{(\mu, k)}_{c,X} \right)\rangle,
\end{eqnarray}
where $J_c^{(\mu)} = J^{(\mu)}_{+}+J^{(\mu)}_{-}$ is the classical (Keldysh) component of the fermionic current, which couples to the quantum gauge field. This shows that the $n$th Rényi entropy can be obtained as an expectation value for the functional integral Eq.~\eqref{eq:contTrafo0} (including integration over the fermions, and noise average).

Equipped with this insight, we apply bosonization; in particular, $J^{(0, k)}_{\sigma,X}=-\frac{1}{\pi}\partial_x\phi^{(k)}_{\sigma,X}$, $J^{(1, k)}_{\sigma,X}=\frac{1}{\pi}\partial_t\phi^{(k)}_{\sigma,X}$ in the bosonic theory. This yields Eq.~\eqref{eq:ZA} from the main text.

\section{Boson Entanglement Entropy for a noisy Gaussian system}\label{app:entro}
In the main text, we discuss the entanglement entropy of the fermion wave function, which we express in terms of the bosonized correlation functions. Here, we discuss the entanglement entropy of the boson degrees of freedom in the limit where the boson theory is Gaussian. Then we obtain the entanglement entropy by applying the techniques put forward in Ref.~\cite{Vidal2003,Peschel_2003,Cramer2006}. The entanglement entropy of a free problem -- completely described by a Gaussian density matrix -- is then computed from its covariance matrix. 

As for the fermions, the boson R\'enyi entropies are defined via Eq.~\eqref{eq:entro}. The difference compared to the main text is that the density matrix $\hat\rho$ is {\it first} expressed in terms of bosons and {\it then} the partial trace and the matrix multiplication are performed. In the path integral, the matrix multiplication is then implemented by the boundary conditions for the {\it boson fields} $\phi_{+,x,t_f}^{(l)}=\phi_{-,x,t_f}^{(l+1)}$ if $x\in A$ and $t_f$ is the time at which $\hat\rho_A^{(c)}$ is evaluated \cite{Calabrese_2009, Calabrese_2004}.

To make Eq. \eqref{eq:entro} amenable to evaluation, we proceed in three main steps, which we first summarize below and then discuss in detail afterwards: 

(i) The boundary conditions and the coupling to the noise in $Z_A(n,\{dW\})$ simplify in the Fourier representation of the replica modes $\phi^{(k)}_x$ in Eq.~\eqref{eq:FourierRep}. This splits the action into a sum as in Eq.~\eqref{eq:linact}. We  show below that the boundary conditions do not modify the action of the $k=0$ mode.
This yields
\begin{align}
    Z_A(n,\{dW\})= Z_A^{(r)}(n)%\times \frac{Z_A(1,\{\sqrt{n}dW\})}{Z_A^{(r)}(1)}=Z_A^{(r)}(n)
    \times z_{\sqrt{n}dW}.\label{eq:trace}
\end{align}
Here, $Z_A^{(r)}(n)=Z_A(n,0)$ is the reduced partition function for $n$ replicas with vanishing noise $dW=0$. The label $r$ indicates that $Z^{(r)}_A$ is determined exclusively by the relative modes with $k>0$. In contrast, $z_{\sqrt{n}dW}$ is a noise dependent quantity, which does not depend on $A$. Importantly, we thus find a factorization of $Z_A(n,\{dW\})$ into a term $Z_A^{(r)}(n)$ that is $A$ dependent, but noise independent, and a term $z_{\sqrt{n}dW}$ which is $A$ independent, but noise dependent. The scaling of the entropy with the bipartition size therefore is entirely encoded in the auxiliary object $Z_A^{(r)}(n)$. Up to a constant, the Rényi entropy in Eq.~\eqref{eq:trace} is determined by
\begin{align}
    S_n(L)=\frac{1}{1-n}\log Z_A^{(r)}(n).\label{eq:Renyi}
    \end{align}
    Here, we will focus on the von Neumann entanglement entropy, obtained by taking the limit $n\rightarrow1$.
 
(ii) The auxiliary object $Z_A^{(r)}(n)$ obtains from Eqs.~(\ref{unitaction},\ref{measact}) by setting the noise $dW\equiv 0$. Then for $dW=0$, each replica (including the absolute mode with $k=0$) evolves according to a non-Hermitian Hamiltonian. This means that each replica approaches a dark state $\rho^D=|\psi_D\rangle\langle\psi_D|$, and $Z^{(r)}_A(n)$ can be obtained from products $\rho^D_A$. The factor $z_{\sqrt{n}dW}$ represents then the $A$-independent normalization. By analogy to the entanglement entropy of Hermitian free bosons, the latter formulation equips us with the expectation that if $|\psi_D\rangle$ is close to an effective ground state of the Hermitian problem, the entanglement entropy in Eq.~\eqref{eq:Renyi} should either grow logarithmically with the system size, or saturate at a finite length scale set by the mass gap. In turn, if $|\psi_D\rangle$ corresponds to a highly excited state, we expect a volume law entanglement entropy.

(iii) For the practical evaluation, the usual tools of conformal field theory are not immediately applicable, since the theory is non-Hermitian. Instead, a feasible approach is to compute 
the entanglement entropy directly from the covariance matrix of the subsystem $A$~\cite{Vidal2003,Peschel_2003,Cramer2006}, evaluated at $t=t_f$
\begin{align}\label{eq:LuttCov}
    \tilde C_{x,y}=\left\langle\left(\begin{array}{cc}\phi^{(r)}_{\sigma,x}\phi_{\sigma,y}^{(r)} & \frac{1}{2\pi}\left\{\phi^{(r)}_{\sigma,x},\partial_t\phi_{\sigma,y}^{(r)}\right\} \\ \frac{1}{2\pi}\left\{\phi^{(r)}_{\sigma,x},\partial_t\phi_{\sigma,y}^{(r)}\right\}& \frac{1}{\pi^2}\partial_t\phi_{\sigma,x}^{(r)}\partial_t\phi_{\sigma,y}^{(r)}\end{array}\right)\right\rangle_{S^{(r)}},
\end{align}
with $x,y\in A$ and $\sigma$ either $+$ or $-$. Here the index $S^{(r)}$ indicates that the expectation value is taken with respect to the Keldysh action in Eq.~\eqref{eq:znh}, with $r$ being any $k>0$. As per the discussion in (ii), the covariance matrix can be obtained equivalently from an average with respect to the dark state $|\psi_D\rangle$ (see Appendix~\ref{app:darkstate}) after replacing $(\phi^{(r)}_{\sigma,x,t}, \frac{1}{\pi}\partial_t\phi_{\sigma,x,t}^{(r)})\rightarrow(\hat\phi^{(r)}_x, \partial_x\hat\theta^{(r)}_x )$~\footnote{The covariance matrix has to be evaluated always at the coordinates corresponding to the operator $\hat\phi_{x}$ and its conjugate $\partial_x\hat\theta_x$. In the Lagrangian formulation of the path integral $\partial_x\hat\theta_x \rightarrow\frac{K_\theta}{\pi}\partial_t\phi_{X}$, where $K_\theta$ is determined in the Hamiltonian formulation.}.

For free bosons in the ground state, the scaling behavior of the entanglement entropy with the size $L$ of the subsystem $A$ can already be inferred from the scaling of the upper left block of $\tilde C_{x,y}$, $\sim\langle \hat\phi^{(r)}_x\hat\phi^{(r)}_{x+L}\rangle$. We verify that this is also true here by evaluating the covariance matrix $\tilde{C}_{x,y}$ on a grid, and determining its eigenvalues $\{\lambda_i \}$. 
The von Neumann entanglement entropy $S_{\text{vN}}$ can then readily be computed from the eigenvalues (see e.g.~\cite{Cramer2006,Peschellong,Adesso} for a detailed discussion of bosonic systems) according to 
\begin{align}
    S_{\text{vN}}(L)=\lim_{n\rightarrow 1}S_n(L)=\sum_i\left[ f\left(\frac{\lambda_i+1}{2}\right)-f\left(\frac{\lambda_i-1}{2}\right)\right].\label{eq:entrolutt1}
\end{align}
Here, $f(x)=x\log x$ and we sum over all $\lambda_i>0$. We discuss the entropies separately for the gapless ($\gamma\ll v$) and gapped ($\gamma\gg v$) cases, as well as for the limit of zero measurements $\gamma=0$.

{\it Step (i)}\ -- Here we provide a derivation for Eqs.~\eqref{eq:trace} and \eqref{eq:Renyi}, which is based on the path integral for the partition reduced function $Z_A(n, \{dW\})$. In the bosonized framework, the latter is obtained in two steps. First, we bosonize the path integral for the general $n$-replica partition function in Eq.~\eqref{eq:nReppf} at the Gaussian fixed point. This yields
\begin{align}
    Z(n,\{dW\})=\int\mathcal{D}[\{\phi^{(l)}_X\}]\exp(i S_n[\{\phi^{(l)}_X, dW\}]),
\end{align}
with the action 
\begin{align}
    S_n[\{\phi^{(l)}_X, dW\}]=&-\frac{1}{2\pi}\int_X \sum_{l=1}^n\sum_{\sigma=\pi}\left(\sigma\phi^{(l)}_{\sigma,X}\partial^2\phi^{(l)}_{\sigma,X}-i(D\phi^{(l)}_{\sigma,X})^2\right)\nonumber\\
    &-\frac{i}{2\pi} \int_X dW_t\left(\sum_{\sigma=\pm}\sum_{l=1}^nD\phi^{(l)}_{\sigma,X}\right)
\end{align}
and $D=\partial_x , \pi m$ depending on whether we consider the scale invariant or the gapped Gaussian fixed point. 

In a second step, we implement the boundary condition for the matrix product in $\text{tr}[(\rho^{(c)}_A)^n]$ in Eq.~\eqref{eq:entro}. Consider the trace and the partial trace to be evaluated after some evolution time $t_f$ at which the density matrix has reached a steady state. This yields the boundary conditions for the two-contour path integral \cite{Calabrese_2009,Cardy_2007} 
\begin{align}
    \phi^{(l)}_{+,t_f,x}=\left\{\begin{array}{cc}  \phi^{(l)}_{-,t_f,x}& \text{ if } x\notin A\\ \phi^{(l+1)}_{-,t_f,x}& \text{ if } x\in A\end{array}\right. ,\label{eq:boundary}
\end{align}
where we set $\phi^{(n+1)}_{-,t_f,x}=\phi^{(1)}_{-,t_f,x}$.
The first condition is the common boundary condition for the partition function of a single bosonic system evaluated at time $t=t_f$. It is realized for fields outside the subset $A$. The second condition, which applies for fields with $x\in A$ implements the multiplication of the reduced density matrices. In the Lagrangian formulation of the partition function, we need to apply the same boundary conditions as in Eq.~\eqref{eq:boundary} also for the temporal derivatives $\partial_t\phi^{(l)}_X$. 

The additional replica index translations in \eqref{eq:boundary} are commonly implemented via so-called {\it branch point twist fields} $T$ \cite{Cardy_2007, Calabrese_2009}, which translate the fields by one replica index $T_A\phi^{(l)}_{+,X}=\phi^{(l+1)}_{+,X}$ if $x\in A$ and $t=t_f$, and else act as the identity $T_A\phi^{(l)}_{+,X}=\phi^{(l)}_{+,X}$. We will not elaborate further on the branch point twist fields, but we will use two important properties of $T$. First, since $T$ is a translation operator on the replicas, it is diagonal in the replica Fourier basis
\begin{align}
    T_A\phi^{(k)}_{+,X}=\exp(i\varphi^{(k)}_X)\phi^{(k)}_{+,X},
\end{align}
where $\varphi^{(k)}_X=2\pi k/n$ if $x\in A$ and $t=t_f$ and $\varphi^{(k)}_X=0$ else. Second, it follows immediately from the properties of $\varphi^{(k)}_X$ that $T_A$ acts as the identity on the absolute replica mode with $k=0$. The absolute mode therefore remains coupled to the noise $dW$, but decouples from the boundary conditions of the reduced density matrix. Conversely, the $k>0$ modes decouple from the noise but are subject to the boundary conditions. 

This decoupling allows us to express the partition function $Z_A(n,\{dW\})$ in Eq.~\eqref{eq:entro} via 
\begin{align}
    Z_A(n,\{dW\})=\prod_{k=1}^{n-1} \langle T_A\rangle_{S^{(k)}}\times Z_A(1,\{\sqrt{n}dW\}),\label{eq:partred}
\end{align}
where 
\begin{align}
 \langle T_A\rangle_{S^{(k)}}=\int\mathcal{D}[\phi^{(k)}_X] T_A\exp(iS^{(k)}[\phi^{(k)}])
\end{align}
with $S^{(k)}[\phi^{(k)}]$ as defined in Eq.~\eqref{eq:linact}, and $Z_A(1,\{\sqrt{n}dW\})$ is the reduced partition function for the absolute mode. It is therefore identical to the reduced partition function of a single replica, but with an increased measurement noise $dW\rightarrow \sqrt{n} dW$.

{\it Step (ii) }-- The analogue of the expression in Eq.~\eqref{eq:partred} for Hermitian systems has been derived in several previous works~\cite{Calabrese_2009, Cardy_2007, Casini_2005, Calabrese_2009}. In our case,  it grants an additional important insight, namely the decoupling of the branch point twist fields from the measurement noise. This makes the entanglement entropy a noise insensitive measure for the Gaussian fixed points. The direct computation of the expectation values of $T_A$ is, however, difficult in most cases. The fields $T_A$ impose a phase to the fields in the interval $x\in A$, which removes the translational invariance of the original action and makes the evaluation of the partition function a more complicated task. We therefore modify Eq.~\eqref{eq:partred} by multiplying it with $1=\frac{Z_A(1,0)}{Z_A(1,0)}$, where $Z_A(1,0)$ is a shorthand for the reduced partition function of an absolute mode with zero measurement noise. We can then reverse the steps leading to Eq.~\eqref{eq:partred} and arrive at
\begin{align}
    Z_A(n,\{dW\})=\underbrace{Z_A(n,0)}_{\equiv Z^{(r)}_A(n)}\times \underbrace{\frac{Z(1,\{\sqrt{n}dW\})}{Z(1,0)}}_{\equiv z_{\sqrt{n}dW}}.\label{eq:partred2}
\end{align}
In the last equation, we removed the index $A$ from the reduced partition function, since for a single replica $Z_A$ is just a number, i.e., the normalization, which is independent of $A$. This proves that the reduced partition function for a Gaussian theory of $n$ replicas with measurement noise is identical to the reduced partition function of the equivalent system with zero measurement noise, multiplied by a bipartition independent number.

The boson action corresponding to $Z_A(n,0)$ is obtained by setting $dW=0$ in Eqs.~\eqref{measact}, \eqref{unitaction} and then applying the bosonization. The elimination of $dW$ removes the additional coupling term of the absolute ($k=0$) mode. This yields an action 
$S=\sum_{l=1}^nS^{(l)}$, which is composed of $n$ identical, non-Hermitian theories
\begin{align}
S^{(l)}[\phi^{(l)}]=-\frac{1}{2\pi}\int_X  \sum_{\sigma=\pm}\left[\sigma \phi_{\sigma,X}^{(l)}\partial^2\phi_{\sigma,X}^{(l)}-\frac{\gamma i}{\pi v}(D\phi_{\sigma,X}^{(l)})^2\right].
\end{align}
All of these non-Hermitian theories evolve towards the same dark state $|\psi_D\rangle$, which is the same state as the dark state for the set of $n-1$  $k>0$-modes. Therefore the boson entanglement entropy is determined by the entanglement entropy in the pure state $|\psi_D\rangle$.
This yields Eq.~\eqref{eq:trace} and we can proceed readily with step (iii) as outlined above.

The entanglement entropy for the boson degrees of freedom shows the same functional dependence as for the fermion degrees of freedom, i.e., in the weak monitoring phase it grows logarithmically with the system size and it saturates for strong monitoring. However, there is a subtle difference between the fermion and the boson entanglement entropy: in the weak monitoring phase, the prefactor of the logarithmic growth increases as a function of the measurement strength. We will discuss this discrepancy in the following paragraph.

If the nonlinearity is irrelevant, the renormalization group flow approaches a Gaussian fixed point at which $\lambda=0$, and the theory is fully parameterized by the fixed point value of $K$. In this case, solving Eqs.~(\ref{eq:LuttCov}, \ref{eq:entrolutt1}) predicts the entanglement entropy and a monotonously growing prefactor
\begin{eqnarray}
S_{\text{vN}}(L)=\frac{c(\gamma)}{3}\log(L), \quad c(\gamma ) =1+\Delta c(\gamma),\label{eq:centralchargeB}
\end{eqnarray}
with $\Delta c(\gamma)\ge0$. In the limit $\gamma, \lambda \rightarrow 0$, i.e., in the absence of measurements, we have $K\rightarrow 1$ and we obtain the free fermion entanglement entropy $S=\frac{1}{3}\log(L)$ with $c(\gamma\rightarrow 0)\rightarrow 1$.

We thus approach the entanglement entropy of Dirac fermions in the ground state. For $\lambda=0$ but $\gamma>0$, the system is already initialized at a Gaussian fixed point with $K=\sqrt{1-i\frac{2\gamma}{\pi v}}$. In this case, $\Delta c(\gamma)$ behaves $\sim \gamma^2$ for small $\gamma\ll v$ and then slowly approaches an upper bound $\Delta c(\gamma\rightarrow\infty)\approx\frac{1}{3}$. For the general case $\gamma>0, \lambda>0$, the nonlinear terms lead to a renormalization of the Gaussian part of the theory, i.e., of an effective measurement strength entering $K$ at the fixed point. Generally, we observe a growth of the parameter $K$ during the renormalization group flow. The computed growth of the prefactor is very different from the predictions that we obtain from implementing the boundary conditions for the fermions first or from numerical results of similar lattice problems.

\bibliography{EntEnt}
\end{document}